# Hypergraphes aléatoires et algorithmiques

<Tsiriniaina Andriamampianina> tsr_ray@yahoo.fr
Laboratoire d'Informatique de Paris Nord (LIPN)

20 juin 2008

Je remercie Vlady Ravelomanana et Christian Lavault de leur confiance et de m'avoir offert l'opportunité de faire une thèse.

Je remercie mes parents, ma soeur et mon frère.

Je remercie Alfredo Viola pour son aide, son appui et ses conseils justes, pesés et sincères.

Je remercie Christophe Fouqueré.

Je remercie mes amis. Je remercie Raluca Andreea Schumacher qui m'a insufflé de l'entrain et plus de dynamisme dans la rédaction de ma thèse. Je remercie Sujeevan Aseeveratham pour ce défi lancé qu'est l'aboutissement d'une vraie première version de la thèse en un temps record (impossible) d'une semaine et pour toutes les aides qu'il m'a apportées, que je ne peux que mettre sous la lumière de ma très grande reconnaissance. Je remercie Mourad Hakem mon ami collègue de bureau, mon surveillant :) sur qui je peux toujours compter. Je remercie Lionnel Falempe pour sa très grande patience qui lui a permis de me donner tort de mon entêtement sur un calcul de probabilité : j'ai gagné alors en compréhension. Je remercie Jalila Sadki mon amie collègue de bureau qui me supportait anonymement et pour tous ses encouragements.

Je remercie tous ceux qui m'ont accordé une pensée pour l'aboutissement de ma thèse qui finalement n'est pas mienne.

# Table des matières

# Chapitre 1

# Introduction

Les hypergraphes, une généralisation des graphes [9] , sont des structures discrètes. Ils permettent une description et un niveau d'abstraction nécessaires pour la conception et l'analyse en algorithmique [26, 30] . Nous notons, dans ces références, le recours aux séries génératrices pour faire de l'analyse énumérative. Nous souhaitons obtenir des caractéristiques quantitatives sur les hypergraphes, pour cela nous utilisons pour l'essentiel les séries génératrices qui permettent l'énumération exacte et asymptotique selon la taille de ces structures. Karoński M. et Łuczak T. , dans [20], étudient les hypergraphes, nous nous distinguons de leur travail par l'utilisation des séries génératrices offrant une concision aux preuves.

Le plan de la thèse est le suivant :

- Dans le second chapitre, nous parlons d'énumération exacte des hypergraphes connexes.
- Dans un troisième chapitre, nous procédons à l'énumération asymptotique de ces structures.
- Et dans le quatrième chapitre, nous établissons quelques caractéristiques des hypergraphes déduites de la performance de l'algorithme glouton d'hypercouplage ou déduites du processus d'hypergraphe évoluant.

# Chapitre 2

# Énumération exacte

Dans ce chapitre, nous adoptons deux manières d'énumérer des composantes classées selon l'*excès* une relation liant le nombre d'hyperarêtes et le nombre de sommets. Une première manière est l'énumération bijective : en mettant en évidence une bijection entre les structures. Une seconde manière de procéder à l'énumération des hypergraphes est celle que nous qualifions de récursive. Une telle distinction est aussi adoptée par Wright E.M dans [22] et [23] pour les graphes, lui permettant d'un côté de justifier la forme des séries génératrices et de l'autre d'automatiser le calcul de ces séries. Avant de procéder à ces énumérations, nous précisons quelques définitions et notions.

## 2.1 Définitions et notions

**Définition 2.1.1.** Un *hypergraphe* est un couple $(V, \mathcal{E})$ , un ensemble $V$ de sommets et un ensemble $\mathcal{E}$ d'*hyperarêtes* soit de sous ensemble de $V$ .

La plupart du temps, sauf mention contraire, un hypergraphe dans cette thèse est *b-uniforme*, c'est à dire que chacune de ses hyperarêtes contient $b$ sommets. Ainsi, la structure d'hypergraphe généralise la structure de graphe qui est alors vue comme un hypergraphe 2-uniforme.

**Définition 2.1.2.** L'excès d'un hypergraphe $\mathcal{H} = (V, \mathcal{E})$ est

$$\text{exces}(\mathcal{H}) = \sum_{e_i \in \mathcal{E}} (|e_i| - 1) - |V| \,. \tag{2.1}$$

L'excès d'un hypergraphe $\mathcal{H} = (V, \mathcal{E})$ ($b$-uniforme) est

$$\text{exces}(\mathcal{H}) = (b-1)|\mathcal{E}| - |V| \,. \tag{2.2}$$

Cette notion d'excès, utilisée dans [21] , permet de classer les hypergraphes. Par exemple, nous avons les définitions suivantes :

**Définition 2.1.3.** Un *hyperarbre* est une *composante* ou hypergraphe connexe d'excès $-1$ , valeur minimum.

**Définition 2.1.4.** Un *hypercycle* est une composante d'excès $0$.

**Définition 2.1.5.** Une composante est dite *complexe* si elle est d'excès $\ell \geq 1$.

## 2.2 Énumération bijective

Dans ce chapitre, pour énumérer les hypergraphes, nous choisissons de nous focaliser aux structures connexes et de distinguer les structures selon leur excès, nous procédons alors à l'énumération des structures connexes des plus "simples" aux plus "complexes" dans le sens où les plus simples sont les hyperarbres d'excès $-1$, viennent ensuite les hypercycles d'excès 0 puis les composantes complexes d'excès $\ell \geq 1$ donné dans l'ordre croissant de ce dernier. Dans cette section, nous adoptons un point de vue bijectif en exhibant clairement une bijection ou en explicitant à travers les séries génératrices une telle bijection.

### 2.2.1 Vue bijective des hyperarbres

Dans le cas des graphes, il y a plusieurs manières d'énumérer les arbres (voir [1, 29, 12, 2, 16]) en particulier via le code de Prüfer que nous généralisons ici afin d'énumérer les forêts d'hyperarbres (ou d'arbres) enracinés.

**Définition 2.2.1.** Une forêt d'hyperarbres enracinés est un ensemble non ordonné d'hyperarbres enracinés.

**Définition 2.2.2.** Une *feuille* est un groupe de $(b-1)$ sommets (non racine dans le cas de structure marquée) de degré 1 dans une même hyperarête.

La connaissance du nombre des forêts d'hyperarbres enracinés est un outil clé pour énumérer les structures qui peuvent être décrites de manière "concise", c'est à dire que les structures sont simplifiées en élaguant récursivement les feuilles. En particulier, nous serons amenés à considérer les structures ainsi élaguées selon leur taille qui sera le nombre d'hyperarbres enracinés contenus dans la forêt définie par l'élagage.

Pour énumérer les forêts à $(k+1)$ hyperarbres enracinés, ayant $n$ sommets et $s$ hyperarêtes, nous procédons à leur codage comme un quadruplet $(R, r, \mathbb{P}, \vec{N})$ où

$$\begin{cases} R : & \text{un ensemble de } (k+1) \text{ sommets parmi } \{1, \ldots, n\} \\ r : & \text{un sommet de } R \\ \mathbb{P} : & \text{un partitionnement non ordonné de } s \text{ sous-ensembles,} \\ & \text{chacun de taille } (b-1)\text{, de } \{1, \ldots, n\} \backslash R \\ \vec{N} : & \text{un } (s-1)\text{-uplet de } \{1, \ldots, n\}^{s-1}. \end{cases} \tag{2.3}$$

Dans le code d'une forêt d'hyperarbres enracinés, nous pouvons facilement lire :
- le nombre de composantes qui n'est autre que le nombre de racines soit $|R|$,
- un sommet racine $r$ qui rattache la feuille de la dernière hyperarête dans le processus d'élagage,
- le nombre $|\mathbb{P}| = \dim(\vec{N}) + 1$ d'hyperarêtes,

**Algorithme 1** : Codage d'une forêt d'hyperarbres enracinés.

**Entrées** : Une forêt, de $(k+1)$ hyperarbres enracinés, ayant $s$ hyperarêtes et $n = s(b-1) + k + 1$ sommets.

**Sorties** : Codage $(R, r, \mathbb{P}, \vec{N})$ défini par (2.3).

**début**

$(R, r, \mathbb{P}, \vec{N}) \leftarrow (\{\text{racine}\}, r, \{\}, ())$

**répéter**

Ajouter l'ensemble des sommets de la plus petite (dans l'ordre alphabétique) feuille dans le partionnement $\mathbb{P}$ et placer le sommet qui le relie dans le tirage $\vec{N}$.

Redéfinir la forêt sans les sommets de la plus petite feuille.

**jusqu'à** *plus aucune hyperarête dans la forêt*

(Le dernier sommet placé dans le tirage est nécessairement une racine).

Définir $r$ comme le dernier sommet placer dans le tirage.

**retourner** $(R, r, \mathbb{P}, \vec{N})$

**fin**

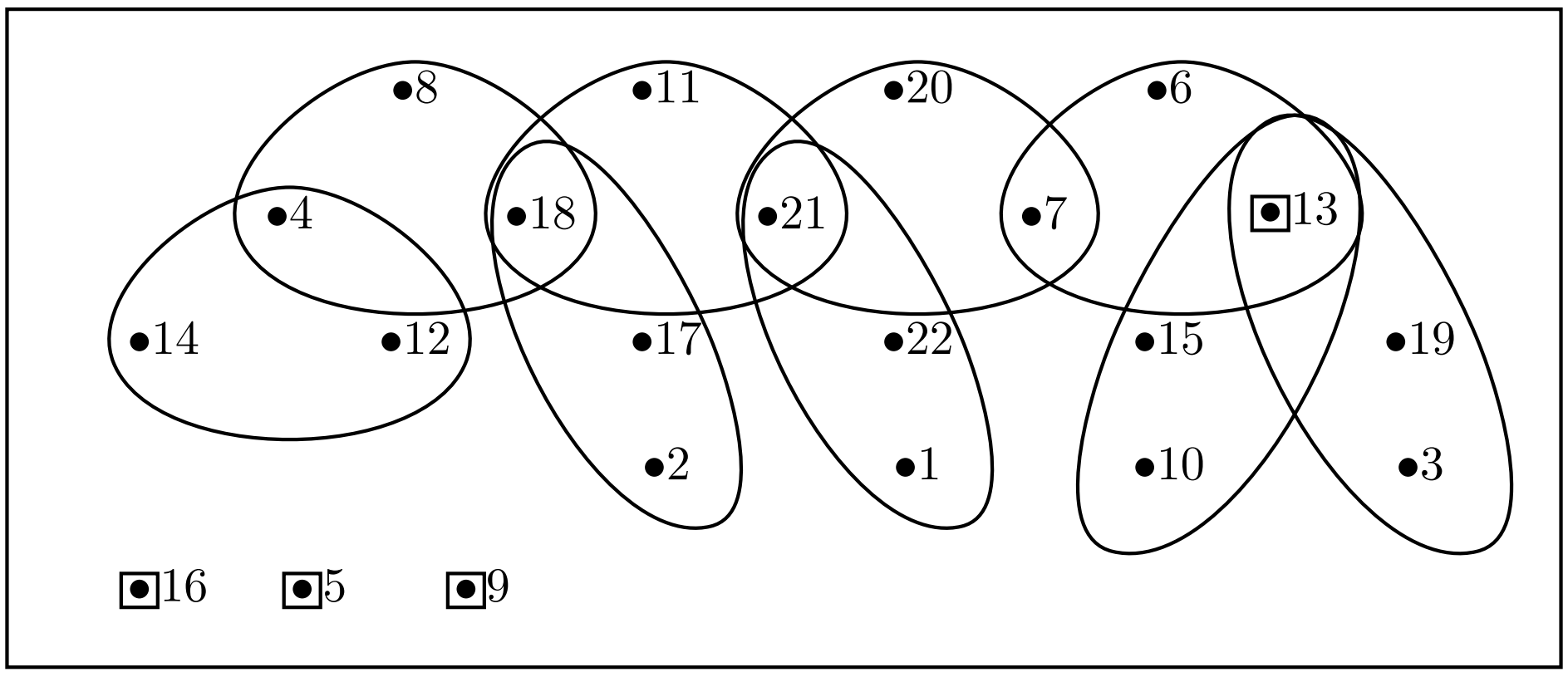

$(R, r, \mathbb{P}, \vec{N})$ :

$$\begin{cases} R = \{5, 9, 13, 16\} \\ r = 13 \\ \mathbb{P} = \{\{1, 22\}, \{2, 17\}, \{3, 19\}, \{4, 8\}, \{6, 7\}, \{10, 15\}, \{11, 18\}, \{12, 14\}, \{20, 21\}\} \\ \vec{N} = (21, 18, 13, 13, 4, 18, 21, 7)\,. \end{cases}$$

FIG. 2.1 – Une forêt d'hyperarbres enracinés et son code.

– le nombre d'hyperarbres non réduit à leur racine correspondant au nombre des racines distinctes apparaissant dans $(\vec{N}, r)$.

Dans l'algorithme 1 de codage, dans la boucle, parmi les feuilles, le choix de la plus petite feuille dans l'ordre alphabétique nous permet de retrouver sans ambigüité la forêt.

En effet, nous avons l'algorithme 2 de décodage qui retourne une forêt car lorsque les hyperarêtes sont itérativement formées, les hyperarbres existants se grandissent, se rejoignent et finissent par s'accrocher à une racine en gardant leur structure d'hyperarbre. Plus précisément, un sommet $x$ de $\vec{N}$ accroche un ensemble de la partition de $\mathbb{P}$ avec la garantie que $x$ soit plus proche d'une racine par rapport aux sommets, de l'ensemble de la partition, qui nécessairement finissent par être connectés à une racine (éventuellement $x$ est une racine).

**Algorithme 2** : Décodage en forêt d'hyperarbres enracinés.

**Entrées** : Des entiers naturels $n$, $k$ et $s$ vérifiant $n = s(b-1) + k + 1$ codage $(R, r, \mathbb{P}, \vec{N})$ défini par (2.3).

**Sorties** : Forêt d'hyperarbre enraciné, dont les sommets racines sont les $(k+1)$ sommets de $R$.

**début**

  **répéter**

    Former une hyperarête avec le premier sommet du tirage $\vec{N}$ et avec les sommets du premier (l'ordre étant celui induit par les étiquettes) ensemble du partitionnement $\mathbb{P}$ ne contenant aucun sommet qui apparaît encore dans le tirage restant.

    Supprimer du partitionnement l'ensemble utilisé et supprimer de la liste du tirage le premier sommet.

  **jusqu'à** *liste du tirage vide*

  Former avec le dernier sous-ensemble du partitionnement et avec le sommet $r$ le $s$-ième hyperarête.

  **retourner** *la forêt obtenue*

**fin**

**Théorème 2.2.3.** *Le nombre de forêts de $(k+1)$ hyperarbres enracinés ayant $s$ hyperarêtes est :*

$$\binom{n}{k+1}(k+1)\frac{(n-k-1)!}{[(b-1)!]^s s!}n^{s-1}\,, \tag{2.4}$$

*avec le nombre de sommets $n = n(s) = s(b-1) + k + 1$.*

*Preuve.* La preuve découle directement de la bijection que définit l'algorithme 1 de codage de telles forêts par les ensembles de quadruplets $(R, r, \mathbb{P}, \vec{N})$ défini en (2.3). ♢

En fixant $k = 0$ dans ce théorème, nous obtenons

**Corollaire 2.2.4.** *Le nombre d'hyperarbres enracinés ayant $s$ hyperarêtes :*

$$\frac{(n-1)!}{[(b-1)!]^s s!}n^s\,, \tag{2.5}$$

*avec le nombre de sommets $n = n(s) = s(b-1) + 1$.*

Il est immédiat que ce résultat généralise le résultat dans le cas des arbres en prenant $b = 2$.

**Corollaire 2.2.5** (Cayley)**.** *Le nombre d'arbres enracinés ayant $n$ sommets est $n^{n-1}$.*

Une avantage que présente une telle démonstration bijective est la possibilité d'effectuer une génération aléatoire afin d'apprendre quelques caractéristiques des structures étudiées. Nous laissons cela comme perspective, pour le moment. Grâce au théorème 2.2.3, nous sommes aussi en mesure de donner une expression explicite du nombre des hypercycles.

### 2.2.2 Vue bijective des hypercycles

Les hypercycles sont des composantes les plus simples après les hyperarbres dans le sens où ces structures sont d'excès $0$. L'énumération des graphes unicycles est associée à Alfred RÉNYI dans [3]. Dans le cas plus général des hypergraphes unicycles, pour procéder à l'énumération de ces structures (la même idée est applicable pour les composantes complexes à excès $\ell \geq 1$ fixé), afin d'utiliser le théorème 2.2.3, les structures sont récursivement élaguées jusqu'à ce qu'il n'y ait plus aucune feuille et obtenir ainsi des *structures lisses*. Ainsi, à un hypercycle ayant $n$ sommets, nous ferons correspondre de manière naturelle un couple $(\mathcal{B}, \mathcal{F})$ avec $\mathcal{B}$ une structure lisse de taille $(k+1)$ et $\mathcal{F}$, une forêt de $(k+1)$ hyperarbres enracinés ayant $n$ sommets, obtenue à partir du processus d'élagage. Les étiquettes de $\mathcal{B}$ seront canonisées dans $\{1, \ldots, k+1\}$ en respectant l'ordre croissante des étiquettes.

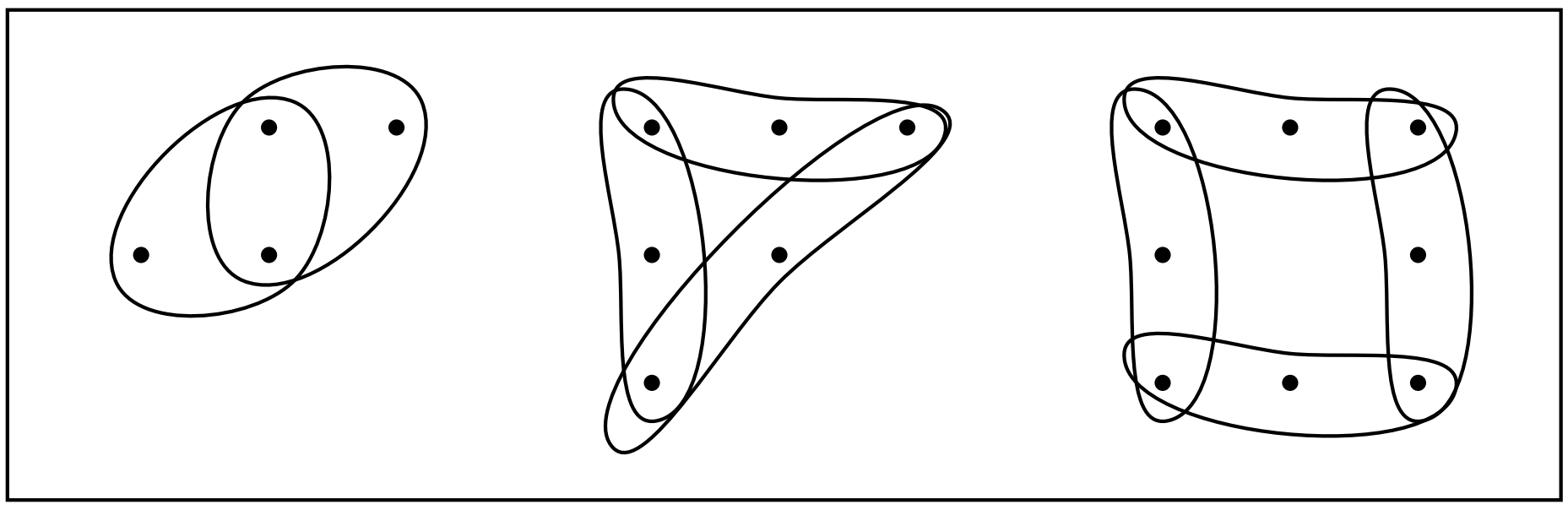

FIG. 2.2 – Hypercycles 3-uniformes lisses non étiquetés de longueur respectivement $2$, $3$ et $4$.

**Remarque 2.2.6.** Comme le processus d'élagage consiste à supprimer les sommets des feuilles ainsi que les hyperarêtes qui les accrochaient, il produit une structure de même excès.

Nous retrouvons le résultat, associé à Selivanov dans [17], pour l'énumération des hypercycles :

**Théorème 2.2.7.** *Le nombre d'hypercycles ayant s hyperarêtes est*

$$\frac{n!n^{s-1}(b-1)}{2[(b-1)!]^s}\sum_{j=2}^{s}\frac{j}{s^j(s-j)!}\,, \tag{2.6}$$

*avec le nombre de sommets* $n = n(s) = s(b-1)$ .

*Preuve.* Un hypercycle de longueur de cycle $j$ correspond à une forêt de $j(b-1)$ hyperarbres enracinés à un arrangement près de ces derniers pour les différentes façons de former le cycle. La forêt aurait $(s-j)$ hyperarêtes et $j(b-1)$ composantes à arranger en cycle. Le nombre d'hypercycles ayant une longueur $j$ de cycle et ayant $s$ hyperarêtes est donc

$$\left\{\binom{n}{j(b-1)}j(b-1)\frac{[(s-j)(b-1)]!}{[(b-1)!]^{s-j}(s-j)!}n^{s-j-1}\right\}\left\{\frac{1}{2}\frac{[j(b-1)]!}{[(b-2)!]^j}\right\}, \tag{2.7}$$

avec $n = n(s) = s(b-1)$ : le premier facteur entre accolade dénombre des forêts d'hyperarbres enracinés et le second dénombre les hypercycles lisses étiquetés avec $\{1,\ldots,j(b-1)\}$ ayant $j$ hyperarêtes. En simplifiant cette équation, nous trouvons le terme de la sommation du théorème, et comme la longueur de cycle peut prendre toute valeur entre 2 et $s$ , nous obtenons le résultat en sommant sur $j$ . ♢

Ainsi, pour faire la preuve du théorème nous étions amenés à distinguer les hypercycles selon la longueur du cycle. Une question à poser est : pour un nombre de sommets $n = s(b-1)$ , quelle est la longueur $j$ de cycle de la classe (selon $j$) qui contribue le plus au nombre des hypercycles ? Nous laisserons cette question en perspective.

Par rapport au résultat d'énumération des hyperarbres (voir des forêts) , l'expression du nombre des hypercycles est beaucoup plus complexe car requiert la sommation. Néanmoins, l'expression de ce nombre reste explicite. Expliciter le nombre des composantes complexes paraît être une tâche ardue.

Dans la suite, nous précisons comment déterminer le nombre de composantes, en décrivant la bijection via les séries génératrices qui permettent une lecture directe des opérations combinatoires sur des structures.

### 2.2.3 Introduction aux séries génératrices exponentielles

À une séquence de nombre $(a_n)_{n\in\mathbb{N}}$ , nous associons la série génératrice exponentielle (SGE)

$$A(z) = \sum_{n=0}^{\infty} a_n \frac{z^n}{n!}\,. \tag{2.8}$$

Les SGEs servent pour l'énumération de structures avec une étiquette propre à chaque sommet et l'indice $n$ , dans l'écriture ci-dessus, définit la taille de la structure étiquetée. Une grande avantage de l'utilisation des SGEs est la facilité de lecture qu'elles offrent : à partir des opérations sur les séries, nous sommes en mesure d'intérpréter en terme d'opérations sur les structures et inversement. [26] donne le dictionnaire suivant pour faire cette lecture :

| Opérations sur les structures | SGE correspondant |
| --- | --- |
| $\mathcal{A} \cup \mathcal{B}$ | $A(z) + B(z)$ |
| $\mathcal{A} \times \mathcal{B}$ | $A(z) \cdot B(z)$ |
| Substituer dans $\mathcal{A}$ par $\mathcal{B}$ | $A \circ B(z)$ |
| Séquence de $\mathcal{A}$ | $(1 - A(z))^{-1}$ |
| Groupe de k $\mathcal{A}$ | $\frac{A(z)^k}{k!}$ |
| Ensemble de $\mathcal{A}$ | $\exp(A(z))$ |
| Cycle de $\mathcal{A}$ | $-\ln\left(\sqrt{1 - A(z)}\right)$ |
| Marquage de $k$ sommets de $\mathcal{A}$ | $\frac{z^k}{k!}\frac{\mathrm{d}^k}{\mathrm{d}\,z^k}A(z)$ |

Nous nous servirons aussi de l'opérateur sur les séries :

$$[z^n]\,A(z) = \frac{a_n}{n!}\,, \tag{2.9}$$

avec $A(z)$ défini à (2.8). Familiarisons nous avec l'utilisation des SGEs à travers les exemples de structures rencontrées jusqu'ici.

**Définition 2.2.8.** La SGE $T$ des hyperarbres enracinés est

$$T(z) = \sum_{s \geq 0} \frac{(n-1)!}{[(b-1)!]^s s!} n^s \frac{z^n}{n!}\,, \tag{2.10}$$

où $n = n(s) = s(b-1) + 1$.

Nous y lisons le nombre des hyperarbres enracinés à $s$ hyperarêtes

$$n!\,[z^n]\,T(z) = \frac{(n-1)!}{[(b-1)!]^s s!} n^s\,. \tag{2.11}$$

**Remarque 2.2.9.** La SGE qui énumère les hyperarbres s'écrit sous la forme $H_{-1} \circ T(z)$. Nous réservons la justification pour plus tard.

Un marquage d'un sommet différencie les hyperarbres des hyperarbres enracinés :

$$T(z) = z \frac{\mathrm{d}}{\mathrm{d}\,z} H_{-1} \circ T(z)\,. \tag{2.12}$$

Un cycle de sommets de longueur $\geq 2$ admet la SGE

$$-\ln(\sqrt{1-z}) - z/2\,. \tag{2.13}$$

Pour un cycle de sommet avec un ensemble de $(b-2)$, structure de SGE $z^{b-1}/(b-2)!$, la substitution dans l'équation précédente donne

$$-\ln\left(\sqrt{1 - \frac{z^{b-1}}{(b-2)!}}\right) - \frac{z^{b-1}}{2(b-2)!}\,. \tag{2.14}$$

Et comme c'est la SGE des hypercycles lisses, pour obtenir la SGE des hypercycles, il faut faire la substitution par la SGE des hyperarbres enracinés. De cette manière, nous obtenons :

**Remarque 2.2.10.** Si $b \geq 3$, la SGE des hypercycles est $H_0 \circ T(z)$ avec

$$H_0(t) = -\ln\left(\sqrt{1 - \frac{t^{b-1}}{(b-2)!}}\right) - \frac{t^{b-1}}{2(b-2)!}. \tag{2.15}$$

Nous pouvons lire en sens inverse la bijection qui nous avait permis de procéder à l'énumération des hypercycles dans la fonction $H_0$. Cette fonction offre aussi la possibilité de déterminer le nombre $n!\,[z^n]\,H_0 \circ T(z)$ d'hypercycles de taille $n$.

Comme il a été déjà évoqué, il est possible d'appliquer, comme dans le cas des hypercycles, un raisonnement proche pour faire de l'énumération des composantes complexes. C'est ce que nous allons développer dans la prochaine sous-section.

### 2.2.4 Vue bijective des composantes complexes

Dans cette sous-section, pour énumérer les composantes complexes, nous renonçons à établir une expression explicite comme dans le cas des forêts d'hyperarbres enracinés ou comme dans le cas des hypercycles. Nous y établissons la SGE des composantes complexes d'excès donné, par des arguments bijectifs comme pour (2.15). Dans le cas de l'énumération des composantes de graphe complexes, l'énumération des composantes ayant deux cycles (d'excès 1) est associé à Bagaev dans [25]. Plus généralement, un résultat pour les composantes de graphes d'excès $\ell \geq 1$ est associé à Wright dans [22].

Ici, nous énumérons des composantes (d'hypergraphe) complexes d'excès $\ell \geq 1$. Pour procéder à cette énumération, une première étape consiste à se ramener à des composantes lisses par élagage des feuilles des structures à énumérer. Les composantes lisses ainsi obtenues ont le même excès que les structures de départ.

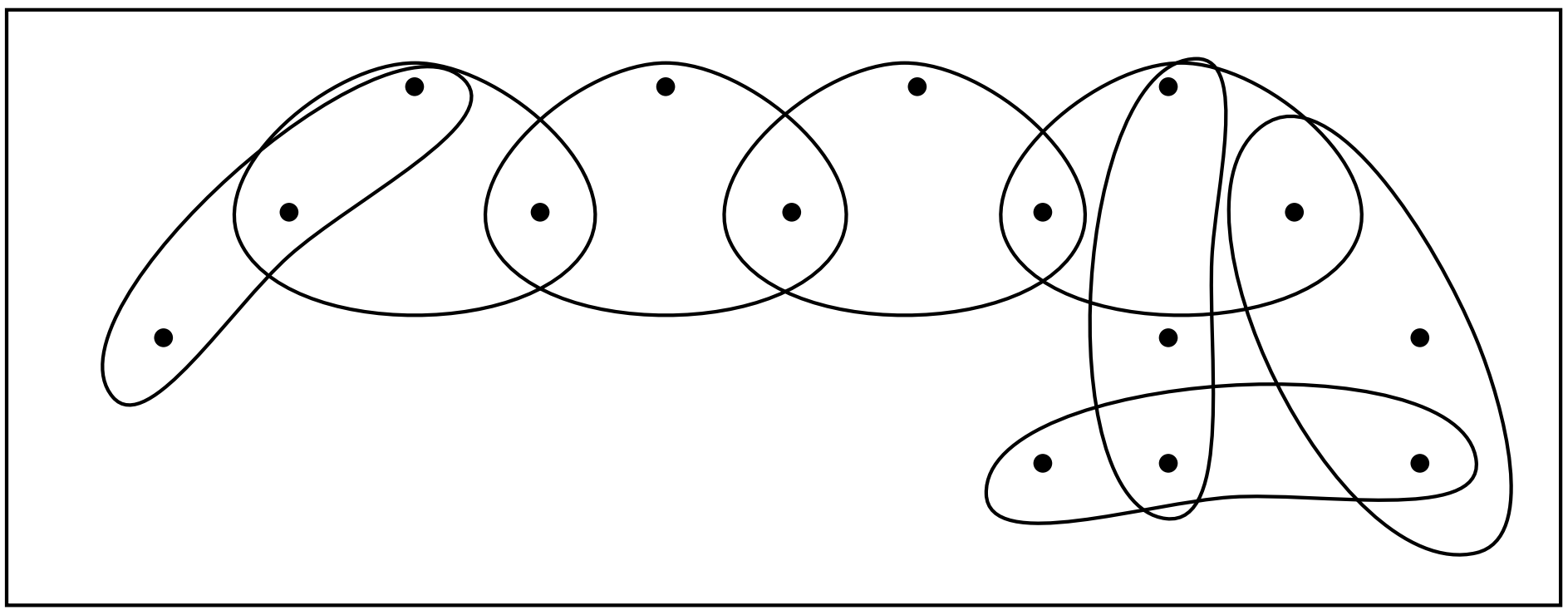

FIG. 2.3 – Une composante lisse non étiquetée d'excès $\ell = 1$.

Nous convenons d'utiliser la SGE avec la variable $t$ pour énumérer des structures lisses, comme dans (2.15). Et dans le but d'alléger les équations, notons

$$\tau(t) = \frac{t^{b-1}}{(b-2)!}. \tag{2.16}$$

(2.15) devient alors

$$H_0(t) = -\ln\left(\sqrt{1-\tau(t)}\right) - \frac{\tau(t)}{2}\,. \tag{2.17}$$

Les composantes complexes lisses d'excès $\ell$ peuvent avoir toute taille positive (entière). C'est aussi le cas pour les hypercycles et nous avions eu recours à la fonction ln du dictionnaire pour exprimer la structure cyclique. Dans le cas des composantes complexes, c'est l'opération de séquence du dictionnaire qui nous permettra de décrire les structures de manière "plus concise" car pour un excès $\ell$ donné nous aurons seulement un nombre fini de structures à considérer pour tout énumérer.

**Définition 2.2.11.** Une *chaîne* est une séquence de sommets avec un groupe de $(b-2)$ sommets. La séquence peut être vide et admet pour SGE

$$\frac{1}{1-\tau(t)}\,. \tag{2.18}$$

Une chaîne, dans cette thèse, est d'abord une sous-structure d'une composante lisse. Dans le cas des graphes, Wright E.M dans [22] distingue trois types de chaîne

- celle qui boucle sur un même sommet,
- celle qui, si brisée, déconnecte la structure en deux,
- celle qui ne boucle pas sur un même sommet et, qui si brisée, ne rompt pas la connexité de la composante.

Une chaîne du premier (respectivement du second (respectivement du troisième)) type pour des graphes sans multi-arête admet au moins deux arêtes (respectivement une arête (respectivement deux arêtes)).

**Définition 2.2.12.** Une *structure basique* est une structure lisse non étiquetée ayant ses chaînes de longueur finie.

Rappelons ici l'idée clé dans [22] de l'énumération bijective des graphes connexes d'excès $\ell \geq 1$ : à partir des composantes, une réduction est faite par élagage récursive, puis les chaînes sont effondrées jusqu'à leur taille minimale et enfin, les étiquettes des sommets sont ignorées pour obtenir un nombre fini de composantes basiques distinctes (structures non étiquetées) qui permettent de procéder à une énumération bijective (une approche très rapidement impraticable, en particulier à cause du calcul des nombres d'automorphismes).

**Remarque 2.2.13.** Comme pour l'élagage, effondrer une hyperarête d'une chaîne en un de ses sommets de degré $\geq 2$ ne change pas l'excès de la structure.

**Définition 2.2.14.** Une *hyperarête spéciale* est une hyperarête contenant au moins 3 sommets au moins de degré 2.

**Remarque 2.2.15.** Une hyperarête spéciale ne peut pas faire partie d'une chaîne.

**Définition 2.2.16.** Un *sommet spécial* est soit un sommet d'une hyperarête spéciale soit un sommet de degré au moins 3.

Ici, nous ne distinguerons dans les structures lisses que deux types de chaînes :

**Définition 2.2.17.** Une *$\alpha$-chaîne* est une chaîne qui boucle sur un même sommet. Une telle chaîne admet au moins deux hyperarêtes.

**Définition 2.2.18.** Une *$\beta$-chaîne* est une chaîne qui lie deux sommets spéciaux. Une telle chaîne admet au moins une hyperarête.

Soit une composante $B$ basique donnée avec les caractéristiques suivantes :

$$\begin{cases} \text{– d'excès } \ell \geq 1 \\ \text{– ayant } m \text{ sommets (non étiquetés)} \\ \text{– ayant } s \text{ hyperarêtes, nécessairement } s \geq \lfloor \frac{\ell+1}{b-1} + 1 \rfloor \\ \text{– ayant } c_\alpha\ \alpha\text{-chaînes} \\ \text{– ayant } c_\beta\ \beta\text{-chaînes} \\ \text{– ayant } p = c_\alpha + c_\beta \text{ chaînes} \\ \text{– de nombre (rationnel) d'automorphismes } g\,. \end{cases} \tag{2.19}$$

Alors les hypergraphes lisses (étiquetés) contenant la composante $B$ à des effondrements près de ses chaînes, admettent la SGE

$$J(t) = \frac{1}{g} \frac{t^m}{(1-\tau(t))^p}\,. \tag{2.20}$$

La lecture de cette équation est la suivante :

1. Fixer un étiquetage des $m$ sommets, c'est le facteur $t^m$ .
2. Regrouper les sommets ainsi étiquetés dans la forme de la structure $B$, c'est le facteur $1/g$ .
3. Éventuellement, rallonger les $p$ chaînes dans l'ordre induite des étiquetages, c'est le facteur $(1-\tau(1))^{-p}$ .

Avec les caractéristiques de $B$ , comme $m = s(b-1) - \ell$, une réécriture de (2.20) nous obtenons :

**Lemme 2.2.19.** *La SGE des composantes lisses (étiquetés) contenant la composante $B$ basique à des effondrements près de ses chaînes :*

$$J(t) = \frac{[(b-2)!]^s}{g} \frac{\tau(t)^s}{t^\ell(1-\tau(t))^p}\,. \tag{2.21}$$

L'énumération des composantes lisses complexes d'excès donné se réduit, ainsi grâce à ce lemme, à la détermination d'un ensemble de structures basiques permettant de les générer sans ambigüité, c'est à dire qu'une composante ne peut être générée qu'à partir d'une unique structure basique. Wright E.M obtient un ensemble de structures qui convient en considérant l'ensemble fini de toutes les structures basiques d'excès $\ell$ n'ayant que des chaînes de longueur minimale selon leur type, et précise ainsi la forme des SGEs des graphes connexes complexes lisses d'excès $\ell$ . Ici, nous faisons le choix d'un autre ensemble de structures basiques pour montrer que

**Théorème 2.2.20.** *La SGE $H_\ell$ des composantes lisses d'excès $\ell$ est :*

$$H_\ell(t) = \frac{\tau(t)^{r_\ell}}{t^\ell} \sum_{p=0}^{3\ell} A_{\ell,p} \left( \frac{\tau(t)}{1-\tau(t)} \right)^p , \tag{2.22}$$

*avec les coefficients $A_{\ell,p}$ , des fractions rationnelles positives de $b$ à coefficients entiers,*

$$r_\ell = \lfloor \frac{\ell+1}{b-1} + 1 \rfloor , \tag{2.23}$$

$$\tau(t) = \frac{t^{b-1}}{(b-2)!} . \tag{2.24}$$

La forme (2.22) de la SGE $H_\ell$ des composantes complexes d'excès $\ell$ donne une bonne intuition d'un terme principal qui contribue le plus au nombre asymptotique de ces structures. En particulier, ce terme est lié à l'indice $p = 3\ell$ et, entre autres références, dans [19] , il est lié aux structures qualifiées de "clean" maximisant le nombre de chaînes. Notons dans ce théorème, le résultat combinatoire de la positivité des coefficients $A_{\ell,p}$ : la preuve est bijective et une preuve par induction sur $(\ell, p)$ est restée hors de notre portée.

D'abord une conséquence immédiate du lemme 2.2.19 est que

**Remarque 2.2.21.** La SGE $H_\ell$ des hypergraphes connexes lisses d'excès $\ell$ s'écrit sous la forme

$$H_\ell(t) = \frac{f_\ell \circ \theta(t)}{t^\ell} , \tag{2.25}$$

avec

$$\theta(t) = 1 - \tau(t) . \tag{2.26}$$

et $f_\ell$ un polynôme de Laurent.

Ensuite pour prouver le théorème 2.2.20 ,

**Lemme 2.2.22.** *Dans une composante complexe basique d'excès $\ell$ , le nombre $p$ de chaînes est au plus $3\ell$ soit $p \leq 3\ell$ .*

*Preuve.* Il suffit de voir la preuve dans le cas d'une composante n'ayant que des chaînes de longueur minimale. Soit donc $B$ une telle composante complexe avec les caractéristiques suivantes :

$$\begin{cases} \text{– d'excès } \ell \geq 1 \\ \text{– ayant } s_0 \text{ hyperarêtes spéciales} \\ \text{– ayant } m_0 \text{ sommets spéciaux} \\ \text{– ayant } c_\alpha\ \alpha\text{-chaînes (toutes de longueur 2)} \\ \text{– ayant } c_\beta\ \beta\text{-chaînes (toutes de longueur une)} \\ \text{– ayant } p = c_\alpha + c_\beta \text{ chaînes.} \end{cases} \tag{2.27}$$

Par définition de l'excès,

$$m_0 + 2(b-1)c_\alpha - c_\alpha + (b-2)c_\beta + \ell = (b-1)(s_0 + 2c_\alpha + c_\beta) \tag{2.28}$$

soit

$$m_0 - s_0(b-1) + \ell = c_\alpha + c_\beta = p\,. \tag{2.29}$$

Notons $B_0$ l'hypergraphe induit par les sommets spéciaux et soit $\ell_0$ son excès alors l'équation précédente s'écrit :

$$-\,\ell_0 + \ell = p\,. \tag{2.30}$$

La valeur de l'excès $\ell_0$ est minimale dans le cas où $B_0$ est une forêt maximisant le nombre de composantes, $B_0$ devant donc se composer de sommets isolés de degré 3 ou d'hyperarêtes ayant exactement 3 sommets de degré 2 dans l'hypergraphe basique dont ils sont issus. Comme il existe un hypergraphe basique d'excès $\ell$ (voir la figure 2.4 ) ayant $2\ell$ sommets spéciaux et sans aucune hyperarête spéciale pour laquelle la valeur $\ell_0 = -2\ell$ est minimale, nous déduisons

$$-\,\ell_0 + \ell \geq 2\ell + \ell = 3\ell \geq p\,. \tag{2.31}$$

Et cette borne supérieure est atteinte. ♢

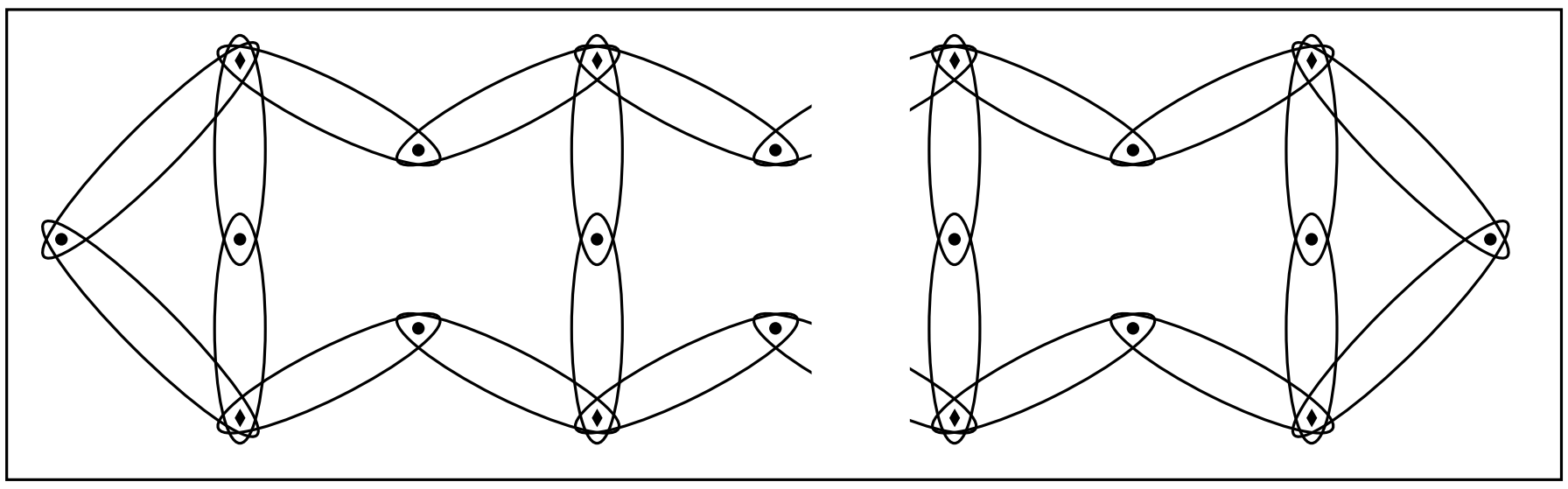

FIG. 2.4 – Structure d'un hypergraphe sans hyperarête spéciale ayant le nombre $3\ell$ maximal de chaînes dans le cas d'excès $\ell$. Seuls les sommets de degré $\geq 3$ y sont représentés.

Ainsi, dans (2.22) , l'indice $p$ est liée au nombre de chaînes d'une structure basique. Quelle serait une interprétation du degré maximal du polynôme $f_\ell$ de Laurent dans la remarque 2.2.21 ? Dans l'équation (2.22) , une interprétation intuitive du degré de $\tau(t)$ est qu'il compte le nombre minimal d'hyperarêtes dans une composante lisse.

**Lemme 2.2.23.** *Le degré maximum du polynôme $f_\ell$ de Laurent, dans la remarque 2.2.21 , est au plus*

$$r_\ell = \lfloor \frac{\ell+1}{b-1} + 1 \rfloor\,. \tag{2.32}$$

*Preuve.* La démonstration passe par la décomposition des composantes lisses d'excès $\ell$ suivante : À une composante d'excès $\ell$ donné, supprimer la plus petite hyperarête dans l'ordre alphabétique induite par l'étiquetage, et obtenir ainsi des composantes éventuellement non lisses avec les caractéristiques $(\ell_i, k_i)$ c'est à dire d'excès $\ell_i$ et ayant $k_i$ sommets appartenant à la plus petite hyperarête supprimée. Des nombres de cycles nous déduisons

$$\ell + 1 = \sum_i (\ell_i + k_i)\,. \tag{2.33}$$

Notons alors les divisions euclidiennes par $(b-1)$ suivantes :

$$\ell_i + k_i = q_i(b-1) + r_i \tag{2.34}$$

et

$$\ell + 1 = q(b-1) + r\,. \tag{2.35}$$

Ainsi,

$$q(b-1) + r = \sum_i \{q_i(b-1) + r_i\} \tag{2.36}$$

et

$$q + 1 \geq \sum_i q_i + 1\,. \tag{2.37}$$

En remarquant que le degré maximum de $f_\ell$ s'interprète comme le nombre minimum d'hyperarêtes, nous avons la démonstration par induction car le second membre de l'équation précédente maximise le degré de $f_\ell$ en prenant une composante ayant un nombre minimum d'hyperarêtes. ♢

La forme du SGE $H_\ell$ (2.22) suggère l'existence d'un ensemble de composantes basiques, d'excès $\ell$ ayant $s = r_\ell, \ldots, r_\ell + 3\ell$ hyperarêtes et dans laquelle seules $s - r_\ell$ chaînes peuvent être insérées, permettant alors de générer toutes les structures lisses. Le choix de prendre les composantes basiques n'ayant que des chaînes de longueur minimale, bien que ce soit la justification immédiate de la remarque 2.2.21 , ne permet pas d'aboutir au résultat car le nombre d'hyperarêtes peut excéder strictement le nombre des chaînes plus $r_\ell$.

La condition qu'une $\beta$-chaîne doit au moins contenir une hyperarête est motivée par la lecture plus facile des emplacements où les chaînes seront introduites. Aussi, en considérant l'ensemble des composantes basiques n'ayant que des $\alpha$-chaînes qui elles mêmes sont de longueur 2, toutes les composantes lisses peuvent être générées sans ambigüité. Toutefois, la condition d'avoir autant d'hyperarêtes que de chaînes plus $r_\ell$ n'est pas une conditionalité respectée à priori dans la façon canonique dont est faite la génération avec cet ensemble. Un plus apporté par la considération d'un tel ensemble est qu'il garantit maintenant un nombre suffisamment peu élevé d'hyperarêtes dans les composantes considérées d'après les 2 lemmes qui suivent.

**Lemme 2.2.24.** *Pour une composante basique d'excès $\ell$ ayant $s_0$ hyperarêtes spéciales et $c_\alpha$ $\alpha$-chaînes*

$$2c_\alpha + s_0 \leq 3\ell + 1. \tag{2.38}$$

*Il existe une composante basique telle que l'égalité est atteinte.*

*Preuve.* Pour maximiser la quantité $2c_\alpha + s_0$ dans une composante basique d'excès $\ell$ donné, $c_\alpha$ est à maximiser en priorité puis seulement après $s_0$. La relation qui lie le nombre des $\alpha$-chaînes et l'excès est $c_\alpha \leq \ell + 1$ et parmi les composantes basiques telles que $c_\alpha = \ell + 1$ au plus $s_0 = \ell - 1$, nous déduisons ainsi le lemme. ♢

Une conséquence directe de ce lemme est :

**Lemme 2.2.25.** *Dans une composante basique d'excès $\ell$ n'ayant que des $\alpha$-chaînes toutes de longueur 2 , le nombre d'hyperarêtes est au plus* $(3\ell + 1)$ *.*

*Preuve.* (Théorème 2.2.20) . Les composantes lisses peuvent être générées sans ambigüité par des composantes basiques ayant $s = r_\ell, \ldots, 3\ell + r_\ell$ hyperarêtes, $c_\alpha$ $\alpha$-chaînes toutes de longueur 2 et $c_\beta$ $\beta$-chaînes toutes de longueur 1 et $s - c_\alpha - c_\beta$ marquages indiquant les emplacements où d'autres chaînes peuvent être insérées. Notons que la quantité $(s - c_\alpha - c_\beta)$ doit être positive et que les emplacements correspondent à des sommets où une $\beta$-chaîne aurait pu être effondrée. Notons aussi que ces composantes basiques ne doivent pas être toutes prises car dans ce cas, une même composante lisse étiquetée peut être obtenue par différentes composantes basiques. En bref, l'idée clé de la preuve : D'un coté, il y a l'ensemble de composantes basiques avec les chaînes de longueur minimale pour leur type et, de l'autre coté, il y a l'ensemble de composantes basiques avec que des $\alpha$-chaînes de longueur 2 qui permettent de générer de manière canonique toutes les composantes lisses. Entre les deux, il y a plusieurs choix d'ensemble de composantes basiques pouvant servir à générer toutes les composantes lisses . En particulier, il est possible de générer les composantes lisses à partir de composantes basiques ayant $s = r_\ell, \ldots, 3\ell + r_\ell$ hyperarêtes en faisant grandir $s - r_\ell$ chaînes aux emplacements des $\alpha$-chaînes toutes de longueur 2, des $\beta$-chaînes toutes de longueur 1 et des sommets sur lesquels une $\beta$-chaîne aurait pu être effondrée. ♢

Dans cette section, nous avons vu le nombre des hyperarbres enracinés ainsi que celui des hypercycles. Quant au nombre des composantes complexes, nous pouvons le déterminer du moins théoriquement car nous savons déterminer les SGEs $H_\ell \circ T(z)$ des composantes d'excès $\ell$ donné : nous connaissons $T(z)$ et nous savons comment obtenir $H_\ell(t)$ . Cependant, la manière vue jusqu'ici pour déterminer ces SGEs est très rapidement rebutante et très difficile à automatiser, particulièrement à cause de la détermination des composantes basiques dont pour chacune il faudrait déterminer son nombre d'automorphismes ce qui nous paraît peu concevable et surtout parce qu'il faut aussi pouvoir garantir avoir considéré toutes les composantes basiques nécessaires, ni une de plus ni une de moins. Nous tirerons profit de la forme des SGEs $H_\ell$ apprise dans cette section pour justifier comment les déterminer récursivement dans la section qui suit.

## 2.3 Énumération récursive

Ayant perçu la puissance d'expression de l'outil que sont les SGEs, dans la présente section une étape supplémentaire est franchie pour l'énumération des composantes en établissant une récurrence dont la résolution permet le calcul automatisable des SGEs $H_\ell$ des composantes d'excès $\ell$ .

Une brève aperçue du contenu de cette section est la suivante :
- elle commence par une vision récursive de l'énumération des hyperarbres enracinés ;
- un outil pour la détermination des coefficients des SGEs, la formule d'inversion de Lagrange, est présenté avec une interprétation combinatoire ;
- puis nous faisons une introduction aux SGEs bivariées qui servent pour une lecture facile de

– la décomposition des composantes d'excès $\ell$ qui mène à une récurrence sur les SGEs univariées
– ensuite la détermination des $H_\ell$ ou la résolution de la récurrence est présentée ;
– et enfin nous terminons par la présentation de deux identités combinatoires permettant d'obtenir les $H_\ell$ sous la forme, permettant une lecture bijective, annoncée dans la section précédente.

### 2.3.1 Énumération récursive des hyperarbres enracinés

Les hyperarbres enracinés sont des structures clés pour les hypergraphes : les SGEs des composantes connexes d'excès $\ell$ s'expriment en fonction de la SGE $T(z)$ des hyperarbres enracinés. Le dictionnaire, introduit dans la section précédente, permet de traduire une décomposition naturelle des hyperarbres enracinés comme étant composés d'une racine et d'un ensemble de groupes de $(b-1)$ hyperarbres enracinés. Cette décomposition est une définition récursive des hyperarbres enracinés. Effectivement, un hyperarbre se définit à partir d'hyperarbres de taille strictement plus petite. Cette définition récursive des hyperarbres enracinés se lit dans une définition implicite de SGE via le dictionnaire.

**Définition 2.3.1.** La SGE $T(z)$ des hyperarbres enracinés :

$$T(z) = z \exp\left(\frac{T(z)^{b-1}}{(b-1)!}\right). \tag{2.39}$$

La lecture de cette définition est la suivante : un hyperarbre enraciné est
– une racine, c'est le facteur $z$
– et un ensemble (c'est la fonction exp ) non ordonné de groupes de $(b-1)$ hyperarbres enracinés (c'est l'exposant $\frac{T(z)^{b-1}}{(b-1)!}$ ), bref c'est le terme

$$\exp\left(\frac{T(z)^{b-1}}{(b-1)!}\right). \tag{2.40}$$

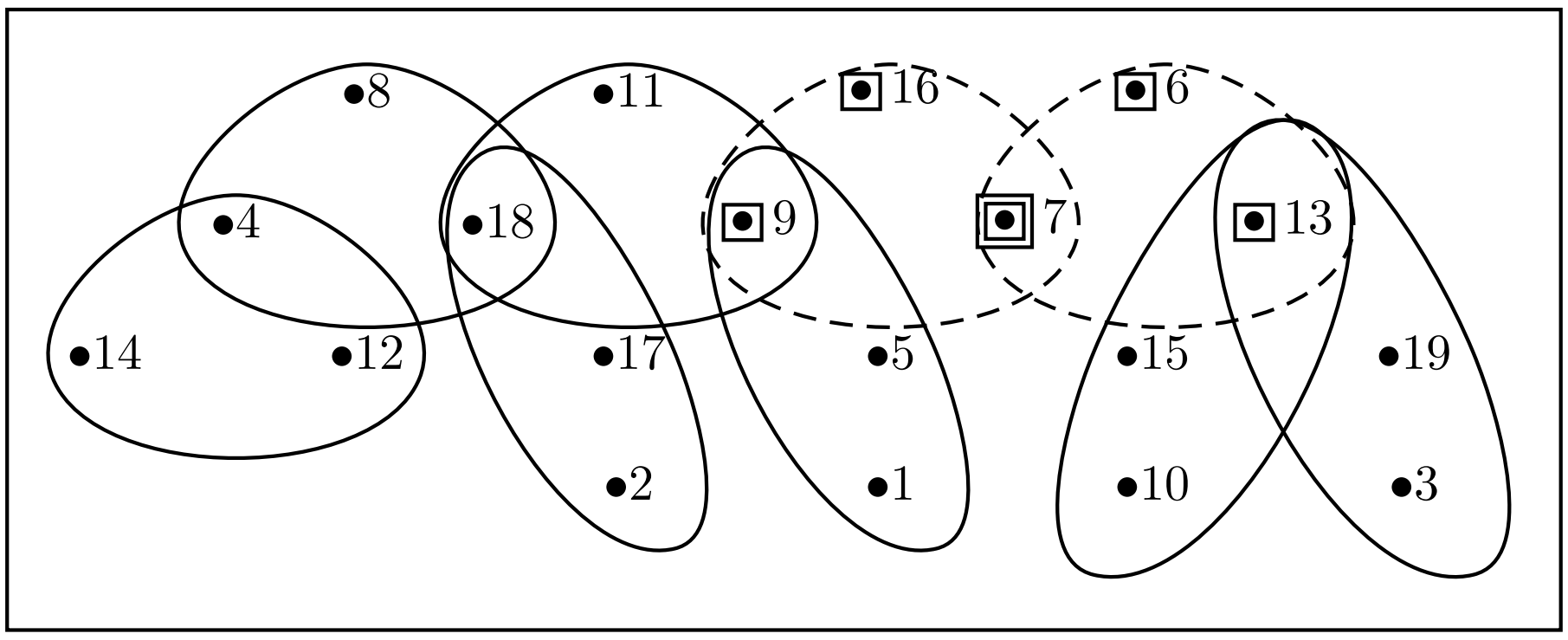

Fig. 2.5 – Un hyperarbre enraciné décomposé.

Par cette définition, il est possible de déterminer récursivement les coefficients $[z^n]\,T(z)$ qui sont les nombres des hyperarbres enracinés à un facteur $n!$ près. Évidement, cette façon de calculer est à proscrire dans ce cas puisque les coefficients sont connus de manière explicite : l'expression mathématique (2.5) pouvant être retrouvée par la formule d'inversion de Lagrange via cette définition implicite de $T(z)$ .

### 2.3.2 Interprétation combinatoire de la formule d'inversion de Lagrange

La formule d'inversion de Lagrange est un outil puissant pour faire de l'énumération de structures arborescentes. La preuve de la formule d'inversion de Lagrange dans [30, page 167] est une preuve analytique : l'extraction des coefficients des SGEs y est exprimée avec la formule intégrale de Cauchy. Cette vision analytique de la formule d'inversion de Lagrange sera notre point de départ pour obtenir des équivalents asymptotiques des coefficients dans le chapitre 3 . Ici, nous adoptons un regard combinatoire dans la preuve du théorème :

**Théorème 2.3.2.** *Notons la SGE G*

$$G(t) = \sum_{k=0}^{\infty} g_k \frac{t^k}{k!}\,, \tag{2.41}$$

*alors par la définition de la SGE $T(z)$, la formule d'inversion de Lagrange s'écrit*

$$[z^n]\,G \circ T(z) = \frac{1}{n}\left[t^{n-1}\right] \exp\left(n\frac{t^{b-1}}{(b-1)!}\right)\frac{\mathrm{d}}{\mathrm{d}\,t}G(t)\,. \tag{2.42}$$

*Preuve.* Notons

$$\Phi(t) = \exp\left(\frac{t^{b-1}}{(b-1)!}\right)\,, \tag{2.43}$$

et

$$\Phi(t)^n = \sum_{m=0}^{\infty} \phi_m(n)\frac{t^m}{m!}\,. \tag{2.44}$$

Nous obtenons alors

$$\left[t^{n-1}\right]\left\{\Phi(t)^n \frac{\mathrm{d}}{\mathrm{d}\,t}G(t)\right\} = \tag{2.45}$$

$$= \sum_{k,m\geq 0 | k+m=n-1} \phi_m(n) g_{k+1}\frac{1}{k!m!} = \tag{2.46}$$

$$= \sum_{k\geq 0} \phi_{n-k-1}(n) g_{k+1}\frac{1}{k!(n-k-1)!}\,, \tag{2.47}$$

$$(n-1)!\,\left[t^{n-1}\right]\left\{\Phi(t)^n\frac{\mathrm{d}}{\mathrm{d}\,t}G(t)\right\} = \tag{2.48}$$

$$= \sum_{k\geq 0}\phi_{n-k-1}(n)g_{k+1}\frac{(n-1)!}{k!(n-k-1)!} = \tag{2.49}$$

$$= \sum_{k\geq 0}\phi_{n-k-1}(n)g_{k+1}\binom{n-1}{k} = \tag{2.50}$$

$$= \sum_{k\geq 0}\phi_{n-k-1}(n)g_{k+1}\frac{k+1}{n}\binom{n}{k+1}. \tag{2.51}$$

Observons que

$$\Phi(t)^n = \sum_{s\geq 0}\frac{n^s[s(b-1)]!}{[(b-1)!]^s s!}\frac{t^{s(b-1)}}{[s(b-1)]!}, \tag{2.52}$$

soit

$$\phi_{n-k-1}(n) = \frac{n^s(n-k-1)!}{[(b-1)!]^s s!}, \tag{2.53}$$

avec $n = n(s) = s(b-1)+k+1$. À partir de (2.53), nous identifions le nombre de forêts de $(k+1)$ hyperarbres enracinés à $n$ sommets du théorème 2.2.3 :

$$\binom{n}{k+1}\frac{k+1}{n}\phi_{n-k-1}(n). \tag{2.54}$$

L'interprétation combinatoire est alors évidente en multipliant (2.42) par $n!$. Nous retrouvons alors nos marques de la section précédente, pour énumérer les structures de taille $n$ :

- les $g_{k+1}$ structures lisses à $(k+1)$ sommets sont croisées, dans l'ordre croissante (uniquement définie) des étiquettes, avec
- les forêts de $(k+1)$ hyperarbres enracinés à $n$ sommets au nombre de

$$\binom{n}{k+1}\frac{k+1}{n}\phi_{n-k-1}(n). \tag{2.55}$$

$\diamondsuit$

**Exemple 2.3.3.** Nous retrouvons le nombre des hyperarbres enracinés en prenant le plus simple des fonctions $G$, dans la formule d'inversion de Lagrange, c'est à dire la fonction identité. En effet,

$$n!\,[z^n]\,T(z) = (n-1)!\,\left[t^{n-1}\right]\exp\left(n\frac{t^{b-1}}{(b-1)!}\right) \tag{2.56}$$

$$= (n-1)!\,\left[t^{n-1}\right]\left(\sum_{j\geq 0}\frac{n^j}{[(b-1)!]^j j!}t^{j(b-1)}\right) \tag{2.57}$$

$$= (n-1)!\frac{n^s}{[(b-1)!]^s s!}, \tag{2.58}$$

avec $n = n(s) = s(b-1)+1$.

**Exemple 2.3.4.** Le nombre des forêts de $(k+1)$ hyperarbres enracinés de SGE $\frac{T(z)^{k+1}}{(k+1)!}$ est

$$\begin{aligned} n!\,[z^n]\,\frac{T(z)^{k+1}}{(k+1)!} &= (n-1)!\,\left[t^{n-1}\right]\left(\frac{t^k}{k!}\exp\left(n\frac{t^{b-1}}{(b-1)!}\right)\right) && (2.59)\\ &= \frac{(n-1)!}{k!}\left[t^{n-1}\right]\left(\sum_{j\geq 0}\frac{n^j}{[(b-1)!]^j j!}t^{j(b-1)+k}\right) && (2.60)\\ &= \frac{(n-1)!}{k!}\frac{n^s}{[(b-1)!]^s s!}, && (2.61)\end{aligned}$$

avec $n = n(s) = s(b-1)+k+1$, soit en accord avec le théorème 2.2.3

$$n!\,[z^n]\,\frac{T(z)^{k+1}}{(k+1)!} = \binom{n}{k+1}(k+1)\frac{(n-k-1)!}{[(b-1)!]^s s!}n^{s-1}. \tag{2.62}$$

**Exemple 2.3.5.** De l'exemple précédent, nous déduisons une formule explicite du nombre des forêts d'hyperarbres enracinés :

$$n!\,[z^n]\exp(T(z)) = (n-1)!\sum_{k=1}^{n-1}\frac{n^{s_k}}{[(b-1)!]^{s_k} s_k!}, \tag{2.63}$$

avec $n = n(s_k) = s_k(b-1)+k+1$ pour tout $k = 1,\ldots,n-1$.

Ainsi, par rapport à la section précédente, nous disposons maintenant d'un outil puissant cachant toute une machinerie bijective permettant de traduire, à partir des SGEs des structures lisses, les structures éventuellement non lisses correspondant. Ceci nous motive encore plus à automatiser la détermination des SGEs des structures lisses. Pour y parvenir, dans un souci de clarté, nous présentons brièvement les SGEs bivariées.

### 2.3.3 Introduction aux séries génératrices exponentielles bivariées

À une séquence de nombre $(a_{n,m})_{n,m\in\mathbb{N}}$ est associée une série génératrice exponentielle bivariée

$$A(w,z) = \sum_{n,m\in\mathbb{N}} a_{n,m}w^m\frac{z^n}{n!}. \tag{2.64}$$

Dans les hypergraphes, il y a deux notions essentielles à savoir la notion de sommet et la notion d'hyperarête. Soulignons que dans cette thèse, les séries sont exponentielles en la variable $z$ relative aux sommets qui sont étiquetés et elles sont ordinaires en la variable $w$, dans le cas de SGEs bivariées, relative aux hyperarêtes. Si jusqu'ici, nous avons choisi de travailler uniquement avec des SGEs univariées, c'est parce que seul le paramètre de la taille de la structure nous est pertinent et l'utilisation de la notion d'excès simplifie souvent de beaucoup les expressions des SGEs : c'est souligné dans [18] par la richesse des résultats énumératifs qui y sont obtenus.

L'introduction des SGEs bivariées nous permettra, en particulier, de souligner l'effet de décompositions dans les structures étudiées. Le dictionnaire de la section précédente s'enrichit du marquage d'une hyperarête ; les opérations de marquage se traduisent en terme de dérivation partielle :

| Opération sur les structures | SGE correspondant |
|---|---|
| Marquage de k sommets de $\mathcal{A}$ | $\frac{z^k}{k!}\frac{\partial^k}{\partial z^k}A(w,z)$ |
| Marquage d'une hyperarête de $\mathcal{A}$ | $w\frac{\partial}{\partial w}A(w,z)$ |

**Exemple 2.3.6.** Pour énumérer les graphes étiquetés ayant 5 sommets et exactement 3 arêtes formant une clique d'ordre 3 nous pouvons recourir à la SGE bivariée :

$$\left(\frac{w^0 z^2}{2!}\right)\left(\frac{w^3 z^3}{3!}\right) = \binom{5}{2}\frac{w^3 z^5}{5!}\,. \tag{2.65}$$

Dans le membre gauche de cette équation :

- le premier facteur correspond à la SGE des graphes ayant 2 sommets et sans arête,
- le second facteur correspond à la SGE des graphes ayant 3 sommets et 3 arêtes et le produit s'interprète à partir du dictionnaire pour justifier l'énumération.

La SGE bivariée de ces graphes avec une arête distinguée est

$$w\frac{\partial}{\partial w}\binom{5}{2}\frac{w^3 z^5}{5!} = 3\binom{5}{2}\frac{w^3 z^5}{5!}\,. \tag{2.66}$$

**Définition 2.3.7.** La SGE bivariée des composantes d'excès $\ell$ s'écrit

$$\hat{H}_\ell(w,z) = w^{\ell/(b-1)} H_\ell \circ T(w^{1/(b-1)}z) = \sum_{s,n} h_\ell(s,n) w^s \frac{z^n}{n!}\,, \tag{2.67}$$

avec $n = s(b-1) - \ell$ et $h_\ell(s,n)$ le nombre de composantes d'excès $\ell$ ayant $s$ hyperarêtes et $n$ sommets.

**Exemple 2.3.8.** La SGE bivariée des hyperarbres enracinés est

$$\frac{T(w^{1/(b-1)z})}{w^{1/(b-1)}}\,. \tag{2.68}$$

Rappelons la définition implicite de la SGE $T(z)$

$$T(z) = z\exp\left(\frac{T(z)^{b-1}}{(b-1)!}\right)\,. \tag{2.69}$$

Ainsi,

$$\begin{aligned} \mathrm{d}\,T(z) &= \left(\mathrm{d}\,z + z\frac{T(z)^{b-2}}{(b-2)!}\mathrm{d}\,T(z)\right)\exp\left(\frac{T(z)^{b-1}}{(b-1)!}\right) && (2.70)\\ &= T(z)\frac{\mathrm{d}\,z}{z} + \frac{T(z)^{b-1}}{(b-2)!}\mathrm{d}\,T(z)\,, && (2.71)\end{aligned}$$

$$\frac{\mathrm{d}\,T(z)}{T(z)} = \frac{1}{\left(1 - \frac{T(z)^{b-1}}{(b-2)!}\right)} \frac{\mathrm{d}\,z}{z} \,. \tag{2.72}$$

Soit

$$z\frac{\mathrm{d}}{\mathrm{d}\,z}T(z) = \frac{T(z)}{\left(1 - \frac{T(z)^{b-1}}{(b-2)!}\right)} \tag{2.73}$$

qui se lit comme l'équivalence des structures d'hyperarbre enraciné ayant un second marquage (la racine peut être éventuellement re-marquée) et une chaîne orientée. Cette relation (2.73) donne

$$w\frac{\partial}{\partial w}T(w^{1/(b-1)}z) = \frac{1}{b-1}\frac{T(w^{1/(b-1)}z)}{\left(1 - \frac{T(w^{1/(b-1)}z)^{b-1}}{(b-2)!}\right)} \tag{2.74}$$

et

$$z\frac{\partial}{\partial z}T(w^{1/(b-1)}z) = \frac{T(w^{1/(b-1)}z)}{\left(1 - \frac{T(w^{1/(b-1)}z)^{b-1}}{(b-2)!}\right)} \,. \tag{2.75}$$

Nous en déduisons

$$w\frac{\partial}{\partial w}\hat{H}_\ell(w,z) = \tag{2.76}$$

$$= w\frac{\partial}{\partial w}\left(w^{\ell/(b-1)}H_\ell \circ T(w^{1/(b-1)}z)\right) \tag{2.77}$$

$$= \frac{\ell w^{\ell/(b-1)}}{b-1}H_\ell \circ T(w^{1/(b-1)}z) + \tag{2.78}$$

$$+\frac{w^{\ell/(b-1)}}{b-1}H_\ell{}' \circ T(w^{1/(b-1)}z)\left(\frac{T(w^{1/(b-1)}z)}{1 - \frac{T(w^{1/(b-1)}z)^{b-1}}{(b-2)!}}\right) \tag{2.79}$$

$$= \frac{\ell w^{\ell/(b-1)}}{b-1}H_\ell \circ T(w^{1/(b-1)}z) + \tag{2.80}$$

$$+\frac{w^{\ell/(b-1)}}{b-1}H_\ell{}' \circ T(w^{1/(b-1)}z)\left(z\frac{\partial}{\partial z}T(w^{1/(b-1)}z)\right) \tag{2.81}$$

$$= \frac{\ell w^{\ell/(b-1)}}{b-1}H_\ell \circ T(w^{1/(b-1)}z) + \tag{2.82}$$

$$+\frac{1}{b-1}z\frac{\partial}{\partial z}\left(w^{\ell/(b-1)}H_\ell \circ T(w^{1/(b-1)}z)\right) \,, \tag{2.83}$$

soit la transcription d'un marquage d'une hyperarête, en terme de marquage d'un sommet pour les composantes d'excès $\ell$, suivante :

$$w\frac{\partial}{\partial w}\hat{H}_\ell(w,z) = \frac{1}{b-1}\left(\ell\hat{H}_\ell(w,z) + z\frac{\partial}{\partial z}\hat{H}_\ell(w,z)\right) \,. \tag{2.84}$$

Là réside l'intérêt de classer les structures selon leur excès car les décompositions exprimées en SGEs bivariées se ramènent alors à des SGEs univariées en réexprimant les SGEs avec un marquage d'hyperarête et en fixant $w = 1$.

**Exemple 2.3.9.** Nous avons pour les hyperarbres enracinés

$$\left.\frac{T(w^{1/(b-1)}z)}{w^{1/(b-1)}}\right|_{w=1} = T(z)\,. \tag{2.85}$$

**Exemple 2.3.10.** Et pour les composantes d'excès $\ell$

$$\left.\left\{w^{\ell/(b-1)} H_\ell \circ T(w^{1/(b-1)}z)\right\}\right|_{w=1} = H_\ell \circ T(z)\,. \tag{2.86}$$

**Exemple 2.3.11.** Les composantes d'excès $\ell$ ayant une hyperarête marquée sont énumérées par la SGE univariée

$$\left.\left\{w\frac{\partial}{\partial w}\hat{H}_\ell(w,z)\right\}\right|_{w=1} = \frac{1}{b-1}\left(\ell H_\ell \circ T(z) + z\frac{\mathrm{d}}{\mathrm{d}\,z}H_\ell \circ T(z)\right)\,. \tag{2.87}$$

### 2.3.4 Récurrence des SGEs $H_\ell$

Dans cette sous-section, l'excès $\ell$ est fixé et à partir d'une décomposition des composantes d'excès $\ell$ est établie une récurrence des SGEs $H_\ell$. Pour décrire la décomposition, les SGEs bivariées sont d'abord utilisées car elles permettent une lecture directe des opérations, sur les structures, qui y sont suggérées. Illustrons l'idée principale de la décomposition, menant à la récurrence, à partir de l'exemple de la figure 2.6 .

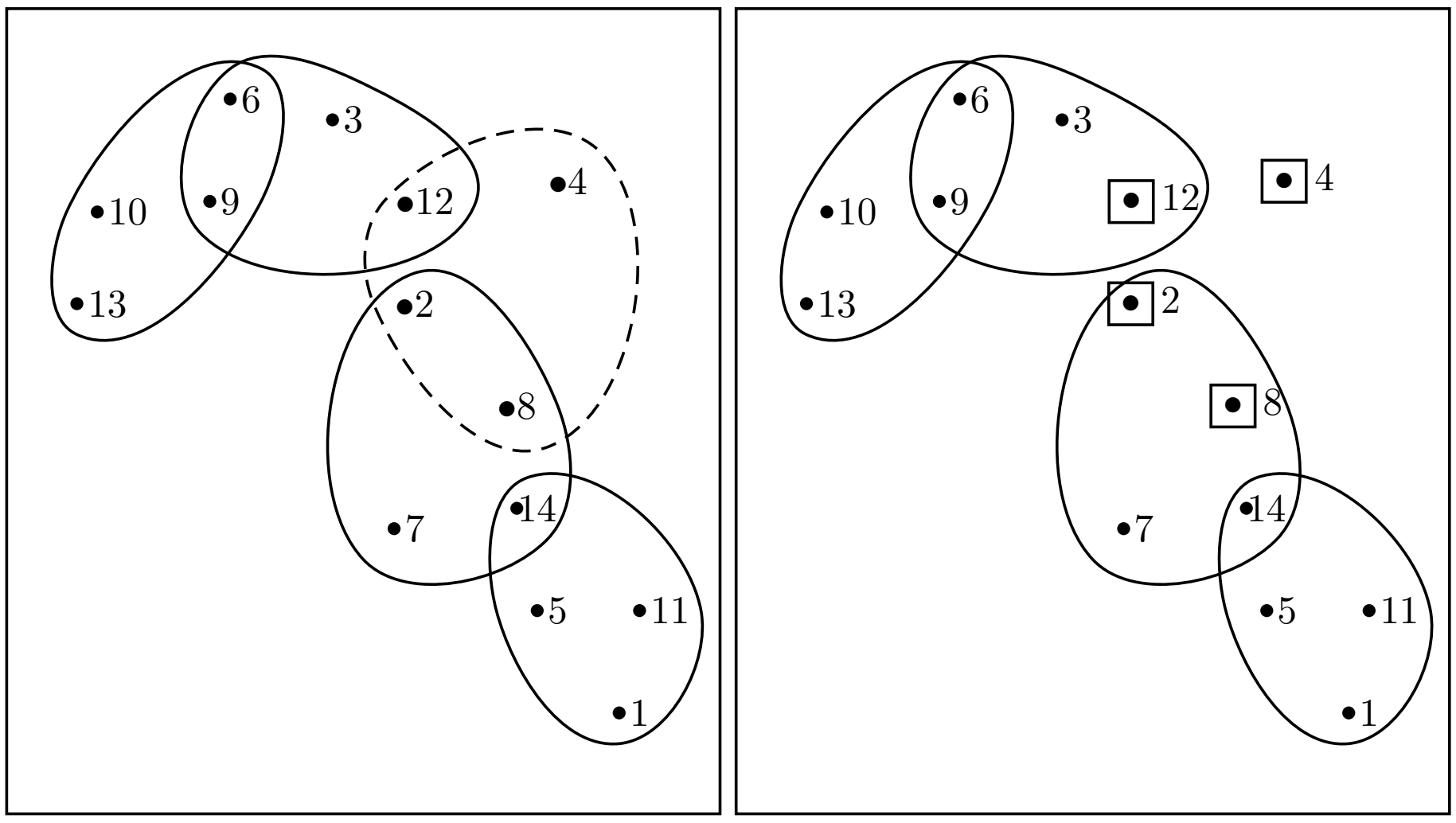

FIG. 2.6 – La décomposition d'une composante sur un exemple.

Dans cette figure, la figure à gauche représente une composante 4-uniforme ayant 14 sommets et 5 hyperarêtes dont une d'elle est marquée à savoir l'hyperarête $\{2,4,8,12\}$ . La figure à droite représente un hypergraphe 4-uniforme ayant aussi 14 sommets mais une hyperarête de moins que la composante à gauche. L'hypergraphe à droite n'est pas connexe : il y a 3 composantes telles que chacune admet au moins un sommet marqué. Les deux figures sont équivalentes l'une l'autre et représentent la même structure, seule le regard qui y est porté change : d'un coté c'est une hyperarête qui est marquée, de l'autre ce sont des sommets de cette hyperarête qui sont marqués. Il est clair que marquer une hyperarête dans une composante revient à marquer au moins un sommet dans chacune d'un certain nombre de composantes et inversement, en prenant soin qu'il y ait autant de sommets marqués que nécessaire pour former une hyperarête et qu'ils n'appartiennent pas tous déjà à une même hyperarête.

Il nous faut cependant faire attention pour un excès $\ell$ donné que les composantes avec chacune au moins un sommet marqué se recombinent bien en un hypergraphe $b$-uniforme d'excès $\ell$. Pour cela nous faisons le lemme suivant :

**Lemme 2.3.12.** *Considérons une famille de composantes d'excès $j_i$ et ayant $k_i$ sommets marqués, $-1 \leq j_i \leq \ell$ et $1 \leq k_i \leq b$. De plus, considérons que cette famille n'est pas celle constituée de seulement une composante avec $b$ sommets marqués tous appartenant déjà à une même hyperarête. Alors si une hyperarête est créé à partir des sommets marqués des différentes composantes, l'hypergraphe obtenu est connexe et $b$-uniforme et d'excès $\ell \geq -1$ si et seulement si*

$$\begin{cases} \sum_i k_i = b \\ \sum_i (j_i + k_i) = \ell + 1\,. \end{cases} \tag{2.88}$$

*Preuve.* Il est clair que l'hypergraphe obtenu est connexe et qu'il est $b$-uniforme si et seulement si le nombre total des sommets marqués dans la famille considérée est égale à $b$ :

$$\sum_i k_i = b\,. \tag{2.89}$$

Notons par $N$ le nombre de composantes dans la famille considérée et par $n$ le nombre total de sommets de ces composantes. Dans la $i$-ième composante, notons $n_i$ le nombre de ses sommets et $s_i$ le nombre de ses hyperarêtes. Alors, l'excès de la composante obtenue à partir de la famille vaut $\ell$ si et seulement si

$$\sum_{i=1}^{N} s_i(b-1) + (b-1) - n = \ell\,. \tag{2.90}$$

Cette dernière équation est équivalente aux suivantes :

$$\sum_{i=1}^{N} s_i(b-1) + b - n = \ell + 1 \tag{2.91}$$

$$\sum_{i=1}^{N} s_i(b-1) + \sum_{i=1}^{N} k_i - \sum_{i=1}^{N} n_i = \ell + 1 \tag{2.92}$$

$$\sum_{i=1}^{N}\left(\{s_i(b-1)-n_i\}+k_i\right)=\ell+1 \tag{2.93}$$

$$\sum_{i=1}^{N}(j_i+k_i)=\ell+1\,. \tag{2.94}$$

♢

La décomposition, suggérée par le marquage d'une hyperarête dans les composantes d'excès $\ell \geq -1$ et illustrée par la figure 2.6 , se traduit en terme de récurrence de SGEs bivariées dans le théorème suivant :

**Théorème 2.3.13.** *Pour tout $\ell \geq -1$ , la SGE bivariée $\hat{H}_\ell$ des composantes d'excès $\ell$ satisfait la relation :*

$$w\frac{\partial}{\partial w}\hat{H}_\ell(w,z)= \tag{2.95}$$

$$=w\left[U^b\mathrm{Cyc}^{\ell+1}\right]\exp\left(\sum_{k=1}^{b}\sum_{j=-1}^{\ell}\mathrm{Cyc}^{j+k}\frac{(Uz)^k}{k!}\frac{\partial^k}{\partial z^k}\hat{H}_j(w,z)\right)+ \tag{2.96}$$

$$-w\frac{\partial}{\partial w}\hat{H}_{\ell-b+1}(w,z)\,, \tag{2.97}$$

*où $\hat{H}_j \equiv 0$ si $j \leq -2$ .*

*Preuve.* La relation dans ce théorème traduit en terme de SGEs bivariées la décomposition. Il faut y lire :

- dans le premier membre, la SGE bivariée des composantes d'excès $\ell$ avec une hyperarête marquée (pour suppression) ($w\frac{\partial}{\partial w}\hat{H}_\ell(w,z)$)
- dans le second membre, la SGE bivariée des familles de composantes ayant $k$ sommets marqués et d'excès $j$ , lesquelles familles, par création d'une hyperarête définie par les sommets marqués des composantes, produisent une composante $b$-uniforme d'excès $\ell$ . Le terme avec un signe négatif contraint les marquages des $b$ sommets à ne pas appartenir à une hyperarête déjà existante (à ne pas recréer l'hyperarête, les structures considérées ici n'admettant pas d'hyperarête multiple) . L'extraction du coefficient de $U^b$ garantit d'avoir marqué $b$ sommets pour définir une hyperarête et l'extraction du coefficient de $\mathrm{Cyc}^{\ell+1}$ garantit de former une composante d'excès $\ell$ d'après le lemme précédent. Dans le premier terme de ce second membre, le facteur $w$ vient confirmer qu'il s'agit bien de créer une hyperarête de plus.

♢

Ce théorème illustre la puissance d'expression des SGEs pour décrire les bijections ou des décompositions entre des structures combinatoires. En effet, dans ce théorème est établie une relation de récurrence entre les composantes d'excès $\ell$ d'un coté et les composantes d'excès strictement plus petit d'un autre. Cependant, l'utilisation des SGEs bivariées pour cette relation de récurrence complique les notations car pour une valeur de l'excès $\ell$ donnée, la SGE $\hat{H}_\ell(w,z)$ bivariée se déduit de la SGE $H_\ell \circ T(z)$ et inversement bien sûr. En effet, rappelons (2.86) et (2.67) :

– à partir de la SGE bivariée se déduit la SGE univariée (il suffit de fixé la valeur de la variable $w = 1$)

$$\left\{ w^{\ell/(b-1)} H_\ell \circ T(w^{1/(b-1)} z) \right\} \Big|_{w=1} = H_\ell \circ T(z)\,, \tag{2.98}$$

– à partir de la SGE univariée se déduit la SGE bivariée pour un excès donné (par définition de l'excès, le nombre de sommets est lié à celui des hyperarêtes)

$$\hat{H}_\ell(w, z) = w^{\ell/(b-1)} H_\ell \circ T(w^{1/(b-1)} z)\,. \tag{2.99}$$

Nous pouvons transcrire le théorème 2.3.13 en terme de SGE univariée, les dérivées partielles deviennent des dérivées droites :

**Théorème 2.3.14.** *La SGE $H_\ell \circ T(z)$ des composantes d'excès $\ell$ satisfait la relation*

$$\frac{1}{b-1} \left( \ell H_\ell \circ T(z) + z \frac{\mathrm{d}}{\mathrm{d}\, z} H_\ell \circ T(z) \right) = \tag{2.100}$$

$$= \left[ U^b \mathrm{Cyc}^{\ell+1} \right] \exp \left( \sum_{k=1}^{b} \sum_{j=-1}^{\ell} \mathrm{Cyc}^{j+k} \frac{(Uz)^k}{k!} \frac{\mathrm{d}^{\,k}}{\mathrm{d}\, z^k} H_j \circ T(z) \right) + \tag{2.101}$$

$$- \frac{1}{b-1} \left( (\ell - 1 + 1) H_{\ell-b+1} \circ T(z) + z \frac{\mathrm{d}}{\mathrm{d}\, z} H_{\ell-b+1} \circ T(z) \right)\,. \tag{2.102}$$

*Preuve.* Avant de fixer $w = 1$, nous observons que le premier membre de l'équation du théorème 2.3.13 se réécrit (rappel de (2.84) )

$$w \frac{\partial}{\partial w} \hat{H}_\ell(w, z) = \frac{1}{b-1} \left( \ell \hat{H}_\ell(w, z) + z \frac{\partial}{\partial z} \hat{H}_\ell(w, z) \right)\,. \tag{2.103}$$

Et ainsi, nous déduisons le théorème. ♢

Par ce théorème, nous disposons d'une récurrence des SGEs $H_\ell \circ T(z)$ sous la forme d'une équation différentielle d'ordre un (dans la mesure où les termes du second membre qui apparaissent avec $H_j$ où $j < \ell$ ont été explicités dans une résolution antérieure) .

Nous avons affirmé sans justification, dans la remarque 2.2.9 , que la SGE des hyperarbres non enracinés s'écrit sous la forme $H_{-1} \circ T(z)$ , soit en fonction de la SGE $T(z)$ des hyperarbres enracinés. La justification est ici faite par l'application de ce théorème pour la valeur $\ell = -1$ , soit le cas de la SGE des hyperarbres non enracinés :

**Théorème 2.3.15.** *La SGE $H_{-1} \circ T(z)$ des hyperarbres non enracinés est telle que*

$$H_{-1}(t) = t - \frac{(b-1)t^b}{b!}\,. \tag{2.104}$$

Cette écriture est valable dans le cas où $b = 2$ , c'est à dire dans le cas des arbres non enracinés.

L'écriture du théorème 2.3.14 se simplifie en une équation différentielle en la "variable" $T(z)$ :

**Théorème 2.3.16.** *Pour $\ell \geq 0$, notons qu'il existe une fonction $J_\ell$ telle que*

$$J_\ell \circ T(z) = \tag{2.105}$$

$$= \left[U^b \mathrm{Cyc}^{\ell+1}\right] \exp\left(\sum_{k=1}^{b}\sum_{j=-1}^{\ell-1} \mathrm{Cyc}^{j+k}\frac{(Uz)^k}{k!}\frac{\mathrm{d}^k}{\mathrm{d}z^k}H_j \circ T(z)\right) + \tag{2.106}$$

$$-\frac{1}{b-1}\left((\ell-1+1)H_{\ell-b+1}\circ T(z) + z\frac{\mathrm{d}}{\mathrm{d}z}H_{\ell-b+1}\circ T(z)\right). \tag{2.107}$$

*Ainsi,*

$$\frac{1}{b-1}\left(\ell H_\ell(t) + t\frac{\mathrm{d}}{\mathrm{d}t}H_\ell(t)\right) = J_\ell(t). \tag{2.108}$$

*Preuve.* Dans l'équation (2.100), en regroupant les termes avec l'indice, relative à l'excès, égale à $\ell$ dans le premier membre nous obtenons :

– d'un coté

$$\frac{1}{b-1}\left(\ell H_\ell \circ T(z) + \left(1 - \frac{T(z)^{b-1}}{(b-2)!}\right) z\frac{\mathrm{d}}{\mathrm{d}z}H_\ell \circ T(z)\right), \tag{2.109}$$

soit par (2.73)

$$\frac{1}{b-1}\left(\ell H_\ell \circ T(z) + T(z)H_\ell{}' \circ T(z)\right) \tag{2.110}$$

– de l'autre, l'intervalle de l'indice $j$ devient $j = -1, \ldots, \ell-1$ dans l'exposant de l'exp,

$$\left[U^b \mathrm{Cyc}^{\ell+1}\right] \exp\left(\sum_{k=1}^{b}\sum_{j=-1}^{\ell-1} \mathrm{Cyc}^{j+k}\frac{(Uz)^k}{k!}\frac{\mathrm{d}^k}{\mathrm{d}z^k}H_j \circ T(z)\right) + \tag{2.111}$$

$$-\frac{1}{b-1}\left((\ell-1+1)H_{\ell-b+1}\circ T(z) + z\frac{\mathrm{d}}{\mathrm{d}z}H_{\ell-b+1}\circ T(z)\right) \tag{2.112}$$

qui vaut (2.110) et comme ce dernier s'exprime en $T(z)$, nous déduisons l'existence de $J_\ell$ telle que

$$J_\ell \circ T(z) = \tag{2.113}$$

$$= \left[U^b \mathrm{Cyc}^{\ell+1}\right] \exp\left(\sum_{k=1}^{b}\sum_{j=-1}^{\ell-1} \mathrm{Cyc}^{j+k}\frac{(Uz)^k}{k!}\frac{\mathrm{d}^k}{\mathrm{d}z^k}H_j \circ T(z)\right) + \tag{2.114}$$

$$-\frac{1}{b-1}\left((\ell-1+1)H_{\ell-b+1}\circ T(z) + z\frac{\mathrm{d}}{\mathrm{d}z}H_{\ell-b+1}\circ T(z)\right). \tag{2.115}$$

En substituant $T(z)$ par la variable $t$, on obtient l'équation différentielle du théorème :

$$\frac{1}{b-1}\left(\ell H_\ell(t) + t\frac{\mathrm{d}}{\mathrm{d}t}H_\ell(t)\right) = J_\ell(t). \tag{2.116}$$

◊

**Remarque 2.3.17.** Une variante (plus bijective) de la preuve de ce théorème est que la décomposition décrite dans $J_\ell$ se comprend grâce aux structures lisses : les structures obtenues par le marquage de sommets dans une composante peuvent être classées en des composantes lisses. Bref, $z^k \frac{\mathrm{d}^k}{\mathrm{d}\,z^k} H_j \circ T(z)$ s'exprime en fonction de $T(z)$ .

### 2.3.5 Résolution de la récurrence des SGEs $H_\ell$

Ainsi, nous disposons, dans le théorème 2.3.16 , d'une récurrence des SGEs $H_\ell$ des composantes lisses d'excès $\ell$ sous la forme d'une équation différentielle d'ordre un suivante :

$$\frac{1}{b-1}\left(\ell H_\ell(t) + t\frac{\mathrm{d}}{\mathrm{d}\,t}H_\ell(t)\right) = J_\ell(t)\,, \tag{2.117}$$

avec $J_\ell$ défini à partir de (2.105) . Dans cette sous-section, nous résolvons ces équations différentielles itérativement en l'excès $\ell$ . À partir de là peut se faire l'automatisation du calcul des SGEs $H_\ell$ des composantes lisses d'excès $\ell \geq 0$ . La solution est uniquement déterminée en considérant la condition initiale

$$H_\ell(t)|_{t=0} = 0\,. \tag{2.118}$$

Adoptons la notation, rencontrée dans la remarque 2.2.21 ,

$$\theta(t) = 1 - \tau(t)\,. \tag{2.119}$$

**Remarque 2.3.18.** Le choix de fixer cette notation, bien que $\theta$ n'admette pas de lecture combinatoire directe, s'est imposé à partir de l'article [22] puis, à posteriori, parce qu'il est commode de manipuler les polynômes de Laurent $f_\ell$ de la remarque 2.2.21 pour définir les SGEs $H_\ell$ . Notons tout de même que les séries, en $t$ , $1/\theta(t)$ et $(1-\theta(t))$ admettent chacune une lecture combinatoire à partir desquelles les SGEs exprimées avec $\theta(t)$ peuvent être comprises.

Rappelons la définition des $f_\ell$.

**Définition 2.3.19.** $f_\ell$ est un polynôme de Laurent défini par l'intermédiaire du SGE $H_\ell$ :

$$H_\ell(t) = \frac{f_\ell \circ \theta(t)}{t^\ell}\,. \tag{2.120}$$

**Remarque 2.3.20.** L'équation homogène associée à (2.117) admet pour solution Cste$/t^\ell$ . Ce qui nous suggère de trouver une solution sous la forme

$$H_\ell(t) = \frac{g_\ell(t)}{t^\ell}\,. \tag{2.121}$$

Cette forme découle aussi de (2.120) .

Par le lemme 2.2.19 ,

**Remarque 2.3.21.** Les composantes d'excès $j$ ayant $k$ sommets marqués admettent une SGE qui s'écrit sous la forme

$$\frac{z^k}{k!}\frac{\mathrm{d}^k}{\mathrm{d}\,z^k}H_j \circ T(z) = \frac{f_{jk} \circ \theta \circ T(z)}{T(z)^j}\,, \tag{2.122}$$

avec $f_{jk}$ un polynôme de Laurent. En particulier, $f_{j,0} = f_j$ .

**Exemple 2.3.22.** Pour les hyperarbres, la forme de l'écriture de la définition 2.3.19 est valide :

$$H_{-1}(t) = t\left(\frac{b-1+\theta(t)}{b}\right)\,, \tag{2.123}$$

soit

$$f_{-1} \circ \theta(t) = \frac{b-1+\theta(t)}{b}\,. \tag{2.124}$$

C'est le seul des $f_j$ de degré minimum positif (ici, ce degré est nul) .

Convenons de noter les polynômes de Laurent $f_{jk}$ avec la variable $x$ , alors en évaluant $\frac{\mathrm{d}}{\mathrm{d}\,z}\{\frac{\mathrm{d}^k}{\mathrm{d}\,z^k}H_j \circ T(z)\}$ par dérivation de fonctions composées sachant (2.73)

$$f_{j,k+1}(x) = -(b-1)\frac{f_{jk}{}'(x)}{x} + (b-1)f_{jk}{}'(x) - j\frac{f_{jk}(x)}{x} - kf_{jk}(x)\,. \tag{2.125}$$

**Remarque 2.3.23.** La détermination récursive de $f_\ell$ . Comme

$$J_\ell(t) = \tag{2.126}$$

$$= \left[U^b\mathrm{Cyc}^{\ell+1}\right]\exp\left(\sum_{k=1}^{b}\sum_{j=-1}^{\ell-1}\mathrm{Cyc}^{j+k}U^k\frac{f_{jk}\circ\theta(t)}{t^j}\right) + \tag{2.127}$$

$$-\frac{1}{b-1}\left((\ell-1+1)\frac{f_{\ell-b+1,0}\circ\theta(t)}{t^{\ell-b+1}} + \frac{f_{\ell-b+1,1}\circ\theta(t)}{t^{\ell-b+1}}\right) \tag{2.128}$$

$$= \frac{1}{t^{\ell-b+1}}\left[U^b\mathrm{Cyc}^{\ell+1}\right]\exp\left(\sum_{k=1}^{b}\sum_{j=-1}^{\ell-1}\mathrm{Cyc}^{j+k}U^k f_{jk}\circ\theta(t)\right) + \tag{2.129}$$

$$-(\frac{1}{b-1})\frac{((\ell-1+1)f_{\ell-b+1,0}\circ\theta(t) + f_{\ell-b+1,1}\circ\theta(t))}{t^{\ell-b+1}}\,. \tag{2.130}$$

Et comme en notant $H_\ell$ sous la forme (2.121) , l'équation différentielle (2.117) devient

$$\frac{1}{b-1}\frac{g_\ell{}'(t)}{t^{\ell-1}} = J_\ell(t)\,, \tag{2.131}$$

soit

$$g_\ell{}'(t) = (b-1)t^{\ell-1}J_\ell(t)\,. \tag{2.132}$$

Nous aurons

$$g_\ell{}'(t) = \tag{2.133}$$

$$= (b-1)t^{b-2}\left[U^b \mathrm{Cyc}^{\ell+1}\right] \exp\left(\sum_{k=1}^{b}\sum_{j=-1}^{\ell-1} \mathrm{Cyc}^{j+k} U^k f_{jk}\circ\theta(t)\right) + \tag{2.134}$$

$$-(b-1)t^{b-2}\frac{((\ell-1+1)f_{\ell-b+1,0}\circ\theta(t) + f_{\ell-b+1,1}\circ\theta(t))}{b-1} \tag{2.135}$$

$$= -(b-2)!\frac{\mathrm{d}\,\theta(t)}{\mathrm{d}\,t}\left[U^b \mathrm{Cyc}^{\ell+1}\right] \exp\left(\sum_{k=1}^{b}\sum_{j=-1}^{\ell-1} \mathrm{Cyc}^{j+k} U^k f_{jk}\circ\theta(t)\right) + \tag{2.136}$$

$$-(b-2)!\frac{\mathrm{d}\,\theta(t)}{\mathrm{d}\,t}\frac{((\ell-1+1)f_{\ell-b+1,0}\circ\theta(t) + f_{\ell-b+1,1}\circ\theta(t))}{b-1} \tag{2.137}$$

car

$$\mathrm{d}\,\theta(t) = -\frac{(b-1)t^{b-2}}{(b-2)!}\mathrm{d}\,t\,. \tag{2.138}$$

En identifiant $\theta(t) = \theta$,

$$g_\ell{}'(t)\mathrm{d}\,t = \tag{2.139}$$

$$= -(b-2)!\left[U^b \mathrm{Cyc}^{\ell+1}\right] \exp\left(\sum_{k=1}^{b}\sum_{j=-1}^{\ell-1} \mathrm{Cyc}^{j+k} U^k f_{jk}(\theta)\right) \mathrm{d}\,\theta + \tag{2.140}$$

$$-(b-2)!\frac{((\ell-1+1)f_{\ell-b+1,0}(\theta) + f_{\ell-b+1,1}(\theta))}{b-1}\mathrm{d}\,\theta\,. \tag{2.141}$$

Ce qui, par intégration, en prenant la solution qui vérifie $f_\ell(1) = 0$, détermine

$$f_\ell\circ\theta(t) = g_\ell(t)\,. \tag{2.142}$$

Nous pouvons maintenant résumer la résolution de la récurrence dans le théorème suivant :

**Théorème 2.3.24.** *La SGE $H_\ell$ des composantes lisses d'excès $\ell \geq 0$ est déterminée récursivement sous la forme*

$$H_\ell(t) = \frac{f_\ell\circ\theta(t)}{t^\ell}\,, \tag{2.143}$$

*avec*

$$f_{\ell,0}(x) = f_\ell(x) = \int_1^x \hat{J}_\ell(u)\mathrm{d}\,u\,, \tag{2.144}$$

*où*

$$\hat{J}_\ell(u) = \tag{2.145}$$

$$= -(b-2)!\left[U^b \mathrm{Cyc}^{\ell+1}\right] \exp\left(\sum_{k=1}^{b}\sum_{j=-1}^{\ell-1} \mathrm{Cyc}^{j+k} U^k f_{jk}(u)\right) + \tag{2.146}$$

$$-(b-2)!\frac{((\ell-1+1)f_{\ell-b+1,0}(u) + f_{\ell-b+1,1}(u))}{b-1}\,, \tag{2.147}$$

*avec*

$$f_{j,k+1}(x) = -(b-1)\frac{{f_{jk}}'(x)}{x} + (b-1){f_{jk}}'(x) - j\frac{f_{jk}(x)}{x} - kf_{jk}(x)\,. \tag{2.148}$$

*Notons en particulier que* $f_{-1,1}(x) = 1$ .

Par ce théorème, il est naturel de déterminer les $f_\ell$ sous forme de polynôme de Laurent, une forme simple qui n'admet pas de lecture combinatoire à première vue. Dans la sous-section suivante, nous montrons comment automatiser la détermination de $H_\ell$ sous la forme (2.22) avec laquelle nous disposons d'une lecture combinatoire.

### 2.3.6 Mise en forme des SGEs $H_\ell$

Avant de donner les premiers exemples des SGEs $H_\ell$ , nous donnons deux identités combinatoires qui permettent de mettre les SGEs $H_\ell$ sous la forme de l'équation (2.22) .

**Lemme 2.3.25.** *Pour* $j, a \in \mathbb{N}^*$ *(donc* $j + a > 0$*) ,*

$$\frac{1}{\theta^j} = \sum_{i=0}^{j-1} \binom{j+a}{i} \frac{(1-\theta)^{j+a-i}}{\theta^{j-i}} + \sum_{i=0}^{a} \binom{j+a-i-1}{j-1} (1-\theta)^{a-i}\ . \tag{2.149}$$

**Lemme 2.3.26.** *Si* $a - j \geq 0$ *alors*

$$\theta^j = \sum_{i=0}^{a} \binom{a-j-i-1}{-j-1} (1-\theta)^{a-i}\ , \tag{2.150}$$

*où si* $k \in \mathbb{N}$ *et* $t \in \mathbb{Z}$ *alors*

$$\binom{t}{k} = \binom{t}{t-k} = \frac{t(t-1)\cdots(t-k+1)}{k!}\,. \tag{2.151}$$

Les preuves de ces deux lemmes sont données en annexe.

Par ces deux identités présentées dans ces deux lemmes, les deux formes des SGEs $H_\ell$ se déduisent l'une de l'autre. Rappelons ici ces deux formes :

- avec $f_\ell$ un polynôme de Laurent, nous avons une forme pratique,

$$H_\ell(t) = \frac{f_\ell \circ \theta(t)}{t^\ell} \tag{2.152}$$

- à partir de (2.22) , nous avons une forme combinatoire,

$$H_\ell(t) = \frac{(1-\theta(t))^{r_\ell}}{t^\ell} \sum_{p=0}^{3\ell} A_{\ell,p} \left(\frac{1-\theta(t)}{\theta(t)}\right)^p \tag{2.153}$$

sachant $\tau(t) = 1 - \theta(t) = t^{b-1}/(b-2)!$ .

La forme pratique se déduit de la forme combinatoire directement en développant l'expression de la forme combinatoire :

$$H_\ell(t) = \frac{1}{t^\ell} \sum_{j=-3\ell}^{r_\ell} c_j(\ell, b)\theta(t)^j\,, \tag{2.154}$$

avec

$$r_\ell = \lfloor \frac{\ell+1}{b-1} + 1 \rfloor\,. \tag{2.155}$$

Et la forme combinatoire se déduit de la forme pratique en utilisant les lemmes avec la valeur $a = r_\ell$ pour réexprimer les $\theta(t)^j$.

**Remarque 2.3.27.** Dans ce qui suit, dans le cas des graphes, c'est à dire si $b = 2$, nous adoptons la notation $W_\ell = H_\ell$ de Wright pour les séries génératrices des composantes d'excès $\ell$.

**Définition 2.3.28.** Notons $X$ la SGE des chaînes non vide

$$X(t) = \frac{1-\theta(t)}{\theta(t)}\,. \tag{2.156}$$

Nous sommes maintenant en mesure de donner les SGEs $H_\ell$ des composantes d'excès $\ell$ :

**Théorème 2.3.29.** *La SGE $H_{-1} \circ T(z)$ des hyperarbres non enracinés est telle que :*

$$H_{-1}(t) = t - \frac{(b-1)t^b}{b!} = t\left(\frac{b-1+\theta(t)}{b}\right)\,. \tag{2.157}$$

**Théorème 2.3.30.** *La SGE $H_0 \circ T(z)$ des hypercycles est telle que :*

- *si $b = 2$ (pour les graphes)*

$$W_0(t) = -\ln\left(\sqrt{\theta(t)}\right) - \frac{1-\theta(t)}{2} - \frac{(1-\theta)^2}{4}\,, \tag{2.158}$$

$$W_0(t) = -\ln\left(\sqrt{1-t}\right) - \frac{t}{2} - \frac{t^2}{4}\,. \tag{2.159}$$

- *si $b \geq 3$*

$$H_0(t) = -\ln\left(\sqrt{\theta(t)}\right) - \frac{1-\theta(t)}{2}\,. \tag{2.160}$$

**Théorème 2.3.31.** *La SGE $H_1 \circ T(z)$ des composantes d'excès 1 est telle que :*

- *si $b = 2$ (pour les graphes)*

$$W_1(t) = \frac{5t^5}{24(1-t)^3} + \frac{t^4}{4(1-t)^2}\,. \tag{2.161}$$

- *si $b = 3$*

$$H_1(t) = \frac{(1-\theta(t))^2}{t}\left(\frac{5}{6}X(t)^3 + \frac{19}{12}X(t)^2 + \frac{5}{6}X(t)\right)\,. \tag{2.162}$$

– *si* $b \geq 4$

$$H_1(t) = \frac{1-\theta(t)}{t}\left(\frac{5(b-1)^2X(t)^3}{24}+\right. \tag{2.163}$$

$$\left.+\frac{(7b-12)(b-1)X(t)^2}{24}+\frac{(b-2)^2X(t)}{12}\right). \tag{2.164}$$

**Théorème 2.3.32.** *La SGE* $H_2 \circ T(z)$ *des composantes d'excès* 2 *est telle que :*

– *si* $b = 2$ *(pour les graphes)*

$$W_2(t) = \frac{5t^8}{16(1-t)^6}+ \tag{2.165}$$

$$+\frac{55t^7}{48(1-t)^5}+\frac{73t^6}{48(1-t)^4}+\frac{3t^5}{4(1-t)^3}+\frac{t^4}{24(1-t)^2}. \tag{2.166}$$

– *si* $b = 3$

$$H_2(t) = \frac{(1-\theta(t))^2}{t^2}\left(5X(t)^6+\frac{55}{3}X(t)^5+\right. \tag{2.167}$$

$$\left.+\frac{307}{12}X(t)^4+\frac{199}{12}X(t)^3+\frac{9}{2}X(t)^2+\frac{1}{6}X(t)\right). \tag{2.168}$$

– *si* $b = 4$

$$H_2(t) = \frac{(1-\theta(t))^2}{t^2}\left(\frac{405}{16}X(t)^6+\frac{405}{4}X(t)^5+\right. \tag{2.169}$$

$$\left.+\frac{315}{2}X(t)^4+118X(t)^3+\frac{2017}{48}X(t)^2+\frac{17}{3}X(t)\right). \tag{2.170}$$

– *si* $b \geq 5$

$$H_2(t) = \frac{1-\theta(t)}{t^2}\left(\frac{5(b-1)^4X(t)^6}{16}+\right. \tag{2.171}$$

$$+\frac{5(11b-17)(b-1)^3X(t)^5}{48}+ \tag{2.172}$$

$$+\frac{(2b-3)(38b-65)(b-1)^2X(t)^4}{48}+ \tag{2.173}$$

$$+\frac{(b-1)(48b^3-244b^2+411b-229)X(t)^3}{48}+ \tag{2.174}$$

$$\left.+\frac{(b-2)^2(13b^2-44b+35)X(t)^2}{48}+\frac{(b-2)^2(b-3)^2X(t)}{48}\right). \tag{2.175}$$

**Théorème 2.3.33.** *La SGE* $H_3 \circ T(z)$ *des composantes d'excès* 3 *est telle que :*

– *si* $b = 2$ *(pour les graphes)*

$$W_3(t) = \frac{1105t^{11}}{1152(1-t)^9}+\frac{395t^{10}}{72(1-t)^8}+\frac{15131t^9}{1152(1-t)^7}+ \tag{2.176}$$

$$+\frac{2399t^8}{144(1-t)^6}+\frac{8303t^7}{720(1-t)^5}+\frac{557t^6}{144(1-t)^4}+\frac{3t^5}{8(1-t)^3}. \tag{2.177}$$

– *si* $b = 3$

$$H_3(t) = \frac{(1-\theta(t))^3}{t^3}\left(\frac{1105}{18}X(t)^9 + \frac{4565}{12}X(t)^8 + \right. \tag{2.178}$$

$$+\frac{72793}{72}X(t)^7 + \frac{35963}{24}X(t)^6 + \frac{1937021}{1440}X(t)^5 + \tag{2.179}$$

$$\left. +\frac{7397}{10}X(t)^4 + \frac{42701}{180}X(t)^3 + \frac{4553}{120}X(t)^2 + \frac{41}{24}X(t)\right). \tag{2.180}$$

– *si* $b = 4$

$$H_3(t) = \frac{(1-\theta(t))^3}{t^3}\left(\frac{89505}{128}X(t)^9 + \frac{74925}{16}X(t)^8 + \right. \tag{2.181}$$

$$+\frac{217431}{16}X(t)^7 + \frac{713511}{32}X(t)^6 + \frac{14467311}{640}X(t)^5 + \tag{2.182}$$

$$+\frac{580073}{40}X(t)^4 + \frac{4154279}{720}X(t)^3 + \frac{59588}{45}X(t)^2 + \tag{2.183}$$

$$\left. +\frac{2641}{18}X(t) + \frac{79}{18}\right). \tag{2.184}$$

– *si* $b = 5$

$$H_3(t) = \frac{(1-\theta(t))^2}{t^3}\left(\frac{35360}{9}X(t)^9 + \frac{210280}{9}X(t)^8 + \right. \tag{2.185}$$

$$+\frac{532426}{9}X(t)^7 + 82729X(t)^6 + \frac{8303267}{120}X(t)^5 + \tag{2.186}$$

$$+\frac{12565331}{360}X(t)^4 + \frac{4851107}{480}X(t)^3 + \frac{46821}{32}X(t)^2 + \tag{2.187}$$

$$\left. +\frac{291}{4}X(t)\right). \tag{2.188}$$

– *si* $b \geq 6$

$$H_3(t) = \frac{1-\theta(t)}{t^3}\left(\frac{1105(b-1)^6X(t)^9}{1152} + \right. \tag{2.189}$$

$$+\frac{5(1259b-1922)(b-1)^5X(t)^8}{1152} + \tag{2.190}$$

$$+\frac{(b-1)^4(15106b^2-47108b+36643)X(t)^7}{1152} + \tag{2.191}$$

$$+\frac{(b-1)^3(9867b^3-47368b^2+75592b-40080)X(t)^6}{576} + \tag{2.192}$$

$$+\frac{(b-1)^2(75529b^4-499564b^3+1234860b^2-1351152b+551736)X(t)^5}{5760} + \tag{2.193}$$

$$+\frac{(b-1)(33791b^5-291742b^4+1002991b^3-1715000b^2+1457088b-491520)X(t)^4}{5760} + \tag{2.194}$$

$$+\frac{(b-2)^2(2056b^4-14424b^3+37353b^2-42148b+17404)X^3}{1440} + \tag{2.195}$$

$$+\frac{(b-2)^2(56b^4-514b^3+1753b^2-2623b+1444)X(t)^2}{360} + \tag{2.196}$$

$$\left. +\frac{(b-2)^2(b-3)^2(b-4)^2X(t)}{240}\right). \tag{2.197}$$

**Remarque 2.3.34.** Les SGEs $H_\ell$ des composantes lisses d'excès $\ell$ pour $b \geq \ell + 3$ admettent une même expression fonction de $b$.

**Lemme 2.3.35.** *Les $f_{jk}$ de la remarque 2.3.21 sont des polynômes de Laurent de degré minimum égal à*

$$\begin{cases} -(3j+2k)-1 & si\ (j,k)=(-1,1) \\ -(3j+2k) & si\ (j,k)\neq(-1,1)\,. \end{cases} \tag{2.198}$$

*Preuve.* La preuve est immédiate par la récurrence des $f_{jk}$ du théorème 2.3.24. ♢

**Lemme 2.3.36.** *Considérons les familles de composantes d'excès $j_i$ et ayant $k_i$ sommets marqués, $-1 \leq j_i \leq \ell - 1$ et $1 \leq k_i \leq b$, composantes qui se recombinent en composante $b$-uniforme d'excès $\ell$ et sans hyperarête multiple. Seules les familles ayant exactement $(b-2)$ hyperarbres enracinés suffisent à la détermination du coefficient du terme de degré minimum $-3\ell$ de $f_\ell$. Et ces familles avec celles ayant exactement $(b-3)$ hyperarbres enracinés suffisent à la détermination du coefficient du terme de degré $(-3\ell+1)$ de $f_\ell$.*

*Preuve.* Notons par $m_{-1,1}$ le nombre de hyperarbres d'une famille fixée. Alors le degré minimum des termes (dans $f_\ell$) liées à la famille considérée est

$$-\sum_i (3j_i + 2k_i) - m_{-1,1} - 1 = -(3\ell+3) + (b - m_{-1,1}) - 1\,, \tag{2.199}$$

d'après le lemme précédent et la décomposition relatée dans l'expression $\hat{J}$ du théorème 2.3.24. ♢

Ce dernier lemme nous permet d'expliciter les deux premiers coefficients de $f_\ell$ à savoir celui du terme de degré respectivement $-3\ell$ puis $(-3\ell+1)$ pour les composantes complexes $(\ell \geq 1)$. Notre motivation pour mieux inspecter ces coefficients vient de leur importance dans la littérature pour faire de l'énumération asymptotique comme souligné dans [28, page 42] : le premier coefficient est lié à ce qui est aussi dénommé coefficient de Wright-Stepanove (voir [4]) ou encore coefficient de Wright-Takács-Louchard et son expression asymptotique a fait l'objet d'étude [6, 7] permettant alors d'obtenir des énumérations de composantes d'excès tendant vers l'infini avec le nombre de sommets [24].

**Théorème 2.3.37.** *La SGE $H_\ell$ des composantes d'excès $\ell \geq 1$ est telle que :*

$$H_\ell(t) = \frac{1}{t^\ell}\left(\frac{\lambda_\ell (b-1)^{2\ell}}{3\ell\theta(t)^{3\ell}} - \frac{(\kappa_\ell - \nu_\ell(b-2))(b-1)^{2\ell-1}}{(3\ell-1)\theta(t)^{3\ell-1}} + \sum_{j=-3\ell+2}^{\lfloor \frac{\ell+1}{b-1}+1 \rfloor} c_j(\ell,b)\theta(t)^j\right)\,, \tag{2.200}$$

*avec*

$$\lambda_0 = \frac{1}{2} \tag{2.201}$$

$$\lambda_\ell = \frac{1}{2}\lambda_{\ell-1}(3\ell-1) + \frac{1}{2}\sum_{p=0}^{\ell-1}\lambda_p\lambda_{\ell-1-p} \qquad pour\ \ell = 1, 2 \ldots \tag{2.202}$$

$$\nu_1 = \frac{1}{3!} + \frac{1}{2!}\lambda_0 \quad (2.203)$$

$$\nu_\ell = +\frac{1}{6}\sum_{s=0}^{\ell-2}\sum_{p=0}^{\ell-2-s}\lambda_s\lambda_p\lambda_{\ell-2-s-p} \quad (2.204)$$

$$+\frac{1}{2}\lambda_{\ell-1} + \frac{1}{2}\sum_{p=0}^{\ell-2}(3p+2)\lambda_p\lambda_{\ell-2-p} \quad (2.205)$$

$$+\frac{1}{6}(3\ell-4)(3\ell-2)\lambda_{\ell-2} \qquad \textit{pour } \ell = 2,3,\dots, \quad (2.206)$$

*soit $\mu_0$ tel que :*

$$\mu_0 = b-1 \quad (2.207)$$

*alors*

$$\kappa_\ell = \frac{1}{2}\left((3\ell-2)\mu_{\ell-1} + (3b\ell - b - 2\ell)\,\lambda_{\ell-1}\right) + \sum_{p=0}^{\ell-1}\mu_p\lambda_{\ell-1-p}\,, \quad (2.208)$$

*où $\mu_\ell$ est à calculer au besoin*

$$\mu_\ell = \kappa_\ell - \nu_\ell(b-2) + \lambda_\ell(b-\frac{2}{3}) \qquad \textit{pour } \ell = 1,2\dots \quad (2.209)$$

*(Notons que $\kappa_\ell = \kappa_\ell(b)$).*

Dans ce chapitre, nous avons obtenu des énumérations explicites des hyperarbres enracinés (donc aussi de ceux non enracinés) , des forêts de $(k+1)$ hyperarbres enracinés (donc par déduction des forêts d'hyperarbres enracinés) et des hypercycles. La complexité rencontrée dans les expressions des énumérations explicites nous a mené à expliciter non pas le nombre même des composantes d'excès donné, mais leur série génératrice via la notion des composantes lisses, et permettre ainsi de procéder à l'énumération en utilisant la formule d'inversion de Lagrange. Si la détermination des SGEs a pu être automatisée grâce entre autre à la grande expressivité des SGEs pour décrire des décompositions des structures à énumérer, la complexité des expressions explicites nous force à préférer des équivalents asymptotiques. L'énumération asymptotique est traitée dans le chapitre suivant.

# Chapitre 3

# Énumération asymptotique

Dans ce chapitre, notre but est de fournir un équivalent asymptotique du nombre de composantes d'excès $\ell$ quand la taille, c'est à dire le nombre de sommets, est grande. Notre motivation vient, d'une part, de la difficulté du calcul du nombre exact pour des composantes de grande taille, et d'autre part, de la connaissance des SGEs des composantes lisses permettant de faire de l'estimation asymptotique via la formule d'inversion de Lagrange, la formule intégrale de Cauchy en analyse complexe et, ici, notamment par la méthode du point col pour obtenir nos résultats. Les notions relatives à l'utilisation des séries génératrices pour l'énumération asymptotique apparaissent dans [26], en particulier ici les résultats asymptotiques sont obtenus via la méthode du point col - voir [15, 8] pour plus de précisions sur les méthodes asymptotiques.

Le plan de ce chapitre est le suivant : nous commençons par l'énumération asymptotique des hyperarbres enracinés, puis enchaînons avec celle des hypercycles et démontrons un encadrement des coefficients $[z^n]\, H_\ell \circ T(z)\,,\ \ell \geq 1\,,$ pour finalement après établir le nombre asymptotique des composantes complexes d'excès donné.

## 3.1 Énumération asymptotique des hyperarbres

Dans cette section, étant vue l'importance des hyperarbres enracinés pour l'énumération exacte des hypergraphes, nous procédons à l'énumération asymptotique de ces composantes les plus simples. La méthode du point col, pour faire de l'énumération asymptotique, est ici illustrée et validée parce que le résultat se vérifie aussi par une autre preuve : via la formule de Stirling, connaissant l'expression du nombre des hyperarbres enracinés à un nombre de sommets donné.

Nous commençons par un exemple, en fixant la valeur de $b = 3$, et déterminons le développement asymptotique des coefficients de la SGE des hyperarbres enracinés 3-uniformes.

**Proposition 3.1.1.** *Le nombre des hyperarbres* 3*-uniformes enracinés ayant* $(2s+1)$ *sommets est déterminé à partir du coefficient de* $z^{2s+1}$ *du développement de la SGE*

$$T_3(z) = z \exp\left(\frac{T_3(z)^2}{2}\right) . \qquad (3.1)$$

*Ce coefficient admet le développement asymptotique, quand* $(s \to \infty)$ *, suivant*

$$\left[z^{2s+1}\right] T_3(z) = \frac{e^{s+1/2}}{2s\sqrt{2\pi s}} \left(1 - \frac{17}{24s} + \frac{481}{1152s^2} - \frac{96133}{414720s^3} + \dots\right). \tag{3.2}$$

*Preuve.*(Via Stirling) Par le corollaire 2.2.4 , le nombre des hyperarbres 3-uniformes enracinés ayant $s$ hyperarêtes et $n = n(s) = 2s + 1$ sommets est

$$\frac{n!}{2^s s!} n^{s-1} = \frac{n!}{2^s s!}(2s+1)^{s-1}. \tag{3.3}$$

Le coefficient de $z^{2s+1}$ de la SGE $T_3(z)$ est donc

$$\left[z^{2s+1}\right] T_3(z) = \frac{(2s+1)^{s-1}}{2^s s!}. \tag{3.4}$$

Par l'expansion de Stirling nous obtenons

$$\left[z^{2s+1}\right] T_3(z) = \frac{(1+\frac{1}{2s})^{s-1} e^s}{2s\sqrt{2\pi s}} \frac{1}{\left(1 + \frac{1}{12s} + \frac{1}{288s^2} - \frac{139}{51840s^3} - \dots\right)}. \tag{3.5}$$

Comme

$$(1 + \frac{1}{2s})^{s-1} = e^{1/2}\left(1 - \frac{5}{8s} + \frac{139}{384s^2} - \frac{207}{1024s^3} + \dots\right) \tag{3.6}$$

nous obtenons l'expansion asymptotique du coefficient $z^{2s+1}$ de $T_3(z)$ suivant

$$\left[z^{2s+1}\right] T_3(z) = \frac{e^{s+1/2}}{2s\sqrt{2\pi s}} \left(1 - \frac{17}{24s} + \frac{481}{1152s^2} - \frac{96133}{414720s^3} + \dots\right). \tag{3.7}$$

◊

Nous retrouvons une version de cette proposition 3.1.1 via la formule d'inversion de Lagrange et via la formule intégrale de Cauchy :

**Proposition 3.1.2.** *Le nombre des hyperarbres* 3*-uniformes enracinés ayant* $(2s + 1)$ *sommets est déterminé à partir du coefficient de* $z^{2s+1}$ *du développement de la SGE* $T_3(z)$ *définie par l'équation* (3.1) *. Ce coefficient est, pour* $(s \to \infty)$ *, tel que*

$$\left[z^{2s+1}\right] T_3(z) = \frac{\exp(s+1/2)}{\sqrt{\pi}(2s+1)^{3/2}} \left\{1 + O(\frac{1}{(2s+1)^{1/6}})\right\}. \tag{3.8}$$

*Preuve.*(Méthode du point col) Par la définition implicite (3.1) de la SGE $T_3(z)$ , la formule d'inversion de Lagrange donne :

$$\left[z^{2s+1}\right] T_3(z) = \frac{1}{2s+1} \left[t^{2s}\right] \exp\left((s+1/2)t^2\right). \tag{3.9}$$

Dans le second membre de cette équation, en extrayant le coefficient de $t^{2s}$ par la formule intégrale de Cauchy, nous trouvons

$$\left[z^{2s+1}\right] T_3(z) = \frac{1}{2i\pi(2s+1)} \oint \frac{\exp\left((s+1/2)t^2\right)}{t^{2s+1}} \mathrm{d}\,t, \tag{3.10}$$

où le contour intégral encercle l'origine du plan complexe. Ainsi, le coefficient de la SGE s'écrit

$$\left[z^{2s+1}\right] T_3(z) = \frac{1}{2i\pi(2s+1)} \oint \exp\left((s+1/2)(t^2 - \ln(t^2))\right) \mathrm{d}\, t\,. \tag{3.11}$$

Le paramètre $s$, qui dénombre les hyperarêtes, est destiné à tendre vers l'infini. Nous soupçonnons dans l'équation précédente un contexte dans lequel la méthode du point col peut être appliquée. Notons que la dérivée de l'exposant dans l'intégrande de l'équation s'annule en deux points $t = \pm 1$ :

$$\frac{\mathrm{d}}{\mathrm{d}\, t}(s+1/2)(t^2 - \ln(t^2)) = 2(s+1/2)(t - \frac{1}{t})\,. \tag{3.12}$$

Nous prenons alors comme contour intégral le carré centré en $0$, aux côtés parallèles aux axes et de diagonale $2\sqrt{2}$. Ce carré passe par les points cols $t = \pm 1$. Dans la figure 3.1, en représentant quand $t$ prend ses valeurs sur ce contour la valeur de la partie réelle de $(t^2 - \ln(t^2))$, nous voyons que quand $s$ est grand la contribution, dans le module de l'intégrande, des piques aux deux points cols est très accentuée, alors deux voisinages limités autour de chacun des points cols suffiront pour capturer l'asymptotique du coefficient recherché. Notons l'équation paramétrique du carré $\gamma$

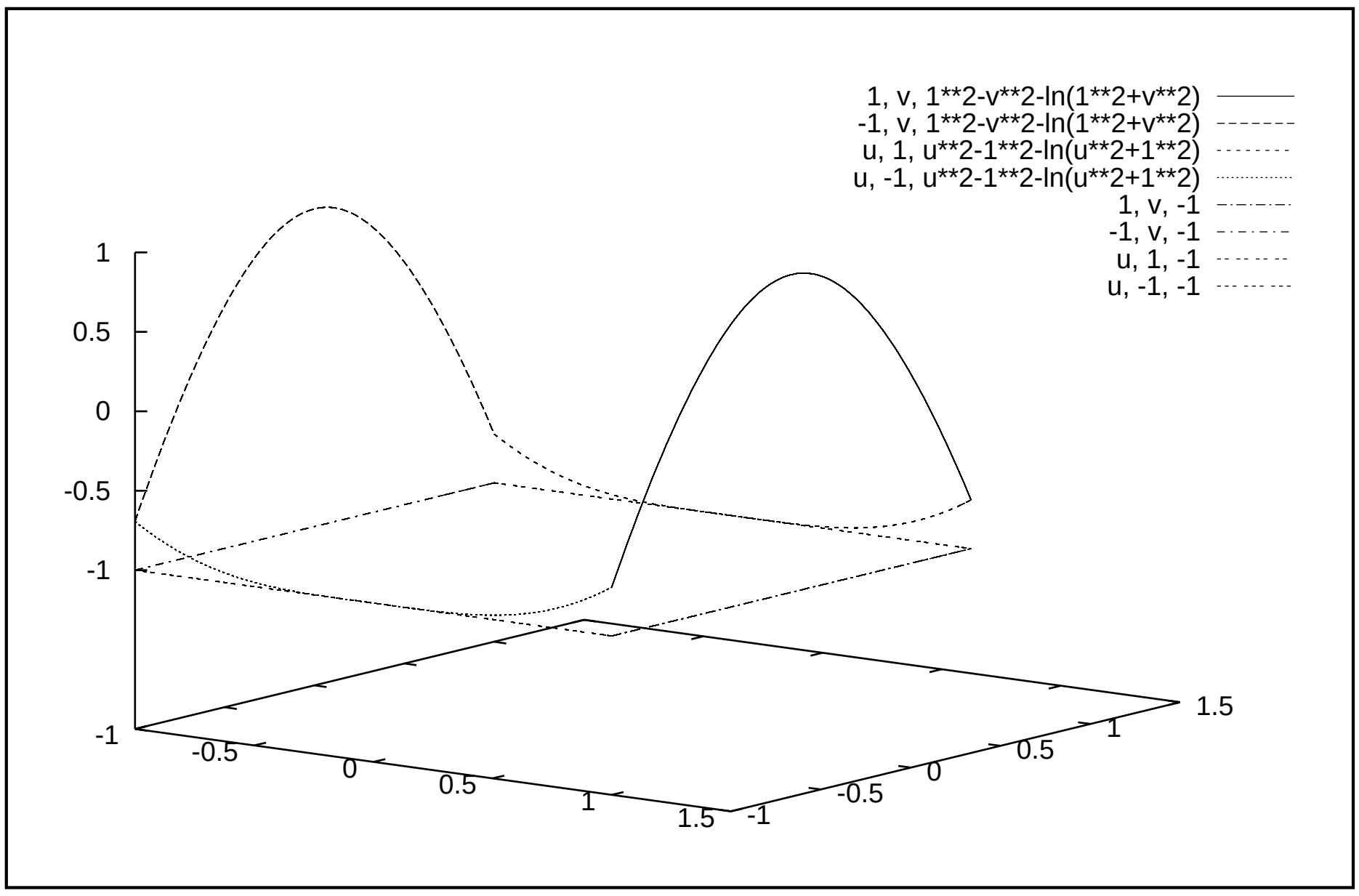

FIG. 3.1 – La valeur de $\Re(t^2 - \ln(t^2))$ sur le contour carré choisi.

$$\gamma : \begin{cases} \gamma_1 : t = 1 + iv\,, & v \uparrow\in [-1, 1] \\ \gamma_{-1} : t = -1 + iv\,, & v \downarrow\in [-1, 1] \\ \gamma_i : u + i\,, & u \downarrow\in [-1, 1] \\ \gamma_{-i} : u + i\,, & u \uparrow\in [-1, 1]\,. \end{cases} \tag{3.13}$$

Soit $\hat{\gamma}_1$ un voisinage de $t = 1$ du chemin $\gamma_1$ . Pour avoir l'ordre de grandeur, nous approchons, dans l'intégrande, l'exposant par

$$\Re((s+1/2)(t^2 - \ln(t^2))) = \Re((s+1/2)(1 + 2(t-1)^2 + O((t-1)^4)))\,. \tag{3.14}$$

Nous considérons alors l'intégrale

$$\begin{aligned} I_1 &= \frac{1}{2i\pi(2s+1)} \int_{\hat{\gamma}_1} \exp\left((s+1/2)(t^2 - \ln(t^2))\right) \mathrm{d}\,t & (3.15)\\ &= \frac{\exp(s+1/2)}{2i\pi(2s+1)} \int_{\hat{\gamma}_1} \exp\left(\Re((2s+1)((t-1)^2 + O((t-1)^4)))\right) \mathrm{d}\,t\,. & (3.16)\end{aligned}$$

En effet, avec le changement de variable

$$\begin{cases} t = 1 + \frac{ir}{\sqrt{2s+1}} \\ \mathrm{d}\,t = \frac{i}{\sqrt{2s+1}} \mathrm{d}\,r\,, \end{cases} \tag{3.17}$$

nous avons

$$\begin{aligned} I_1 &= \frac{\exp(s+1/2)}{2\pi(2s+1)^{3/2}} \int_{-\kappa_s}^{\kappa_s} \exp\left(-r^2 + O(\frac{r^4}{(2s+1)})\right) \mathrm{d}\,r & (3.18)\\ &= \frac{\exp(s+1/2)}{2\pi(2s+1)^{3/2}} \int_{-\kappa_s}^{\kappa_s} \exp\left(-r^2\right)(1 + O(\frac{r^4}{(2s+1)}))\mathrm{d}\,r & (3.19)\\ &= \frac{\exp(s+1/2)}{2\pi(2s+1)^{3/2}} \left\{ \int_{-\kappa_s}^{\kappa_s} \exp\left(-r^2\right) \mathrm{d}\,r + \int_{-\kappa_s}^{\kappa_s} \exp\left(-r^2\right) O(\frac{r^4}{(2s+1)})\mathrm{d}\,r \right\} & (3.20)\end{aligned}$$

$r$ étant une variable réelle,

$$\begin{aligned} \int_{-\kappa_s}^{\kappa_s} \exp\left(-r^2\right) O(\frac{r^4}{(2s+1)})\mathrm{d}\,r &= \int_{-\kappa_s}^{\kappa_s} O(\frac{r^4}{(2s+1)})\mathrm{d}\,r & (3.21)\\ &= O(\frac{{\kappa_s}^5}{(2s+1)})\,. & (3.22)\end{aligned}$$

Ainsi, sur la portion $\hat{\gamma}_1$ du chemin $\gamma_1$ , approcher l'exposant permet d'avoir

$$I_1 = \frac{\exp(s+1/2)}{2\pi(2s+1)^{3/2}} \left\{ \int_{-\kappa_s}^{\kappa_s} \exp\left(-r^2\right) \mathrm{d}\,r + O(\frac{{\kappa_s}^5}{(2s+1)}) \right\}\,. \tag{3.23}$$

Comme

$$\begin{aligned} \int_{\kappa_s}^{\infty} \exp(-r^2)\mathrm{d}\,r &\leq \int_{\kappa_s}^{\infty} \exp(-\kappa_s r)\mathrm{d}\,r & (3.24)\\ &\leq \frac{1}{\kappa_s} \int_{{\kappa_s}^2}^{\infty} \exp(-r)\mathrm{d}\,r & (3.25)\\ &= O(\frac{\exp(-{\kappa_s}^2)}{\kappa_s})\,, & (3.26)\end{aligned}$$

en prenant $\kappa_s = (2s+1)^{1/6}$ pour définir $\hat{\gamma}_1$ , nous trouvons sur ce chemin

$$I_1 = \frac{\exp(s+1/2)}{2\pi(2s+1)^{3/2}} \left\{ \int_{-\infty}^{\infty} \exp\left(-r^2\right) \mathrm{d}\, r + O(\frac{1}{(2s+1)^{1/6}}) \right\} . \tag{3.27}$$

Nous y reconnaissons l'intégrale de Gauss et pouvons écrire que

$$I_1 = \frac{\exp(s+1/2)}{2\sqrt{\pi}(2s+1)^{3/2}} \left\{ 1 + O(\frac{1}{(2s+1)^{1/6}}) \right\} . \tag{3.28}$$

Des lignes de preuve similaires s'appliquent pour la portion $\gamma_{-1}$ du contour. Sur cette portion, en prenant le voisinage $\hat{\gamma}_{-1}$ centré sur le point col $-1$ et de longueur deux fois $\kappa_s$, qui peut être pris égale à $(2s+1)^{1/3}$ , l'exposant de l'intégrande peut être approché et nous obtenons

$$\begin{aligned}
I_{-1} &= \frac{1}{2i\pi(2s+1)} \int_{\hat{\gamma}_{-1}} \exp\left((s+1/2)(t^2 - \ln(t^2))\right) \mathrm{d}\, t && (3.29)\\
&= \frac{\exp(s+1/2)}{2\pi(2s+1)^{3/2}} \left\{ \int_{-\kappa_s}^{\kappa_s} \exp\left(-r^2\right) \mathrm{d}\, r + O(\frac{{\kappa_s}^5}{(2s+1)}) \right\} && (3.30)\\
&= \frac{\exp(s+1/2)}{2\sqrt{\pi}(2s+1)^{3/2}} \left\{ 1 + O(\frac{1}{(2s+1)^{1/6}}) \right\} . && (3.31)
\end{aligned}$$

Soit la portion de $\gamma_1$

$$\gamma_1' : t = 1 + iv, \;\; v \uparrow\in [\kappa_s/\sqrt{2s+1}, 1] . \tag{3.32}$$

Nous approximons l'exposant de l'intégrande sur cette portion au point $t = 1 + i\kappa_s/\sqrt{2s+1}$ .

$$\begin{aligned}
&\left| \frac{1}{2i\pi(2s+1)} \int_{\gamma_1'} \exp\left((s+1/2)(t^2 - \ln(t^2))\right) \mathrm{d}\, t \right| = && (3.33)\\
&\leq \frac{1}{2\pi(2s+1)} \int_{\kappa_s/\sqrt{2s+1}}^{1} \exp\left((s+1/2)(1 - v^2 - \ln(1+v^2))\right) \mathrm{d}\, v = && (3.34)\\
&= \frac{\exp(s+1/2)}{2\pi(2s+1)} \int_{\kappa_s/\sqrt{2s+1}}^{1} \exp\left((s+1/2)(-2v^2 + O(v^4))\right) \mathrm{d}\, v = && (3.35)\\
&= \frac{\exp(s+1/2)}{2\pi(2s+1)^{3/2}} \int_{\kappa_s}^{\sqrt{2s+1}} \exp\left(-r^2 + O(\frac{r^4}{(2s+1)})\right) \mathrm{d}\, r = && (3.36)\\
&= \frac{\exp(s+1/2)}{2\pi(2s+1)^{3/2}} \int_{\kappa_s}^{\sqrt{2s+1}} \exp\left(-r^2\right) (1 + O(\frac{r^4}{(2s+1)})) \mathrm{d}\, r = && (3.37)\\
&= \frac{\exp(s+1/2)}{2\pi(2s+1)^{3/2}} O(\frac{\exp(-{\kappa_s}^2)}{\kappa_s}) . && (3.38)
\end{aligned}$$

L'ordre de grandeur ci-dessus est valide pour $\kappa_s = (2s+1)^{1/6}$ et il souligne que, sur la portion de $\gamma_1$ , la contribution de l'intégrale est exponentiellement petite en dehors de $\hat{\gamma}_1$ .

Soit la portion de $\gamma_i$

$$\gamma_i' : t = u + i, \;\; u \downarrow\in [0, 1] . \tag{3.39}$$

Nous approximons l'exposant de l'intégrande sur cette portion au point $t = 1 + i$.

$$\left| \frac{1}{2i\pi(2s+1)} \int_{\gamma_i'} \exp\left((s+1/2)(t^2 - \ln(t^2))\right) \mathrm{d}\, t \right| = \tag{3.40}$$

$$\leq \frac{1}{2\pi(2s+1)} \int_1^0 \exp\left((s+1/2)(u^2 - 1 - \ln(u^2+1))\right) \mathrm{d}\, v = \tag{3.41}$$

$$= \frac{\exp(-(s+1/2)\ln(2))}{2\pi(2s+1)} \int_1^0 \exp\left((s+1/2)(-(1-u) + O((u-1)^2))\right) \mathrm{d}\, u = \tag{3.42}$$

$$= \frac{\exp(-(s+1/2)\ln(2))}{2\pi(2s+1)} \int_1^0 \exp\left((s+1/2)(-r + O(r^2))\right) \mathrm{d}\, r = \tag{3.43}$$

$$= \frac{\exp(-(s+1/2)\ln(2))}{\pi(2s+1)^2} \int_{s+1/2}^0 \exp\left(-r + O(\frac{r^2}{(s+1/2)})\right) \mathrm{d}\, r = \tag{3.44}$$

$$= \frac{\exp(-(s+1/2)\ln(2))}{\pi(2s+1)^2} \int_{s+1/2}^0 \exp(-r)(1 + O(\frac{r^2}{(s+1/2)}))\mathrm{d}\, r = \tag{3.45}$$

$$= O(\exp(-(s+1/2)\ln(2))) . \tag{3.46}$$

Ainsi, sur les portions $\gamma_i \cup \gamma_{-i}$ du contour $\gamma$, la contribution à l'asymptotique est exponentiellement petite :

$$\frac{1}{2i\pi(2s+1)} \int_{\gamma_i \cup \gamma_{-i}} \exp\left((s+1/2)(t^2 - \ln(t^2))\right) \mathrm{d}\, t = O(\frac{1}{2^{s+1/2}}) . \tag{3.47}$$

De même, sur les portions $(\gamma_1 \backslash \hat{\gamma}_1) \cup (\gamma_{-1} \backslash \hat{\gamma}_{-1})$ du contour $\gamma$, la contribution à l'asymptotique est exponentiellement petite par rapport à celle des portions limitées $\hat{\gamma}_1 \cup \hat{\gamma}_{-1}$ :

$$\frac{1}{2i\pi(2s+1)} \int_{(\gamma_1 \backslash \hat{\gamma}_1) \cup (\gamma_{-1} \backslash \hat{\gamma}_{-1})} \exp\left((s+1/2)(t^2 - \ln(t^2))\right) \mathrm{d}\, t = \tag{3.48}$$

$$= \frac{\exp(s+1/2)}{(2s+1)^{2/3}} O(\frac{\exp(-(2s+1)^{2/6})}{(2s+1)^{1/6}}) . \tag{3.49}$$

Ces deux ordres de grandeur étant exponentiellement petits par rapport à la contribution de l'intégrale sur les portions $\hat{\gamma}_1 \cup \hat{\gamma}_{-1}$, l'intégrale sur le contour $\gamma$ vaut

$$\left[z^{2s+1}\right] T_3(z) = \frac{\exp(s+1/2)}{\sqrt{\pi}(2s+1)^{3/2}} \left\{ 1 + O(\frac{1}{(2s+1)^{1/6}}) \right\} . \tag{3.50}$$

$\diamondsuit$

La méthode du point col, pour la démonstration de cette proposition, permet aussi d'avoir un développement asymptotique complet du coefficient $\left[z^{2s+1}\right] T_3(z)$. Notons que dans les deux propositions, le terme principal est identique mais l'échelle asymptotique est différente : par l'expansion de Stirling, l'échelle est plus précise car c'est en $1/s$, et par la méthode du point col, l'échelle est, à priori, en $1/s^{1/6}$.

La proposition 3.1.2 est un cas particulier du théorème suivant :

**Théorème 3.1.3.** *Le nombre des hyperarbres (b-uniformes) enracinés ayant* $(s(b-1)+1)$ *sommets est déterminé à partir du coefficient de* $z^{s(b-1)+1}$ *du développement de la SGE*

$$T(z) = z \exp\left(\frac{T(z)^{b-1}}{(b-1)!}\right). \tag{3.51}$$

*Ce coefficient est, pour* $(s \to \infty)$ *, tel que*

$$\left[z^{s(b-1)+1}\right] T(z) = \frac{\exp\left(\frac{n}{b-1}\right)}{n[(b-2)!]^s \sqrt{2s\pi}} \left\{1 + O(\frac{1}{s^{1/6}})\right\}, \tag{3.52}$$

*avec* $n = n(s) = s(b-1) + 1$ *.*

*Preuve.* La définition implicite (3.51) donne, par la formule d'inversion de Lagrange avec la formule intégrale de Cauchy, une expression intégrale du coefficient $\left[z^{s(b+1)+1}\right]$ de la SGE $T(z)$ :

$$[z^n]\, T(z) = \frac{1}{2i\pi n} \oint \exp\left(\frac{n}{b-1}\frac{t^{b-1}}{(b-2)!} - s\ln(t^{b-1})\right) \frac{\mathrm{d}\,t}{t}, \tag{3.53}$$

avec $n = n(s) = s(b-1) + 1$ et avec un contour qui encercle l'origine du plan complexe. Cette formulation intégrale se simplifie en adoptant la notation $\tau(t)$ que nous rappelons ci-après

$$\tau(t) = \frac{t^{b-1}}{(b-2)!}. \tag{3.54}$$

En effet, nous obtenons l'expression suivante du coefficient

$$[z^n]\, T(z) = \tag{3.55}$$

$$= \frac{1}{2i\pi n[(b-2)!]^s} \oint \exp\left(\frac{\tau(t)}{b-1}\right) \exp\left(s(\tau(t) - \ln(\tau(t)))\right) \frac{\mathrm{d}\,t}{t} \tag{3.56}$$

$$= \frac{1}{2i\pi n[(b-2)!]^s(b-1)} \oint \exp\left(\frac{\tau(t)}{b-1}\right) \exp\left(s(\tau(t) - \ln(\tau(t)))\right) \frac{\mathrm{d}\,\tau(t)}{\tau(t)}. \tag{3.57}$$

Comme la dérivée logarithmique de la fonction $\tau$ est

$$\frac{\mathrm{d}\,\tau}{\tau} = (b-1)\frac{\mathrm{d}\,t}{t}, \tag{3.58}$$

$\tau$ définit un changement de variable, et pour garder un contour simple, il faut multiplier par $(b-1)$ :

$$[z^n]\, T(z) = \frac{1}{2i\pi n[(b-2)!]^s} \oint \exp\left(\frac{\tau}{b-1}\right) \exp\left(s(\tau - \ln(\tau))\right) \frac{\mathrm{d}\,\tau}{\tau}, \tag{3.59}$$

soit, en gardant la notation $t$ à la place de $\tau$ ,

$$[z^n]\, T(z) = \frac{1}{2i\pi n[(b-2)!]^s} \oint \exp\left(\frac{t}{b-1}\right) \exp\left(s(t - \ln(t))\right) \frac{\mathrm{d}\,t}{t}. \tag{3.60}$$

Le paramètre $s$ dans l'exposant, nous indique que le point col à considérer est $t = 1$, soit la solution de l'équation

$$\frac{\mathrm{d}}{\mathrm{d}\,t}\left(t - \ln(t)\right) = 1 - \frac{1}{t} = 0\,. \tag{3.61}$$

Pour le choix du contour d'intégration passant par ce point col $t = 1$, nous prenons un demi cercle "gauche" $\gamma$ :

$$\gamma : \begin{cases} \gamma_1 : t = 1 + iv\,, & v \uparrow\in [-2, 2] \\ \gamma_0 : t = 1 + 2\exp(i\alpha)\,, & \alpha \uparrow\in [\pi/2, 3\pi/2]\,. \end{cases} \tag{3.62}$$

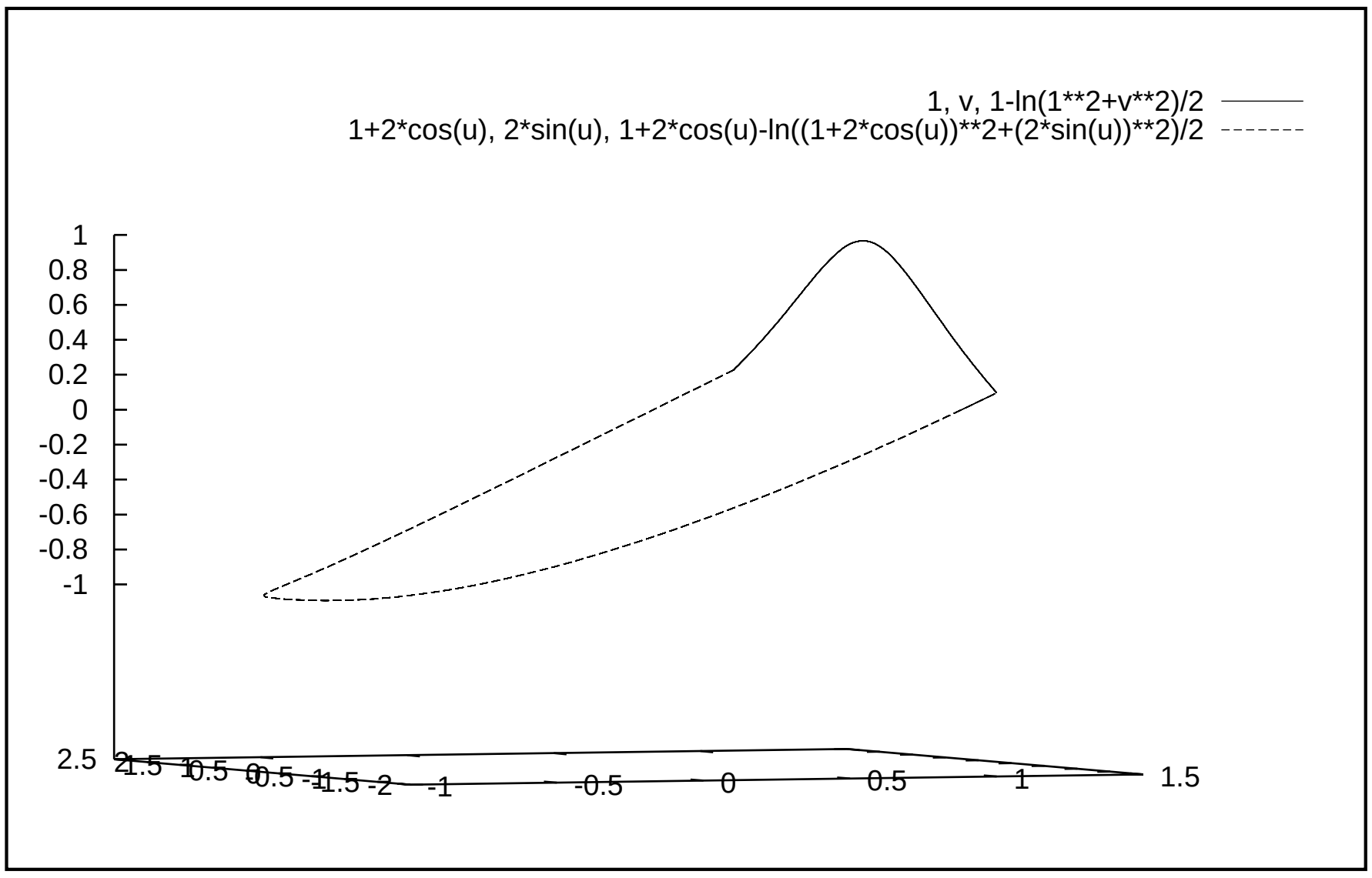

FIG. 3.2 – La valeur de $\Re(t - \ln(t))$ sur le contour (demi cercle gauche) $\gamma$.

Dans la figure 3.2, nous voyons que la contribution, pour le coefficient recherché de la SGE, dans le module de l'intégrande sera très accentuée au point col pour $(s \to \infty)$.

Notons $\kappa_s$ tel que :

$$\hat{\gamma}_1 : t = 1 + iv, \quad v \uparrow\in [-\kappa_s, \kappa_s]\,. \tag{3.63}$$

La contribution, au coefficient, de l'intégrale sur cette portion de chemin de $\gamma$ est

$$I_1 = \tag{3.64}$$

$$= \frac{1}{2i\pi n[(b-2)!]^s} \int_{\hat{\gamma}_1} \exp\left(\frac{t}{b-1}\right) \exp\left(s(t - \ln(t))\right) \frac{\mathrm{d}\,t}{t} = \tag{3.65}$$

$$= \frac{\exp(s + 1/(b-1))}{2i\pi n[(b-2)!]^s} \int_{\hat{\gamma}_1} \exp\left(s((t-1)^2/2 + O((t-1)^4))\right) \mathrm{d}\,t\,, \tag{3.66}$$

soit, avec le changement de variable

$$\begin{cases} t = 1 + \frac{ir\sqrt{2}}{\sqrt{s}} \\ \mathrm{d}\,t = \frac{i\sqrt{2}}{\sqrt{s}}\mathrm{d}\,r\,, \end{cases} \tag{3.67}$$

$$\begin{aligned} I_1 &= \frac{\exp(s+1/(b-1))}{\pi n[(b-2)!]^s\sqrt{2s}} \int_{-\kappa_s}^{\kappa_s} \exp\left(-r^2 + O(\frac{r^4}{s})\right) \mathrm{d}\, r & (3.68)\\ &= \frac{\exp(s+1/(b-1))}{\pi n[(b-2)!]^s\sqrt{2s}} \int_{-\kappa_s}^{\kappa_s} \exp(-r^2)(1+O(\frac{r^4}{s}))\mathrm{d}\, r & (3.69)\\ &= \frac{\exp(s+1/(b-1))}{\pi n[(b-2)!]^s\sqrt{2s}} \left\{ \int_{-\kappa_s}^{\kappa_s} \exp(-r^2)\mathrm{d}\, r + \int_{-\kappa_s}^{\kappa_s} O(\frac{r^4}{s})\mathrm{d}\, r \right\}. & (3.70)\end{aligned}$$

Ainsi, sur la portion $\hat{\gamma}_1$, approcher l'exposant donne l'ordre de grandeur de l'intégrale :

$$I_1 = \frac{\exp(s+1/(b-1))}{\pi n[(b-2)!]^s\sqrt{2s}} \left\{ \int_{-\kappa_s}^{\kappa_s} \exp(-r^2)\mathrm{d}\, r + O(\frac{\kappa_s^5}{s}) \right\}. \quad (3.71)$$

En prenant $\kappa_s = s^{1/6}$ et en complétant le domaine d'intégration, pour obtenir la droite réelle dans l'intégrale et aboutir ainsi à l'intégrale de Gauss, il s'ensuit la contribution de l'intégrale restreinte à la portion $\hat{\gamma}_1$ :

$$I_1 = \frac{\exp(s+1/(b-1))}{n[(b-2)!]^s\sqrt{2s\pi}} \left\{ 1 + O(\frac{1}{s^{1/6}}) \right\}. \quad (3.72)$$

Pour avoir le résultat, il faut encore négliger la contribution de l'intégrale sur le restant du contour défini en (3.62), à savoir $\gamma_1 \backslash \hat{\gamma}_1$ puis $\gamma_0$.

1. Sur la portion $\gamma_1 \backslash \hat{\gamma}_1$ du contour, la contribution au coefficient est

$$\begin{aligned} I_1' &= & (3.73)\\ &= \frac{1}{2i\pi n[(b-2)!]^s} \int_{\gamma_1\backslash\hat{\gamma}_1} \exp\left(\frac{t}{b-1}\right) \exp\left(s(t-\ln(t))\right) \frac{\mathrm{d}\, t}{t} = & (3.74)\\ &= 2\frac{\exp(s+1/(b-1))}{\pi n[(b-2)!]^s\sqrt{2s}} \int_{\kappa_s}^{\sqrt{2s}} \exp(-r^2)(1+O(\frac{r^4}{s}))\mathrm{d}\, r = & (3.75)\\ &= \frac{\exp(s+1/(b-1))}{n[(b-2)!]^s\sqrt{2s}} O(\frac{\exp(-\kappa_s^2)}{\kappa_s}). & (3.76)\end{aligned}$$

2. Sur la portion $\gamma_0$, considérons le chemin

$$\gamma_0' : t = 1 + 2\exp(i\alpha), \quad \alpha \uparrow\in [\pi/2, \pi], \quad (3.77)$$

la contribution sur ce chemin au coefficient est

$$I_0 = \tag{3.78}$$
$$= \frac{1}{2i\pi n[(b-2)!]^s} \int_{\gamma_0'} \exp\left(\frac{t}{b-1}\right) \exp\left(s(t-\ln(t))\right) \frac{\mathrm{d}\,t}{t} = \tag{3.79}$$
$$= O(\frac{\exp(1/(b-1))}{2\pi n[(b-2)!]^s} \int_{\gamma_0'} \exp\left(s(t-\ln(t))\right) \mathrm{d}\,t) = \tag{3.80}$$
$$= O(\frac{e^{1/(b-1)}}{\pi n[(b-2)!]^s} \tag{3.81}$$
$$\int_{\pi/2}^{\pi} e^{s(1+2\cos(\alpha)-\ln((1+\cos(\alpha))^2+4\sin(\alpha)^2)/2)} \mathrm{d}\,\alpha) = \tag{3.82}$$
$$= O(\frac{e^{s-s\ln(\sqrt{5})+1/(b-1)}}{\pi n[(b-2)!]^s} \int_{\pi/2}^{\pi} e^{-\frac{8s}{5}(\alpha-\frac{\pi}{2})} \mathrm{d}\,\alpha) = \tag{3.83}$$
$$= O(\frac{e^{s-s\ln(\sqrt{5})+1/(b-1)}}{\pi n[(b-2)!]^s}) = \tag{3.84}$$
$$= O(\frac{e^{s+1/(b-1)}}{\pi n[(b-2)!]^s 5^{s/2}})\,. \tag{3.85}$$

Nous trouvons le même ordre de grandeur sur la portion $\gamma_0 \backslash \gamma_0'$ :

$$\hat{I}_0 = \tag{3.86}$$
$$= \frac{1}{2i\pi n[(b-2)!]^s} \int_{\gamma_0' \backslash \gamma_0} \exp\left(\frac{t}{b-1}\right) \exp\left(s(t-\ln(t))\right) \frac{\mathrm{d}\,t}{t} = \tag{3.87}$$
$$= O(\frac{e^{s-s\ln(\sqrt{5})+1/(b-1)}}{\pi n[(b-2)!]^s} \int_{\pi}^{3\pi/2} e^{\frac{8s}{5}(\alpha-\frac{3\pi}{2})} \mathrm{d}\,\alpha) = \tag{3.88}$$
$$= O(\frac{e^{s+1/(b-1)}}{\pi n[(b-2)!]^s 5^{s/2}})\,. \tag{3.89}$$

Ainsi, la contribution au coefficient de la portion $\gamma_1 \backslash \hat{\gamma}_1$ avec celle de la portion $\gamma_0$ sont exponentiellement petites par rapport à celle de la seule portion $\hat{\gamma}_1$ du contour $\gamma$. Nous déduisons, l'ordre de grandeur $I_1$ du coefficient de la SGE :

$$\left[z^{s(b-1)+1}\right] T(z) = \frac{\exp\left(\frac{n}{b-1}\right)}{n[(b-2)!]^s \sqrt{2s\pi}} \left\{1 + O(\frac{1}{s^{1/6}})\right\}\,. \tag{3.90}$$

◇

Par définition d'une SGE $A(z)$ énumérant des structures $\mathcal{A}$ étiquetées, le nombre de ces structures de taille $n$ est $n!$ fois le coefficient $[z^n]\,A(z)$. Ainsi, connaissant l'équivalent asymptotique du factoriel par la formule de Stirling, $n! \sim n^n e^{-n}\sqrt{2\pi n}$, et connaissant un équivalent asymptotique du coefficient $[z^n]\,T(z)$ par le théorème 3.1.3 précédent, il se déduit le nombre asymptotique des hyperarbres enracinés donc de celui des hyperarbres non enracinés (conséquence directe de la proposition).

**Théorème 3.1.4.** *Le nombre d'hyperarbres enracinés ayant $s$ hyperarêtes et $n = n(s) = s(b-1)+1$ sommets, pour $(s \to \infty)$, est :*

$$n!\,[z^n]\,T(z) \sim \frac{\sqrt{b-1}s^{s(b-1)}}{\exp\left((b-2)(s+1/(b-1))\right)}\left[\frac{(b-1)^{b-1}}{(b-2)!}\right]^s . \tag{3.91}$$

*Preuve.* Pour prouver ce théorème il suffit de remarquer que

$$n! \sim n^n \exp(-n)\sqrt{2\pi n}\,, \tag{3.92}$$

et que

$$[z^n]\,T(z) \sim \frac{\exp\left(\frac{n}{b-1}\right)}{n[(b-2)!]^s\sqrt{2s\pi}} . \tag{3.93}$$

donc

$$n!\,[z^n]\,T(z) \sim \frac{n^{n-1}\sqrt{n/s}}{[(b-2)!]^s \exp\left(n-\frac{n}{b-1}\right)} . \tag{3.94}$$

Et l'expression (3.91) s'ensuit. ♢

Le nombre des hyperarbres non enracinés ayant $n$ sommets diffère d'un facteur $n$ en moins que celui des hyperarbres enracinés ayant le même nombre de sommets, il devient clair que

**Théorème 3.1.5.** *Le nombre d'hyperarbres ayant $s$ hyperarêtes et $n = n(s) = s(b-1)+1$ sommets, asymptotiquement quand $(s \to \infty)$, est :*

$$n!\,[z^n]\,H_{-1}\circ T(z) \sim \frac{1}{\sqrt{b-1}}\frac{s^{s(b-1)-1}}{\exp\left((b-2)(s+1/(b-1))\right)}\left[\frac{(b-1)^{b-1}}{(b-2)!}\right]^s . \tag{3.95}$$

Ainsi, le nombre asymptotique des hyperarbres non enracinés est déterminé indirectement par la méthode du point col car nous avons appliqué cette méthode pour déterminer l'asymptotique du coefficient $[z^n]\,T(z)$ de la SGE relative aux hyperarbres enracinés. Le point de départ de l'estimation asymptotique étant l'application de la formule d'inversion de Lagrange à la SGE pouvant être exprimée en fonction de $T(z)$, pour avoir une expression intégrale du coefficient à estimer. Ayant à disposition la SGE des hypercycles lisses, l'estimation asymptotique du nombre des hypercycles est aussi accessible et elle est faite dans la section suivante.

## 3.2 Énumération asymptotique des hypercycles

Dans cette section, pour énumérer asymptotiquement les hypercycles, nous estimons le $n$-ième coefficient de la SGE $H_0 \circ T(z)$ que nous avons déterminée dans le chapitre précédent (voir (2.15) ) et qui est rappelée ci-après

$$H_0(t) = -\ln\left(\sqrt{1-\frac{t^{b-1}}{(b-2)!}}\right) - \frac{t^{b-1}}{2(b-2)!} . \tag{3.96}$$

Disposant de la formule d'inversion de Lagrange et de la formule intégrale de Cauchy, par la méthode du point col, nous déterminons un équivalent asymptotique du coefficient de la SGE $H_0 \circ T(z)$.

Nous résumons la méthode du point col, illustrée par des preuves de la section précédente, comme suit :

1. Fixer un contour passant par le point col (susceptible de convenir à la méthode).
2. Établir le terme principal de l'asymptotique dans un voisinage, à préciser, du point col du contour.
3. Négliger la contribution de l'intégrale sur le chemin restant du contour, c'est à dire excluant la portion du voisinage défini plus haut.
4. Conclure.

Par une variante de cette méthode, en autorisant que le contour ne passe pas exactement sur le point col mais dans un voisinage proche, nous trouvons :

**Proposition 3.2.1.** *Le nombre des hypercycles ayant $s(b-1)$ sommets est déterminé à partir du coefficient de $z^{s(b-1)}$ du développement de la SGE $H_0 \circ T(z)$ avec*

$$H_0(t) = -\ln\left(\sqrt{1-\frac{t^{b-1}}{(b-2)!}}\right) - \frac{t^{b-1}}{2(b-2)!}. \tag{3.97}$$

*Ce coefficient est, pour $(s \to \infty)$, tel que*

$$[z^n]\, H_0 \circ T(z) = \frac{(b-1)\exp(s)}{4n[(b-2)!]^s}\left\{1 + O(1/s)\right\}, \tag{3.98}$$

*avec $n = n(s) = s(b-1)$.*

*Preuve.* La SGE $H_0$ (3.97) des hypercycles lisses permet, grâce à la formule d'inversion de Lagrange, d'exprimer le $n$-ième ($n = n(s) = s(b-1)$) coefficient $[z^n]\, H_0 \circ T(z)$ sous forme d'une intégrale :

$$[z^n]\, H_0 \circ T(z) = \frac{b-1}{4i\pi n}\oint \frac{\tau(t)}{1-\tau(t)} \exp\left(n\frac{\tau(t)}{b-1} - n\ln(t)\right)\frac{\mathrm{d}\,t}{t} \tag{3.99}$$

avec un contour intégral qui encercle, dans le sens direct, l'origine du plan complexe et $\tau(t) = t^{b-1}/(b-2)!$. En effet, la dérivée de la SGE $H_0$ est

$$\frac{\mathrm{d}}{\mathrm{d}\,t}H_0(t) = \frac{\tau(t)}{2t(1-\tau(t))}. \tag{3.100}$$

Nous trouvons alors

$$[z^n]\, H_0 \circ T(z) = \tag{3.101}$$
$$= \frac{1}{4i\pi n}\oint \frac{\tau(t)}{1-\tau(t)} \exp\left(\frac{n\tau(t)}{b-1} - \frac{n\ln(\tau(t)(b-2)!)}{b-1}\right)\frac{\mathrm{d}\,\tau(t)}{\tau(t)} = \tag{3.102}$$
$$= \frac{b-1}{4i\pi n[(b-2)!]^s}\oint \frac{t}{1-t}\exp(s(t-\ln(t)))\frac{\mathrm{d}\,t}{t} = \tag{3.103}$$
$$= \frac{b-1}{4i\pi n[(b-2)!]^s}\oint \frac{1}{1-t}\exp(s(t-\ln(t)))\mathrm{d}\,t. \tag{3.104}$$

Sous cette forme intégrale, nous observons que l'intégrande présente un point col qui est déterminé par la racine de la dérivée de l'exposant et un point singulier déterminé par la racine de $(1-t)$. Soit $t=1$ est à la fois un point col et un point singulier. Pour appliquer la méthode du point col, nous adoptons la formulation intégrale suivante du coefficient :

$$[z^n]\, H_0 \circ T(z) = \frac{b-1}{4i\pi n[(b-2)!]^s} \oint \exp\left(s(t - \ln(t) - \frac{\ln(1-t)}{s})\right) \mathrm{d}\, t\,. \tag{3.105}$$

Notons par $h_s$ le facteur du paramètre $s$ dans l'exposant de l'intégrale précédente :

$$h_s(t) = t - \ln(t) - \frac{\ln(1-t)}{s}\,. \tag{3.106}$$

Pour identifier d'éventuel point col, il faut trouver la racine de ${h_s}'(t) = 0$. La dérivée ${h_s}'$ étant :

$${h_s}'(t) = 1 - \frac{1}{t} + \frac{1}{s(1-t)}\,, \tag{3.107}$$

nous identifions le point col, de module strictement inférieur à 1 pour $(s \to \infty)$,

$$t_s = 1 - \frac{\sqrt{4s+1}-1}{2s}\,. \tag{3.108}$$

Nous faisons alors le choix d'un contour $\gamma$, similaire à celui de la preuve du théorème 3.1.3, mais contournant le point col $t_s$ :

$$\gamma : \begin{cases} \gamma_1 : t = t_s + \exp(i\alpha)/2\,, & \downarrow\in [\pi/2, 3\pi/2] \\ \gamma_2 : t = t_s + iv\,, & v \uparrow\in [-3, -1/2] \cup [1/2, 3] \\ \gamma_0 : t = t_s + 3\exp(i\alpha)\,, & \alpha \uparrow\in [\pi/2, 3\pi/2]\,. \end{cases} \tag{3.109}$$

Nous voyons, sur l'exemple de la figure 3.3, que le module de l'intégrande capturera l'essentiel de l'asymptotique pour le paramètre $(s \to \infty)$. Nous qualifions la méthode comme méthode du point col dans la mesure où l'intégration sur le contour $\gamma_1$ est égale à l'intégration sur le segment $t \uparrow\in [t_s - i/2, t_s + i/2]$.

Le contour $\gamma$ d'intégration (3.109) étant fixé, il nous faut déterminer le terme principal de l'asymptotique du coefficient porté par la portion $\gamma_1$ du contour autour du point col $t_s$. Notons $I_1$ la valeur de l'intégrale sur la portion $\gamma_1$, c'est à dire

$$I_1 = \frac{b-1}{4i\pi n[(b-2)!]^s} \int_{\gamma_1} \frac{1}{1-t} \frac{\exp(st)}{t^s} \mathrm{d}\, t\,, \tag{3.110}$$

soit, avec le changement de variable

$$\begin{cases} t \mapsto 1 + \frac{t}{s} \\ \mathrm{d}\, t \mapsto \frac{\mathrm{d}\, t}{s} \\ \gamma_1 \mapsto \bar{\gamma}_1 : t = st_s - s + se^{i\alpha}/2\,, \quad \alpha \downarrow\in [\pi/2, 3\pi/2]\,, \end{cases} \tag{3.111}$$

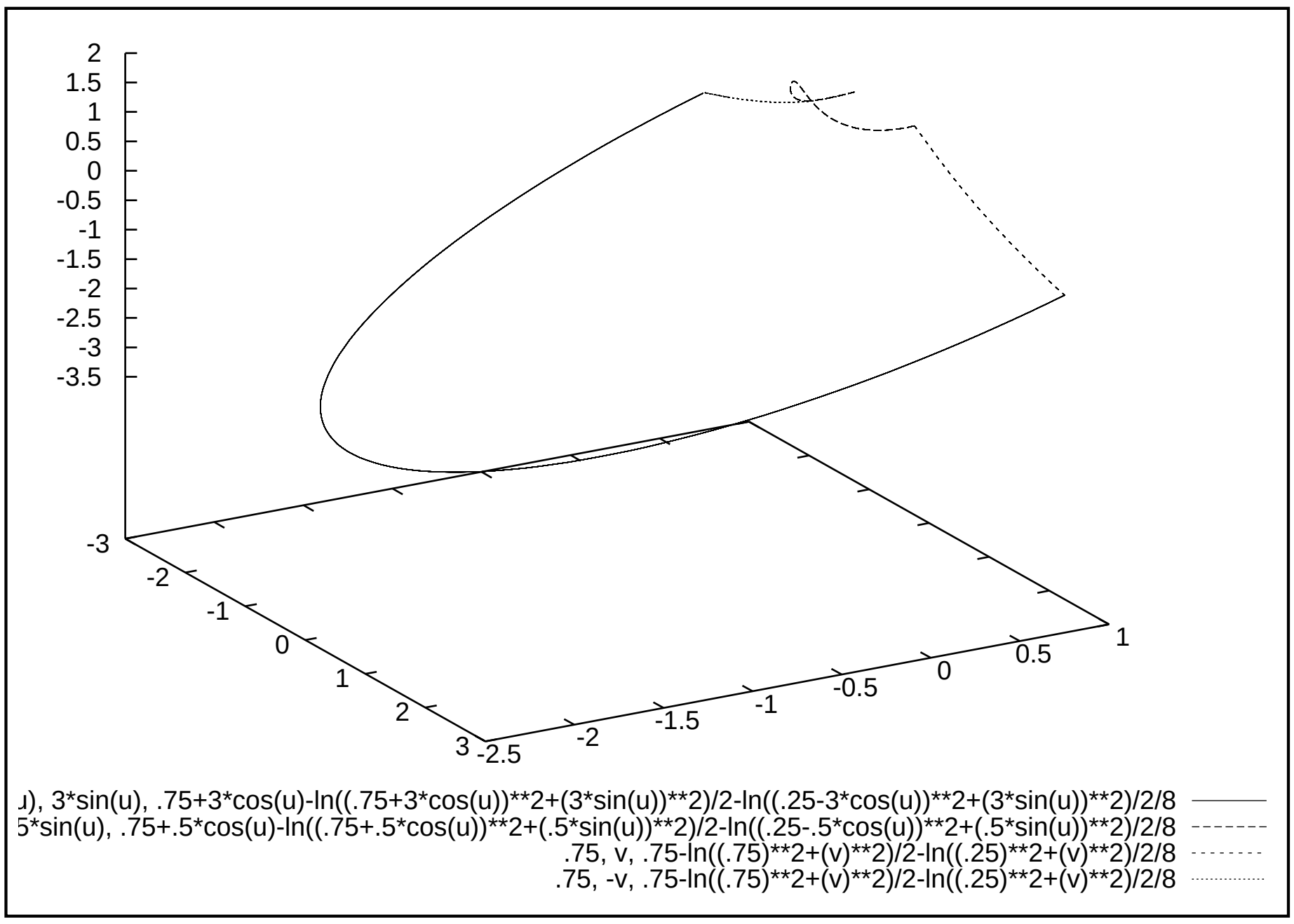

FIG. 3.3 – La valeur de $\Re(t - \ln(t) - \ln(1-t)/s)$ sur le contour $\gamma$ pour $s = 8$.

$$
\begin{aligned}
I_1 &= -\frac{(b-1)\exp(s)}{4i\pi n[(b-2)!]^s} \int_{\bar{\gamma}_1} \exp(O(t^2/s)) \frac{\mathrm{d}\,t}{t} && (3.112)\\
&= -\frac{(b-1)\exp(s)}{4i\pi n[(b-2)!]^s} \int_{\bar{\gamma}_1} \frac{\mathrm{d}\,t}{t}(1 + O(t^2/s)) && (3.113)\\
&= \frac{(b-1)\exp(s)}{4n[(b-2)!]^s} \{1 + O(1/s)\} . && (3.114)
\end{aligned}
$$

Ainsi, la contribution de l'intégrale sur $\gamma_1$ est

$$
I_1 = \frac{(b-1)\exp(s)}{4n[(b-2)!]^s} \{1 + O(1/s)\} . \qquad (3.115)
$$

Il nous reste à négliger la contribution de l'intégrale sur le contour restant, à savoir sur $\gamma_2$ et sur $\gamma_0$. Notons $I_2$ l'intégrale sur la portion $\gamma_2$, alors

$$
\begin{aligned}
I_2 &= && (3.116)\\
&= \frac{b-1}{4i\pi n[(b-2)!]^s} \int_{\gamma_2} \exp(s h_s(t))\mathrm{d}\,t = && (3.117)\\
&= \frac{(b-1)\exp(s)}{4i\pi n[(b-2)!]^s} \int_{\gamma_2} \exp\left(s h_s(t) - s\right) \mathrm{d}\,t . && (3.118)
\end{aligned}
$$

Rappelons l'équation paramétrique de la portion $\gamma_2$ du contour (3.109) :

$$
\gamma_2 : t = t_s + iv\,, \quad v \uparrow\in [-3, -1/2] \cup [1/2, 3] . \qquad (3.119)
$$

Sur $\gamma_2$ , la partie réelle $\Re(h_s(t)-1)$ , pour $(s\to\infty)$ , exprimée avec le paramètre $v$ de l'équation précédente, est telle que : d'un coté, si $v\in[-3,-1/2]$ ,

$$\Re(h_s(t)-1)\lesssim\frac{\ln(8)}{5}(v+3)-\ln(\sqrt{10}) \tag{3.120}$$

et d'un autre coté, si $v\in[1/2,3]$ ,

$$\Re(h_s(t)-1)\lesssim-\frac{\ln(8)}{5}(v-3)-\ln(\sqrt{10})\,. \tag{3.121}$$

La valeur de l'intégrande $I_2$ est donc telle que

$$I_2= \tag{3.122}$$

$$=O(\frac{(b-1)\exp(s)}{4\pi n[(b-2)!]^s10^{s/2}}\int_{-3}^{-1/2}e^{s\ln(8)(v+3)/5}\mathrm{d}\,v)+ \tag{3.123}$$

$$+O(\frac{(b-1)\exp(s)}{4\pi n[(b-2)!]^s10^{s/2}}\int_{-1/2}^{3}e^{-s\ln(8)(v-3)/5}\mathrm{d}\,v)= \tag{3.124}$$

$$=O(\frac{(b-1)\exp(s)}{4\pi n[(b-2)!]^s10^{s/2}})\,. \tag{3.125}$$

Ainsi, la contribution au coefficient de l'intégrale sur la portion $\gamma_2$ du contour $\gamma$ est exponentiellement plus petite que celle $I_1$ sur la portion $\hat{\gamma}_1$ .

Nous trouvons aussi que la contribution de l'intégrale sur la portion $\gamma_0$ , définie en (3.109) , est exponentiellement plus petite que $I_1$ . Pour le montrer, notons $I_0$ cette contribution :

$$I_0= \tag{3.126}$$

$$=\frac{b-1}{4i\pi n[(b-2)!]^s}\int_{\gamma_0}\exp\left(sh_s(t)\right)\mathrm{d}\,t= \tag{3.127}$$

$$=\frac{(b-1)\exp(s)}{4i\pi n[(b-2)!]^s}\int_{\gamma_0}\exp\left(sh_s(t)-s\right)\mathrm{d}\,t\,. \tag{3.128}$$

Rappelons l'équation paramétrique du demi cercle $\gamma_0$

$$\gamma_0:t=t_s+3\exp(i\alpha)\,,\quad \alpha\uparrow\in[\pi/2,3\pi/2]\,. \tag{3.129}$$

Avec cette notation, en prenant $\alpha$ comme variable, la partie réelle du facteur de $s$ de l'exposant $(h_s(t)-1)$ est telle que :

– Dans un voisinage de $t=t_s+3i$ , soit encore dans un voisinage de $\alpha=\pi/2$ ,

$$\Re(h_s(t)-1)=-\ln(\sqrt{10})-\frac{27}{10}(\alpha-\frac{\pi}{2})+O((\alpha-\frac{\pi}{2})^2)\,. \tag{3.130}$$

– Et dans un voisinage de $t=t_s-3i$ , soit encore dans un voisinage de $\alpha=3\pi/2$ ,

$$\Re(h_s(t)-1)=-\ln(\sqrt{10})+\frac{27}{10}(\alpha-\frac{3\pi}{2})+O((\alpha-\frac{3\pi}{2})^2)\,. \tag{3.131}$$

Ainsi, l'intégrale $I_0$ est telle que

$$I_0 = \tag{3.132}$$

$$= O(\frac{(b-1)\exp(s)}{4\pi n[(b-2)!]^s 10^{s/2}} \tag{3.133}$$

$$\left\{\int_{\pi/2}^{\pi} e^{-\frac{27s}{10}(\alpha-\frac{\pi}{2})}\mathrm{d}\,\alpha + \int_{\pi}^{3\pi/2} e^{\frac{27s}{10}(\alpha-\frac{3\pi}{2})}\mathrm{d}\,\alpha\right\}) = \tag{3.134}$$

$$= O(\frac{(b-1)\exp(s)}{4\pi n[(b-2)!]^s 10^{s/2}}) . \tag{3.135}$$

$I_0$ est donc aussi exponentiellement petit par rapport à $I_1$ . Nous concluons alors que $I_1$ est l'ordre de grandeur du coefficient de la SGE et

$$[z^n]\, H_0 \circ T(z) = \frac{(b-1)\exp(s)}{4n[(b-2)!]^s}\left\{1 + O(\frac{1}{s})\right\} . \tag{3.136}$$

◇

Notons que ce résultat est valide dans le cas des graphes en prenant simplement $b = 2$ et ainsi obtenir l'asymptotique du coefficient de la SGE $W_0 \circ T(z)$ des unicycles.

Comme nous avons vu pour les hyperarbres enracinés, le nombre asymptotique des hypercycles se déduit aussi de l'asymptotique du coefficient de sa SGE $H_0 \circ T(z)$ , donné par la proposition 3.2.1 , grâce à la formule de Stirling :

**Théorème 3.2.2.** *Le nombre d'hypercycles ayant $s$ hyperarêtes et $n = n(s) = s(b-1)$ sommets, pour $(s \to \infty)$, est :*

$$n!\,[z^n]\, H_0 \circ T(z) \sim \frac{\sqrt{2\pi(b-1)}}{4}\frac{s^{s(b-1)-1/2}}{\exp(s(b-2))}\left[\frac{(b-1)^{b-1}}{(b-2)!}\right]^s . \tag{3.137}$$

Ainsi, nous avons une fois de plus la confirmation que l'analyse complexe, la méthode du point col pour inspirer la recherche d'un contour d'intégration, convient pour faire de l'énumération asymptotique : par cette méthode a été établi, théorème 3.1.3, le nombre asymptotique des hyperarbres et , théorème 3.2.2 , le nombre asymptotique des hypercycles quand la taille des structures est grande. Ces énumérations asymptotiques ont pu être faites parce que nous disposons des SGEs des structures lisses correspondantes. La méthode du point col, à priori, permettra aussi de trouver un contour aboutissant à un résultat d'énumération asymptotique des composantes complexes d'excès $\ell \geq 1$ donné, sachant que nous disposons d'un moyen "pratique" (un programme) pour déterminer l'expression de la SGE $H_\ell$ des structures lisses correspondantes. Dans la section suivante, nous procédons à l'énumération asymptotique des composantes complexes selon leur excès.

## 3.3 Énumération asymptotique des composantes complexes

Les preuves vues jusqu'ici pour avoir l'asymptotique des hyperarbres et des hypercycles nécessitent la connaissance des SGEs des composantes lisses correspondant. Un schéma de preuve similaire permet aussi d'obtenir des résultats d'énumération asymptotique des composantes complexes, une des difficultés ici étant d'énoncer un résultat générique pour tout excès.

Dans cette section, nous commençons par généraliser un résultat d'encadrement des coefficients des SGEs $W_\ell \circ T(z)$ (cas des graphes) obtenu par Wright E.M dans [22] aux hypergraphes puis nous énonçons un lemme pour l'asymptotique des coefficients de la SGE des séquences de $m$ chaînes, et enfin nous énonçons le résultat d'énumération asymptotique.

### 3.3.1 Encadrement des coefficients de la SGE des composantes complexes

Dans cette sous-section, nous tirons profit de la possibilité, par le théorème 2.3.37 vu dans le chapitre 2, de déterminer les premiers coefficients de la SGE, sous la forme (2.120), des composantes complexes d'excès donné. Bien que dans ce chapitre, un langage analytique domine à des fins d'énumération asymptotique, dans cette sous-section en particulier, nous retrouvons un raisonnement plus combinatoire (lecture bijective de SGE) comme dans le chapitre précédent. Nous soulignons en particulier, l'intérêt de la lecture combinatoire (voir la preuve du théorème 2.3.2 ) de la formule d'inversion de Lagrange comme illustration clé pour avoir l'encadrement des coefficients.

Pour les notations de l'énoncé du théorème qui suit, reprenons ici les notations utilisées pour les déclinaisons des formes des SGEs $H_\ell$ :

$$H_\ell(t) = \tag{3.138}$$

$$= \frac{(1-\theta(t))^{r_\ell}}{t^\ell} \sum_{p=0}^{3\ell} A_{\ell,p} \left(\frac{1-\theta(t)}{\theta(t)}\right)^p = \tag{3.139}$$

$$= \frac{1}{t^\ell} \sum_{j=-3\ell}^{r_\ell} c_j(\ell,b)\theta(t)^j = \tag{3.140}$$

$$= \frac{f_\ell \circ \theta(t)}{t^\ell}, \tag{3.141}$$

avec $r_\ell = \lfloor (\ell+1)/(b-1)+1 \rfloor$ et $\theta(t) = 1 - t^{b-1}/(b-2)!$ .

Dans le théorème suivant, nous énonçons l'encadrement des coefficients de la SGE $H_\ell \circ T(z)$ des composantes complexes d'excès $\ell \geq 1$ .

**Théorème 3.3.1.** *Le nombre $n!\,[z^n]\,H_\ell \circ T(z)$ de composantes complexes d'excès $\ell$ ayant $n$ sommets admet l'encadrement suivant :*
*La majoration*

$$n!\,[z^n]\,H_\ell \circ T(z) \leq (n-1)!\,\left[t^{n+\ell}\right] \left\{ \frac{3\ell(b-1)A_{\ell,3\ell}}{\theta(t)^{3\ell+1}} \Phi(t)^n \right\} \tag{3.142}$$

*et la minoration*

$$(n-1)!\,\left[t^{n+\ell}\right] \left\{ \left( \frac{B_\ell}{\theta(t)^{3\ell+1}} - \frac{C_\ell}{\theta(t)^{3\ell}} \right) \Phi(t)^n \right\} \leq n!\,[z^n]\,H_\ell \circ T(z)\,, \tag{3.143}$$

*où*

$$B_\ell = 3\ell(b-1)A_{\ell,3\ell}\,, \tag{3.144}$$

$$C_\ell = (9\ell^2+\ell)A_{\ell,3\ell} + (b-1)(3\ell-1)(r_\ell A_{\ell,3\ell} - A_{\ell,3\ell-1})\,, \tag{3.145}$$

*et*

$$\Phi(t) = \exp\left(\frac{t^{b-1}}{(b-1)!}\right)\,. \tag{3.146}$$

*En fait,* $B_\ell$ *et* $-C_\ell$ *désignent respectivement les coefficients de* $x^{-3\ell-1}$ *et de* $x^{-3\ell}$ *dans*

$$-(b-1)(1-x){f_\ell}'(x) - \ell f_\ell(x)\,, \tag{3.147}$$

$$f_\ell(x) = (1-x)^{r_\ell}\sum_{p=0}^{3\ell} A_{\ell,p}\left(\frac{1-x}{x}\right)^p\,. \tag{3.148}$$

*Preuve.* Le nombre des composantes d'excès $\ell \geq 1$ ayant $n$ sommets est déterminé grâce à la formule d'inversion de Lagrange :

$$n!\,[z^n]\,H_\ell \circ T(z) = \tag{3.149}$$

$$= (n-1)!\,\left[t^{n-1}\right]\left(\Phi(t)^n \frac{\mathrm{d}}{\mathrm{d}\,t}\frac{f_\ell\circ\theta(t)}{t^\ell}\right) = \tag{3.150}$$

$$= (n-1)!\,\left[t^{n-1}\right]\left(\frac{\Phi(t)^n}{t^{\ell+1}}(-(b-1)(1-\theta(t)){f_\ell}'\circ\theta(t) - \ell f_\ell\circ\theta(t))\right) = \tag{3.151}$$

$$= (n-1)!\,\left[t^{n+\ell}\right]\left(\Phi(t)^n(-(b-1)(1-\theta(t)){f_\ell}'\circ\theta(t) - \ell f_\ell\circ\theta(t))\right)\,. \tag{3.152}$$

Notons pour $\ell \geq 1$,

$$R_\ell(x) = -(b-1)(1-x){f_\ell}'(x) - \ell f_\ell(x) \tag{3.153}$$

alors $R_\ell$ est un polynôme de Laurent de degré minimum $(-3\ell-1)$, de degré maximum borné par $r_\ell = \lfloor(\ell+1)/(b-1)+1\rfloor$, et à coefficients rationnels. Le nombre de composantes ci-dessus s'écrit :

$$n!\,[z^n]\,H_\ell\circ T(z) = (n-1)!\,\left[t^{n+\ell}\right](R_\ell\circ\theta(t)\Phi(t)^n)\,. \tag{3.154}$$

Il s'ensuit que les coefficients de la SGE $R_\ell\circ\theta(t)$ sont tous positifs ou nuls, ce que nous notons pour une série en la variable $t$ comme suit :

$$R_\ell\circ\theta(t) \succeq_t 0\,. \tag{3.155}$$

En effet, la multiplication par $\Phi(t)^n$ qui apparaît dans le second membre de (3.154) souligne la décomposition des structures à énumérer, comme dans la preuve de la formule d'inversion de Lagrange (théorème 2.3.2) : sur les structures lisses sont greffés des hyperarbres enracinés. Ainsi, $R_\ell\circ\theta(t)$ est une SGE énumérant des structures lisses $\mathcal{R}_\ell$. Connaissant les caractéristiques du polynôme de Laurent $R_\ell$, nous sommes assurés de l'existence d'un nombre $a_{3\ell+1}$ tel que

$$\frac{a_{3\ell+1}}{\theta(t)^{3\ell+1}} - R_\ell\circ\theta(t) \succeq_t 0\,. \tag{3.156}$$

En effet, il existe un entier $N$ tel que les SGEs

$$\{\theta(t)^{-3\ell-1}, \theta(t)^{-3\ell}, \ldots, \theta(t)^{-1}, (1-\theta(t)), \ldots, (1-\theta(t))^N\} \tag{3.157}$$

forment une échelle de décomposition de $R_\ell \circ \theta(t) \succeq_t 0$ . Combinatoirement, cette équation (3.156) dit que $a_{3\ell+1}\theta(t)^{-3\ell-1}$ compte des structures lisses combinatoires beaucoup plus nombreuses à un nombre de sommets donné que $\mathcal{R}_\ell$ . Prenons $a_{3\ell+1}$ le plus petit nombre vérifiant cette propriété (3.156) , formellement

$$a_{3\ell+1} = \min\{a \in \mathbb{R}\,, \frac{a}{\theta(t)^{3\ell+1}} - R_\ell \circ \theta(t) \succeq_t 0\}\,. \tag{3.158}$$

L'équation (3.156) souligne alors l'existence de structures combinatoires lisses (à une pondération multiplicative près) $\mathcal{R}_\ell^1$ énumérées par la SGE $R_\ell^1 \circ \theta(t)$ définie par son premier membre :

$$R_\ell^1(x) = \frac{a_{3\ell+1}}{x^{3\ell+1}} - R_\ell(x)\,. \tag{3.159}$$

Comme $a_{3\ell+1}$ est choisi comme étant le plus petit, nous sommes assurés de trouver un nombre $a_{3\ell}$ tel que

$$R_\ell \circ \theta(t) - \left(\frac{a_{3\ell+1}}{\theta(t)^{3\ell+1}} - \frac{a_{3\ell}}{\theta(t)^{3\ell}}\right) \succeq_t 0\,, \tag{3.160}$$

soit

$$\frac{a_{3\ell}}{\theta(t)^{3\ell}} - R_\ell^1 \circ \theta(t) \succeq_t 0\,. \tag{3.161}$$

Prenons ici aussi dans (3.160) $a_{3\ell}$ comme étant le plus petit :

$$a_{3\ell} = \min\{a \in \mathbb{R}\,, \frac{a}{\theta(t)^{3\ell}} - R_\ell^1 \circ \theta(t) \succeq_t 0\}\,. \tag{3.162}$$

Le procédé se réitère et avec un plus petit nombre $a_{3\ell-1}$ donne

$$\left(\frac{a_{3\ell+1}}{\theta(t)^{3\ell+1}} - \frac{a_{3\ell}}{\theta(t)^{3\ell}} + \frac{a_{3\ell-1}}{\theta(t)^{3\ell-1}}\right) - R_\ell \circ \theta(t) \succeq_t 0\,, \tag{3.163}$$

soit

$$\frac{a_{3\ell-1}}{\theta(t)^{3\ell-1}} - R_\ell^2 \circ \theta(t) \succeq_t 0\,, \tag{3.164}$$

avec

$$R_\ell^2(x) = R_\ell(x) - \left(\frac{a_{3\ell+1}}{x^{3\ell+1}} - \frac{a_{3\ell}}{x^{3\ell}}\right)\,. \tag{3.165}$$

Le procédé se réitère et s'arrête après avoir déterminé $a_1$ ou bien quand on obtient des structures $\mathcal{R}_\ell^{j_0}$ énumérées par la SGE $R_\ell^{j_0} \circ \theta(t)$ , $R_\ell^{j_0}$ étant un polynôme. Ainsi, à un signe près, $R_\ell^{j_0}(x)$ vaut

$$R_\ell(x) - \left(\frac{a_{3\ell+1}}{x^{3\ell+1}} - \frac{a_{3\ell}}{x^{3\ell}} + \ldots + (-1)^{j_0-3\ell+1}\frac{a_{j_0}}{x^{j_0}}\right)\,. \tag{3.166}$$

En y faisant une identification des $a_j$ avec les coefficients du polynôme de Laurent $R_\ell$ , nous obtenons en particulier les deux inégalités du théorème définissant l'encadrement comme conséquence immédiate de la propriété de positivité imposée à la définition des $a_j$ (qui par cette identification sont rationnels) :

– la majoration est, par (3.160) ,

$$n!\,[z^n]\,H_\ell \circ T(z) = \tag{3.167}$$
$$= (n-1)!\left[t^{n+\ell}\right](R_\ell \circ \theta(t)\Phi(t)^n) \leq \tag{3.168}$$
$$\leq (n-1)!\left[t^{n+\ell}\right](\frac{a_{3\ell+1}}{\theta(t)^{3\ell+1}}\Phi(t)^n) \tag{3.169}$$

– et la minoration est, par (3.163) ,

$$n!\,[z^n]\,H_\ell \circ T(z) = \tag{3.170}$$
$$= (n-1)!\left[t^{n+\ell}\right](R_\ell \circ \theta(t)\Phi(t)^n) \geq \tag{3.171}$$
$$\geq (n-1)!\left[t^{n+\ell}\right]((\frac{a_{3\ell+1}}{\theta(t)^{3\ell+1}} - \frac{a_{3\ell}}{\theta(t)^{3\ell}})\Phi(t)^n)\,. \tag{3.172}$$

♢

Dans cette preuve, le raisonnement se base sur un procédé d'inclusion exclusion pour compter des structures lisses qui contiennent un nombre fini de chaînes : toutes les structures sont comptées en excès comme si chacune avait $3\ell$ chaînes ; des extra structures seront alors inclues dans le compte ainsi fait et ces extra structures peuvent à leur tour être comptées en excès comme si chacune avait $(3\ell - 1)$ chaînes ; etc.

Nous retiendrons la version de la majoration suivante :

**Théorème 3.3.2.** *Le nombre $n!\,[z^n]\,H_\ell \circ T(z)$ des composantes complexes d'excès $\ell$ ayant $n$ sommets admet la majoration suivante :*

$$n!\,[z^n]\,H_\ell \circ T(z) \leq 3\ell A_{\ell,3\ell}(b-1)(n-1)!\left[t^{n+\ell}\right]\left\{\frac{\tau(t)}{(1-\tau(t))^{3\ell+1}}\Phi(t)^n\right\}\,, \tag{3.173}$$

*avec $\tau(t) = t^{b-1}/(b-2)!$ .*

*Preuve.* La preuve est similaire à celle du théorème d'encadrement précédent en faisant la remarque que les structures comptées ont toutes au moins une hyperarête donc la preuve est valide en faisant le remplacement $R_\ell(x) \leftarrow R_\ell(x)/(1-x)$ dans l'équation (3.153) . ♢

### 3.3.2 La contribution asymptotique de $(m-1)$ chaînes

Dans la justification du théorème 3.3.2 , l'utilité de la classification des structures selon le nombre de chaînes est mise en évidence pour procéder par inclusion exclusion et aboutir ainsi à la conclusion de l'importance, relatée par le théorème précédent, de la contribution des structures à $3\ell$ chaînes dans le nombre des composantes d'excès $\ell$ . La contribution dans l'asymptotique du coefficient de la SGE $H_\ell \circ T(z)$ provient essentiellement du terme avec le facteur $\theta(t)^{-3\ell}$ dans (2.153) . Cette remarque est intuitive par la forme même de la SGE $H_\ell$ et est appuyée bijectivement par le théorème précédent. Cette relation, entre les structures maximisant le nombre de chaînes et l'asymptotique du coefficient de la SGE, a été mise à profit par exemple dans [18] et dans [19] avec les notions de structures qualifiées de "*clean*" ou "*unclean*" . Pour tirer profit de la décomposabilité des structures en chaînes, nous disposons du lemme suivant :

**Lemme 3.3.3.** *Pour* $(s \to \infty)$ *, nous avons l'asymptotique du coefficient suivant*

$$\left[t^{n+\ell}\right]\left\{\frac{\tau(t)}{(1-\tau(t))^m}\Phi(t)^n\right\} = \frac{e^{s+\frac{m}{2}-\frac{\ell}{b-1}}(s/m)^{m/2}}{2\sqrt{s\pi}[(b-2)!]^s}\left\{1+O(\frac{1}{s^{1/2}})\right\}, \tag{3.174}$$

*avec* $n = n(s) = s(b-1)-\ell$ *,* $\tau(t) = t^{b-1}/(b-2)!$ *,* $\Phi(t) = \exp(\tau(t)/(b-1))$ *et* $m \geq 2$ *.*

*Preuve.* La formule intégrale de Cauchy donne

$$\left[t^{n+\ell}\right]\left\{\frac{\tau(t)}{(1-\tau(t))^m}\Phi(t)^n\right\} = \tag{3.175}$$

$$\frac{1}{2i\pi}\oint\frac{\tau(t)}{(1-\tau(t))^m}\exp\left(n\frac{\tau(t)}{b-1}-(n+\ell)\ln(t)\right)\frac{\mathrm{d}\,t}{t}. \tag{3.176}$$

Comme $n = n(s) = s(b-1)-\ell$, l'équation précédente devient

$$\left[t^{n+\ell}\right]\left\{\frac{\tau(t)}{(1-\tau(t))^m}\Phi(t)^n\right\} = \tag{3.177}$$

$$= \frac{1}{2i\pi[(b-2)!]^s}\oint\frac{\tau(t)}{(1-\tau(t))^m}\frac{\exp\left(s\tau(t)-s\ln(\tau(t))\right)}{\exp(\ell\tau(t)/(b-1))}\frac{\mathrm{d}\,t}{t} = \tag{3.178}$$

$$= \frac{1}{2i\pi[(b-2)!]^s(b-1)}\oint\frac{\tau(t)}{(1-\tau(t))^m}\frac{\exp\left(s\tau(t)-s\ln(\tau(t))\right)}{\exp(\ell\tau(t)/(b-1))}\frac{\mathrm{d}\,\tau(t)}{\tau(t)} = \tag{3.179}$$

$$= \frac{1}{2i\pi[(b-2)!]^s}\oint\frac{1}{(1-t)^m}\frac{\exp\left(st-s\ln(t)\right)}{\exp(\ell t/(b-1))}\mathrm{d}\,t, \tag{3.180}$$

où le contour d'intégration encercle, une fois dans le sens direct, l'origine du plan complexe. Avec l'intégrale ainsi représentée, nous remarquons de nouveau que 1 est à la fois point col et point singulier. Pour déterminer un contour d'intégration menant au terme principal de l'asymptotique, nous adoptons l'écriture intégrale suivante :

$$\left[t^{n+\ell}\right]\left\{\frac{\tau(t)}{(1-\tau(t))^m}\Phi(t)^n\right\} = \tag{3.181}$$

$$= \frac{1}{2i\pi[(b-2)!]^s}\oint\frac{\exp\left(s(t-\ln(t)-m\ln(1-t)/s)\right)}{\exp(\ell t/(b-1))}\mathrm{d}\,t. \tag{3.182}$$

Notons alors $h$, le facteur du paramètre $s$ dans l'exposant de cette expression intégrale :

$$h(t) = t-\ln(t)-\frac{m}{s}\ln(1-t). \tag{3.183}$$

Le point col à considérer est racine de $h'(t) = 0$. Comme la dérivée $h'$ est

$$h'(t) = 1-\frac{1}{t}+\frac{1}{(s/m)(1-t)}, \tag{3.184}$$

le point col de plus petit module (strictement inférieur à 1 ) est

$$t_0 = 1 - \frac{\sqrt{4(s/m)+1} - 1}{2(s/m)}\,. \tag{3.185}$$

Nous prenons alors un contour $\gamma$ similaire à (3.62) , celui qui a servi pour l'énumération asymptotique des hyperarbres enracinés :

$$\gamma : \begin{cases} \gamma_1 : t = t_0 + iv/s\,, & v \uparrow\in [-3s, 3s] \\ \gamma_0 : t = t_0 + 3\exp(i\alpha)\,, & \alpha \uparrow\in [\pi/2, 3\pi/2]\,. \end{cases} \tag{3.186}$$

Soit une portion $\hat{\gamma}_1$ du chemin $\gamma_1$ :

$$\hat{\gamma}_1 : t = t_0 + iv/s\,, \quad v \uparrow\in [-\kappa_s, \kappa_s]\,, \tag{3.187}$$

avec $\kappa_s$ , un nombre positif qui sera précisé plus tard. Notons $I_1$ la valeur de l'intégrale sur cette portion $\hat{\gamma}_1$ :

$$I_1 = \frac{1}{2i\pi[(b-2)!]^s} \int_{\hat{\gamma}_1} \frac{\exp(sh(t))}{\exp(\ell t/(b-1))} \mathrm{d}\,t\,, \tag{3.188}$$

soit, avec le changement de variable

$$\begin{cases} t \mapsto 1 + \frac{t}{s} \\ \mathrm{d}\,t \mapsto \frac{\mathrm{d}\,t}{s} \\ \gamma_1 \mapsto \gamma_1' : t = st_0 - s + iv\,, \quad v \uparrow\in [-3s, 3s]\,, \end{cases} \tag{3.189}$$

$$I_1 = \frac{\exp(-\frac{\ell t_0}{b-1})}{2s\pi[(b-2)!]^s} \int_{-\kappa_s}^{\kappa_s} \exp(sf(v))\mathrm{d}\,v\,, \tag{3.190}$$

avec

$$f(v) = h(t_0 + iv/s) - \ell iv/(bs^2 - s^2)\,. \tag{3.191}$$

Comme

$$\Re(f(v)) = \tag{3.192}$$
$$= \Re(h(t_0 + iv/s)) = \tag{3.193}$$
$$= t_0 - \frac{1}{2}\ln(t_0^2 + (v/s)^2) - \frac{m}{s}\ln((1-t_0)^2 + (v/s)^2) = \tag{3.194}$$
$$= 1 + \frac{m}{2s}\ln(\frac{s}{m}) + \frac{m}{2s} + O(\frac{1}{s^{3/2}}) + \tag{3.195}$$
$$-\frac{v^2}{s^2}(1 + \frac{3m^{1/2}}{2s^{1/2}} + \frac{5m}{s} + \frac{11m^{3/2}}{16s^{3/2}} + O(\frac{1}{s^2})) + \tag{3.196}$$
$$+O(\frac{v^4}{s^3})\,, \tag{3.197}$$

$$I_1 = \frac{e^{s+\frac{m}{2}-\frac{\ell t_0}{b-1}}(s/m)^{m/2}}{2s\pi[(b-2)!]^s} \times \tag{3.198}$$

$$\times \int_{-\kappa_s}^{\kappa_s} e^{-\frac{v^2}{s}(1+\frac{3m^{1/2}}{2s^{1/2}}+\frac{5m}{s}+\frac{11m^{3/2}}{16s^{3/2}})}(1+O(\frac{v^2}{s^3}))\mathrm{d}\,v\,. \tag{3.199}$$

Notons

$$C(s) = \frac{1}{s}(1+\frac{3m^{1/2}}{2s^{1/2}}+\frac{5m}{s}+\frac{11m^{3/2}}{16s^{3/2}}) \tag{3.200}$$

alors par le changement de variable

$$\begin{cases} v = \frac{r}{\sqrt{C(s)}} \\ \mathrm{d}\,v = \frac{\mathrm{d}\,r}{\sqrt{C(s)}}\,, \end{cases} \tag{3.201}$$

l'intégrale $I_1$ est telle que

$$I_1 = \tag{3.202}$$

$$= \frac{e^{s+\frac{m}{2}-\frac{\ell t_0}{b-1}}(s/m)^{m/2}}{2s\pi[(b-2)!]^s\sqrt{C(s)}} \int_{-\kappa_s\sqrt{C(s)}}^{\kappa_s\sqrt{C(s)}} e^{-r^2}(1+O(\frac{r^2}{C(s)s^3}))\mathrm{d}\,r = \tag{3.203}$$

$$= \frac{e^{s+\frac{m}{2}-\frac{\ell t_0}{b-1}}(s/m)^{m/2}}{2s\pi[(b-2)!]^s\sqrt{C(s)}} \left\{ \int_{-\kappa_s\sqrt{C(s)}}^{\kappa_s\sqrt{C(s)}} e^{-r^2}\mathrm{d}\,r + O(\frac{{\kappa_s}^3 C(s)^{3/2}}{C(s)s^3}) \right\} = \tag{3.204}$$

$$= \frac{e^{s+\frac{m}{2}-\frac{\ell t_0}{b-1}}(s/m)^{m/2}}{2s\pi[(b-2)!]^s\sqrt{C(s)}} \left\{ \int_{-\kappa_s\sqrt{C(s)}}^{\kappa_s\sqrt{C(s)}} e^{-r^2}\mathrm{d}\,r + O(\frac{{\kappa_s}^3 C(s)^{1/2}}{s^3}) \right\}. \tag{3.205}$$

Comme pour $(s \to \infty)$, $C(s) \sim 1/s$, nous obtenons

$$I_1 = \frac{e^{s+\frac{m}{2}-\frac{\ell t_0}{b-1}}(s/m)^{m/2}\sqrt{s}}{2s\pi[(b-2)!]^s} \left\{ \int_{-\kappa_s/\sqrt{s}}^{\kappa_s/\sqrt{s}} e^{-r^2}\mathrm{d}\,r + O(\frac{{\kappa_s}^3}{s^{7/2}}) \right\}. \tag{3.206}$$

Cette ordre asymptotique est valide, pour $m \geq 1$ fixé, en prenant $\kappa_s = s$ et se réécrit alors :

$$I_1 = \frac{e^{s+\frac{m}{2}-\frac{\ell t_0}{b-1}}(s/m)^{m/2}\sqrt{s}}{2s\pi[(b-2)!]^s} \left\{ \int_{-\sqrt{s}}^{\sqrt{s}} e^{-r^2}\mathrm{d}\,r + O(\frac{1}{s^{1/2}}) \right\}. \tag{3.207}$$

Ainsi, l'équivalent asymptotique de l'intégrale $I_1$ sur le segment $\hat{\gamma}_1$ de longueur deux fois $(\kappa_s = s)$ est :

$$I_1 = \frac{e^{s+\frac{m}{2}-\frac{\ell t_0}{b-1}}(s/m)^{m/2}}{2\sqrt{s\pi}[(b-2)!]^s} \left\{ 1 + O(\frac{1}{s^{1/2}}) \right\}. \tag{3.208}$$

Pour obtenir le résultat, il faut encore négliger la contribution de l'intégrale sur le contour restant.

1. Sur la portion $\gamma_1\backslash\hat{\gamma}_1$, la contribution $I_1'$ est

$$I_1' = \frac{1}{2i\pi[(b-2)!]^s}\int_{\gamma_1\backslash\hat{\gamma}_1}\frac{\exp(sh(t))}{\exp(\ell t/(b-1))}\mathrm{d}\,t\,, \tag{3.209}$$

avec

$$\gamma_1\backslash\hat{\gamma}_1 : t = t_0 + iv/s\,, \quad v \uparrow\in [-3s, -\kappa_s] \cup [\kappa_s, 3s] \tag{3.210}$$

ou bien (comme nous avons pris $\kappa_s = s$)

$$\gamma_1\backslash\hat{\gamma}_1 : t = t_0 + iv\,, \quad v \uparrow\in [-3, -1] \cup [1, 3]\,. \tag{3.211}$$

L'intégrale $I_1'$, pour $(s \to \infty)$, devient alors

$$I_1' = \frac{e^{s+m/2-\frac{\ell t_0}{b-1}}(s/m)^{m/2}}{2\sqrt{s}\pi[(b-2)!]^s}\left\{\int_{-3}^{-1} + \int_1^3\right\}\exp(s\hat{f}(v))\mathrm{d}\,v\,, \tag{3.212}$$

avec la fonction $\hat{f}$ définie comme suit

$$\hat{f}(v) = h(t_0+iv) - \frac{i\ell}{sb-s} - 1 - \frac{m(1+\ln(s/m))}{2s} + \frac{\ln(s)}{2s}\,. \tag{3.213}$$

Sur l'intervalle $v \in [-3, -1]$, nous pouvons caractériser la partie réelle $\Re(\hat{f}(v))$ comme suit :

$$\Re(f(v)) = \tag{3.214}$$

$$= \Re(h(t_0+iv) - 1 - \frac{m(1-\ln(s/m))}{2s} + \frac{\ln(s)}{2s} \lesssim \tag{3.215}$$

$$\lesssim \frac{\ln(2)}{4}(v+1) - \frac{\ln(2)}{2}\,. \tag{3.216}$$

Et sur l'intervalle $v \in [1, 3]$, nous obtenons la caractérisation suivante :

$$\Re(f(v)) = \tag{3.217}$$

$$= \Re(h(t_0+iv) - 1 - \frac{m(1-\ln(s/m))}{2s} + \frac{\ln(s)}{2s} \lesssim \tag{3.218}$$

$$\lesssim -\frac{\ln(2)}{4}(v-1) - \frac{\ln(2)}{2}\,. \tag{3.219}$$

Des deux majorations précédentes, nous en déduisons que l'intégrale $I_1'$ vérifie

$$I_1' = O(\frac{e^{s+m/2-\frac{\ell t_0}{b-1}}(s/m)^{m/2}}{2\sqrt{s}\pi[(b-2)!]^s}\frac{1}{2^{s/2}}\times \tag{3.220}$$

$$\times\left\{\int_{-3}^{-1} e^{s\ln(2)(v+1)/4}\mathrm{d}\,v + \int_1^3 e^{-s\ln(2)(v-1)/4}\mathrm{d}\,v\right\}) \tag{3.221}$$

et l'ordre de grandeur asymptotique de l'intégrale $I_1'$ est

$$I_1' = O(\frac{e^{s+m/2}(s/m)^{m/2}}{2\sqrt{s}\pi[(b-2)!]^s}\frac{1}{2^{s/2}})\,, \tag{3.222}$$

c'est à dire que l'intégrale $I_1'$ est d'un facteur exponentiellement négligeable par rapport à $I_1$. Nous montrons de même pour le contour restant.

2. Sur la portion $\gamma_0$ , définie à l'équation (3.186) , la contribution $I_0$ est

$$I_0 = \tag{3.223}$$
$$= \frac{1}{2i\pi[(b-2)!]^s} \int_{\gamma_0} \frac{\exp(sh(t))}{\exp(\ell t/(b-1))} \mathrm{d}\, t = \tag{3.224}$$
$$= \frac{e^{s+m/2-\frac{\ell t_0}{b-1}}(s/m)^{m/2}}{2\sqrt{s}\pi[(b-2)!]^s} \left\{ \int_{\pi/2}^{\pi} + \int_{\pi}^{3\pi/2} \right\} \exp(s\bar{f}(\alpha))\mathrm{d}\, \alpha\, , \tag{3.225}$$

avec la fonction $\bar{f}$ définie comme suit

$$\bar{f}(\alpha) = h(t_0 + 3e^{i\alpha}) - \frac{3\ell e^{i\alpha}}{sb - s} + i\frac{\alpha}{s} - 1 - \frac{m(1 + \ln(s/m))}{2s} + \frac{\ln(s)}{2s}\, . \tag{3.226}$$

Sur l'intervalle $\alpha \in [\pi/2, \pi]$ , nous avons la partie réelle $\Re(\bar{f}(\alpha))$ telle que

$$\Re(\bar{f}(\alpha)) = \tag{3.227}$$
$$= \Re(h(t_0 + 3e^{i\alpha})) - \frac{3\ell\cos(\alpha)}{sb - s} - 1 - \frac{m(1 + \ln(\frac{s}{m}))}{2s} + \frac{\ln(s)}{2s} = \tag{3.228}$$
$$= -\ln(\sqrt{10}) - (\frac{27}{10} + O(\frac{1}{\sqrt{s}}))(\alpha - \frac{\pi}{2}) + O(\frac{1}{\sqrt{s}}) + O((\alpha - \frac{\pi}{2})^2)\, . \tag{3.229}$$

Et sur l'intervalle $\alpha \in [\pi, 3\pi/2]$ , nous avons la partie réelle $\Re(\bar{f}(\alpha))$ telle que

$$\Re(\bar{f}(\alpha)) = \tag{3.230}$$
$$= -\ln(\sqrt{10}) + (\frac{27}{10} + O(\frac{1}{\sqrt{s}}))(\alpha - \frac{3\pi}{2}) + O(\frac{1}{\sqrt{s}}) + O((\alpha - \frac{3\pi}{2})^2)\, . \tag{3.231}$$

Aussi, l'intégrale $I_0$ est telle que

$$I_0 = \tag{3.232}$$
$$= O(\frac{e^{s+m/2-\frac{\ell t_0}{b-1}}(s/m)^{m/2}}{2\sqrt{s}\pi[(b-2)!]^s} \frac{1}{10^{s/2}} \tag{3.233}$$
$$\left\{ \int_{\pi/2}^{\pi} e^{-27s(\alpha-\pi/2)/10} \mathrm{d}\, \alpha + \int_{\pi}^{3\pi/2} e^{27s(\alpha-3\pi/2)/10} \mathrm{d}\, \alpha \right\}) = \tag{3.234}$$
$$= O(\frac{e^{s+m/2-\frac{\ell t_0}{b-1}}(s/m)^{m/2}}{2\sqrt{s}\pi[(b-2)!]^s} \frac{1}{10^{s/2}})\, . \tag{3.235}$$

$I_0$ est donc aussi exponentiellement petit par rapport à $I_1$ . Nous concluons que $I_1$ est l'ordre de grandeur du coefficient :

$$\left[t^{n+\ell}\right] \left\{ \frac{\tau(t)}{(1 - \tau(t))^m} \Phi(t)^n \right\} = \frac{e^{s+\frac{m}{2}-\frac{\ell t_0}{b-1}}(s/m)^{m/2}}{2\sqrt{s\pi}[(b-2)!]^s} \left\{ 1 + O(\frac{1}{s^{1/2}}) \right\}\, . \tag{3.236}$$

♢

Ce lemme offre la possibilité de faire le saut entre l'énumération exacte de nos structures décomposables en chaînes et leur énumération asymptotique.

### 3.3.3 Énoncé du théorème d'énumération asymptotique

**Proposition 3.3.4.** *Le coefficient* $[z^n]\, H_\ell \circ T(z)$ *du SGE des composantes d'excès* $\ell$ *admet l'équivalent asymptotique suivant*

$$[z^n]\, H_\ell \circ T(z) = \tag{3.237}$$

$$3\ell A_{\ell,3\ell} \frac{(b-1)}{n} \frac{e^{s+\frac{3\ell+1}{2}-\frac{\ell}{b-1}} (\frac{s}{3\ell+1})^{\frac{3\ell+1}{2}}}{2\sqrt{s\pi}[(b-2)!]^s} \left\{1 + O(\frac{1}{s^{1/2}})\right\}, \tag{3.238}$$

*avec* $n = n(s) = s(b-1) - \ell$, *où* $s$ *est le nombre d'hyperarêtes.*

*Preuve.* Par le théorème 3.3.2 , l'équivalent asymptotique du coefficient recherché est porté par

$$I = 3\ell A_{\ell,3\ell} \frac{(b-1)}{n} \left[t^{n+\ell}\right] \left\{ \frac{\tau(t)}{(1-\tau(t))^{3\ell+1}} \Phi(t)^n \right\}, \tag{3.239}$$

avec $\tau(t) = t^{b-1}/(b-2)!$ et $\Phi(t) = \exp(\tau(t)/(b-1))$. Et par le lemme 3.3.3 , avec $m = 3\ell + 1$ , nous obtenons l'équivalent asymptotique de $I$ suivant :

$$I = 3\ell A_{\ell,3\ell} \frac{(b-1)}{n} \frac{e^{s+\frac{3\ell+1}{2}-\frac{\ell}{b-1}} (\frac{s}{3\ell+1})^{\frac{3\ell+1}{2}}}{2\sqrt{s\pi}[(b-2)!]^s} \left\{1 + O(\frac{1}{s^{1/2}})\right\}. \tag{3.240}$$

◇

**Théorème 3.3.5.** *Le nombre* $n!\,[z^n]\, H_\ell \circ T(z)$ *des composantes complexes d'excès* $\ell$ *, ayant* $s$ *hyperarêtes et* $n = n(s) = s(b-1) - \ell$ *sommets, admet l'équivalent asymptotique suivant :*

$$n!\,[z^n]\, H_\ell \circ T(z) = \tag{3.241}$$

$$\frac{3\ell A_{\ell,3\ell}(b-1)(\frac{es}{3\ell+1})^{\frac{3\ell+1}{2}}}{\sqrt{2sn}e^{sb-2s+\ell/(b-1)}} \frac{[s(b-1)]^{s(b-1)-\ell}}{[(b-2)!]^s} \left\{1 + O(\frac{1}{\sqrt{s}})\right\} \tag{3.242}$$

$A_{\ell,3\ell}$ *défini dans la forme (2.22) de la SGE* $H_\ell$ *, se détermine par le théorème 2.3.37.*

*Preuve.* Nous avons par le théorème 3.3.2 et par le lemme 3.3.3 avec $m = 3\ell + 1$

$$n!\,[z^n]\,H_\ell \circ T(z) = \tag{3.243}$$

$$= 3\ell A_{\ell,3\ell}(b-1)(n-1)!\frac{e^{s+\frac{3\ell+1}{2}-\frac{\ell}{b-1}}\frac{s}{3\ell+1}^{\frac{3\ell+1}{2}}}{2\sqrt{s\pi}[(b-2)!]^s}\left\{1+O(\frac{1}{\sqrt{s}})\right\} = \tag{3.244}$$

$$= 3\ell A_{\ell,3\ell}(b-1)\frac{(\frac{n}{e})^n\sqrt{2\pi n}}{n}\frac{e^{s-\frac{\ell}{b-1}}(\frac{es}{3\ell+1})^{\frac{3\ell+1}{2}}}{2\sqrt{s\pi}[(b-2)!]^s}\left\{1+O(\frac{1}{\sqrt{s}})\right\} = \tag{3.245}$$

$$= 3\ell A_{\ell,3\ell}(b-1)(\frac{n}{e})^n\frac{e^{s-\frac{\ell}{b-1}}(\frac{es}{3\ell+1})^{\frac{3\ell+1}{2}}}{\sqrt{2sn}[(b-2)!]^s}\left\{1+O(\frac{1}{\sqrt{s}})\right\} = \tag{3.246}$$

$$= 3\ell A_{\ell,3\ell}(b-1)(\frac{n}{e})^n\frac{e^{s-\frac{\ell}{b-1}}(\frac{es}{3\ell+1})^{\frac{3\ell+1}{2}}}{\sqrt{2sn}[(b-2)!]^s}\left\{1+O(\frac{1}{\sqrt{s}})\right\} = \tag{3.247}$$

$$= 3\ell A_{\ell,3\ell}(b-1)(\frac{n}{e})^n\frac{e^{-\frac{n}{b-1}}(\frac{es}{3\ell+1})^{\frac{3\ell+1}{2}}}{\sqrt{2sn}[(b-2)!]^s}\left\{1+O(\frac{1}{\sqrt{s}})\right\} = \tag{3.248}$$

$$= 3\ell A_{\ell,3\ell}(b-1)(\frac{n}{e^{(b-2)/(b-1)}})^n\frac{(\frac{es}{3\ell+1})^{\frac{3\ell+1}{2}}}{\sqrt{2sn}[(b-2)!]^s}\left\{1+O(\frac{1}{\sqrt{s}})\right\}. \tag{3.249}$$

Comme pour $(s \to \infty)$

$$n^n = (s(b-1)-\ell)^{s(b-1)-\ell} = [s(b-1)]^{s(b-1)-\ell}\exp(-\ell)\,, \tag{3.250}$$

l'équivalent asymptotique recherché s'écrit

$$n!\,[z^n]\,H_\ell \circ T(z) = \tag{3.251}$$

$$= 3\ell A_{\ell,3\ell}(b-1)\frac{[s(b-1)]^{s(b-1)-\ell}e^{-\ell}}{e^{n(b-2)/(b-1)}}\frac{(\frac{es}{3\ell+1})^{\frac{3\ell+1}{2}}}{\sqrt{2sn}[(b-2)!]^s}\left\{1+O(\frac{1}{\sqrt{s}})\right\} = \tag{3.252}$$

$$= \frac{3\ell A_{\ell,3\ell}(b-1)(\frac{es}{3\ell+1})^{\frac{3\ell+1}{2}}}{\sqrt{2sn}e^{sb-2s+\ell/(b-1)}}\frac{[s(b-1)]^{s(b-1)-\ell}}{[(b-2)!]^s}\left\{1+O(\frac{1}{\sqrt{s}})\right\}. \tag{3.253}$$

◇

Dans ce chapitre, qui est l'enchaînement de l'énumération exacte, des énumérations asymptotiques ont été obtenues avec un recours à l'analyse complexe. Nous soulignons en particulier l'utilisation de la méthode du point col pour toutes les énumérations asymptotiques faites : des hyperarbres enracinés aux hypercycles et aux composantes complexes selon leurs excès. Notons que le chemin que nous avons choisi ici dévoile complètement la "magie" de la formule de transfert donnant à partir des SGEs et de leur forme, le terme asymptotique principal voire l'expansion complète de ses coefficients : en particulier, les contours d'intégration choisis peuvent servir pour expliciter des expansions complètes. En perspective immédiate du travail fait dans ce chapitre est la considération de l'énumération asymptotique des composantes complexes d'excès infini et suffisamment peu denses pour que les preuves données ici restent valides - un tel résultat parait dans [21, 27]. Ainsi, la plupart des travaux faits sur les graphes peut trouver une généralisation par une démarche similaire à celle suivie jusqu'ici : passant par une récurrence des SGEs, identifiant la forme des SGEs, puis par la formule d'inversion de Lagrange et la formule intégrale de Cauchy, établir une expression intégrale des coefficients et en déduire alors une portion du contour capturant la contribution principale de l'intégrale.

Dans le chapitre suivant, nous illustrons par les hypergraphes aléatoires qu'effectivement, les travaux et résultats existant sur les graphes peuvent être généralisés aux hypergraphes et ainsi, revisités et compris via les SGEs et l'analyse complexe pour obtenir des caractéristiques asymptotiques des structures.

# Chapitre 4

# Hypergraphes aléatoires

"On s'appuie de l'histoire. Mais notre histoire n'est pas notre code. Nous devons nous défier de prouver ce qui doit se faire par ce qui s'est fait. Car c'est précisément de ce qui s'est fait que nous nous plaignons" (1788).

Dans ce chapitre, nous adoptons une vision dynamique des hypergraphes : dans la prochaine section, les hyperarêtes d'une composante sont récursivement enlevées jusqu'à ce qu'il n'en reste plus et dans une autre section, les hyperarêtes sont ajoutées une par une jusqu'à l'obtention d'une certaine propriété dans la structure. Notre outil principal pour mener notre étude restent les SGEs (voir [30, 26]) qui permettent par exemple dans [13] d'obtenir diverses caractérisations statistiques.

## 4.1 Hypergraphes et hypercouplage glouton

Cette section est une généralisation aux hypergraphes de résultats de [11] sur l'analyse en moyenne de la performance de l'algorithme glouton de couplage.

**Définition 4.1.1.** Un hypercouplage est une collection d'hyperarêtes deux à deux disjointes.

Le problème de trouver dans un graphe un couplage qui maximise le nombre de ses arêtes est connu en informatique. Ce problème admet une version avec les hypergraphes : *maximum hypercouplage*, consistant à trouver un hypercouplage, ayant un nombre maximum d'hyperarêtes, dans un hypergraphe.

L'algorithme 3 est un algorithme glouton et il fournit un résultat qui, de manière générale, n'est pas optimum, mais présente l'avantage de la rapidité et garantit tout de même que l'hypercouplage qu'il retourne soit au moins maximal à défaut d'être maximum

**Algorithme 3** : Hypercouplage glouton sur hypergraphe

**Entrées** : Une composante.
**Sorties** : Un hypercouplage maximal.
**début**
  Initialiser avec un hypercouplage vide.
  **répéter**
    Choisir aléatoirement une hyperarête à placer dans l'hypercouplage.
    Supprimer cette hyperarête et celles qui lui étaient adjacentes.
  **jusqu'à** *il n'y a plus d'hyperarête*
  **retourner** *l'hypercouplage construit*
**fin**

pour l'hypergraphe. Notons aussi que c'est un algorithme non déterministe car le choix d'une hyperarête à ajouter dans l'hypercouplage est aléatoire.

Dans cette section, nous proposons de faire l'analyse de la performance de cet algorithme glouton en déterminant la taille moyenne des hypercouplages qu'il retourne, soit le nombre moyen d'hyperarêtes contenus dans ces hypercouplages. Une telle analyse à été faite dans [11] pour le cas des graphes. Dans la sous-section qui suit, est présenté le formalisme mathématique pour procéder à cette analyse de l'algorithme glouton.

### 4.1.1 Définitions et notions

Dans cette section, notre but est d'apprécier la variable aléatoire qu'est le nombre d'hyperarêtes contenues dans l'hypercouplage produit par l'algorithme 3 glouton, quand l'hypergraphe de départ est une composante d'excès $\ell$ et ayant un grand nombre $s$ d'hyperarêtes (donc un grand nombre de sommets $n = n(s) = s(b-1) - \ell$). Nous fixons ici les notations pour faire l'analyse :

- $Y_n^{(\ell)}$ : la variable aléatoire correspondant à la taille de l'hypercouplage produit par l'algorithme glouton quand l'entrée de l'algorithme est choisie uniformément parmi les composantes d'excès $\ell$, ayant $n$ sommets et $s$ hyperarêtes tels que $n = n(s) = s(b-1) - \ell$.
- $f_n^{(\ell)}$ : la fonction génératrice de probabilité (FGP) associée à la variable aléatoire $Y_n^{(\ell)}$,

$$f_n^{(\ell)}(z) = \sum_{y \geq 0} P(Y_n^{(\ell)} = y) z^y \,, \tag{4.1}$$

  où
- $z$ dénote la variable liée au nombre d'hyperarêtes d'un hypercouplage.

La dérivée de la fonction génératrice de probabilité (4.1) évaluée en $z = 1$ est l'espérance $E_n^{(\ell)}$ de la variable aléatoire $Y_n^{(\ell)}$, soit du nombre d'hyperarêtes dans un hypercouplage produit par l'algorithme glouton quand l'entrée est choisie uniformément parmi les composantes d'excès $\ell$ ayant $n$ sommets. Comme la fonction de probabilité vaut 1 quand elle est évaluée en $z = 1$, pour étudier la récurrence sur $n$ afin d'obtenir une estimation

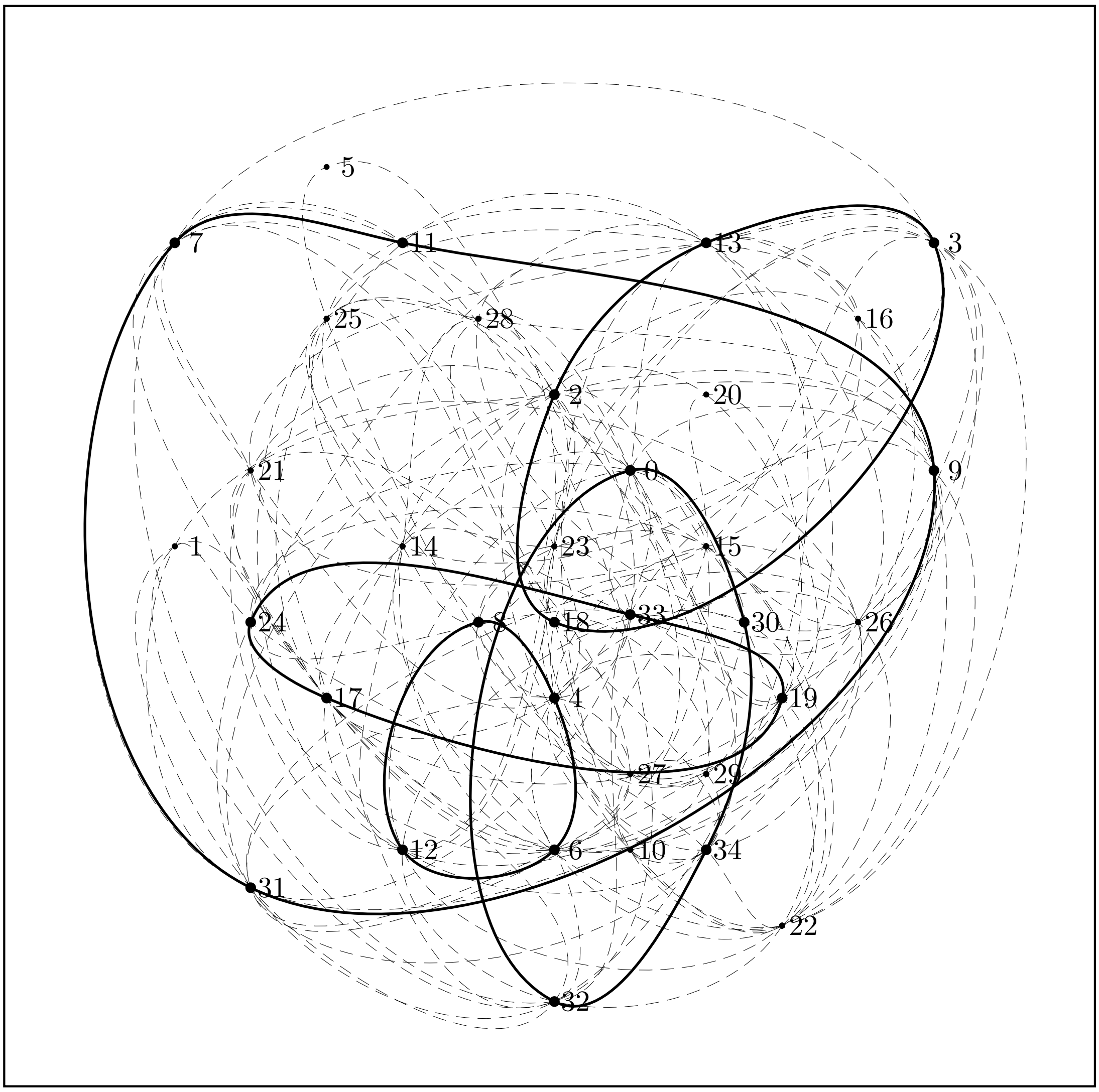

FIG. 4.1 – Une trace d'un déroulement de l'algorithme 3 sur une composante explicitée dans l'annexe.

asymptotique de l'espérance $Y_n^{(\ell)}$, nous introduisons la SGE $G^{(\ell)}$ bivariée

$$G^{(\ell)}(x,z) = \sum_{n\geq 0} c_n^{(\ell)} f_n^{(\ell)}(z) \frac{x^n}{n!}, \tag{4.2}$$

avec

- $c_n^{(\ell)}$ : le nombre de composantes d'excès $\ell$ ayant $n$ sommets,
- $x$ : dénote la variable liée au nombre de sommets de l'hypergraphe d'excès $\ell$ donné en entrée.

Une propriété des FGPs $f_n^{(\ell)}$ nous amène à définir la fonction génératrice de la moyenne :

$$E^{(\ell)}(x) = \frac{\partial}{\partial z} G^{(\ell)}(x,z)\Big|_{z=1} = \sum_{n\geq 0} E_n^{(\ell)} c_n^{(\ell)} \frac{x^n}{n!}\,. \tag{4.3}$$

Nous avons aussi la propriété de la SGE $G(\ell)$ suivante :

$$G^{(\ell)}(x,1) = H_\ell \circ T(x)\,, \tag{4.4}$$

avec $H_\ell \circ T(x)$, la SGE des composantes d'excès $\ell$, et $T(x)$, la SGE des hyperarbres enracinés, dont nous rappelons la définition implicite comme suit :

$$T(x) = x \exp\left(\frac{T(x)^{b-1}}{(b-1)!}\right)\,. \tag{4.5}$$

**Remarque 4.1.2.** Notons la traduction, soit la lecture par rapport aux deux variables, des opérations sur les SGEs (4.2) bivariées dans

$$G^{(j)}(x,z) + G^{(k)}(x,z) = \sum_n (c_n^{(j)} f_n^{(j)}(z) + c_n^{(k)} f_n^{(k)}(z)) \frac{x^n}{n!} \tag{4.6}$$

$$G^{(j)}(x,z) \times G^{(k)}(x,z) = \sum_n \sum_i \binom{n}{i} c_i^{(j)} c_{n-i}^{(k)} f_i^{(j)}(z) f_{n-i}^{(k)}(z) \frac{x^n}{n!}\,. \tag{4.7}$$

La lecture combinatoire pour les sommets étiquetés pour les deux opérations d'addition et de multiplication suit la justification du dictionnaire de [26]. Une lecture combinatoire comme l'union et le produit, est alors immédiate par les propriétés d'une fonction génératrice de probabilité si les variables considérées sont indépendantes. Aussi, pour utiliser les SGEs bivariées $G^{(\ell)}$, il est important de garantir cette indépendance des variables aléatoires considérées dans la décomposition souhaitée.

Dans la prochaine sous-section, nous donnons la décomposition qui nous servira pour l'analyse.

### 4.1.2 Décomposition ou récurrence

Dans le chapitre sur l'énumération exacte, nous avons établi une récurrence des SGEs $H_\ell$ des composantes d'excès $\ell$ en partant de la décomposition suggérée par le marquage d'une hyperarête : c'est la décomposition utilisée par Wright dans [22] pour les graphes. Cette décomposition ne se traduit pas facilement (directement) avec les SGEs bivariées $G^{(\ell)}$ car une fois la décomposition faite, nous perdons la lecture combinatoire dans les composantes dissociées : les variables aléatoires ne correspondent pas aux différentes tailles des hypercouplages produits avec ces composantes en entrée de l'algorithme glouton.

L'algorithme 3 d'hypercouplage glouton suggère une nouvelle décomposition des structures : une composante d'excès $\ell$ ayant une hyperarête marquée, se décompose, non pas seulement en ignorant l'hyperarête mais aussi toutes les hyperarêtes qui lui étaient adjacentes, en des composantes d'excès plus petit. Pour établir une bijection, il nous faut pouvoir recombiner les composantes dissociées, en la composante d'excès $\ell$ de départ. Pour procéder bijectivement à cette recombinaison des composantes dissociées, nous devons marquer un ou plusieurs sommets de ces composantes pour les lier à l'hyperarête marquée en créant d'autres hyperarêtes, chacune pouvant avoir $b_0 = 1, \ldots, b-1$ sommets en commun avec cette hyperarête marquée, $k = 1, \ldots, b - b_0$ sommets (parmi ceux marqués) en commun avec une ou plusieurs composantes dissociées et $b - b_0 - k$ racines d'hyperarbres.

La décomposition est donc descriptible par des ensembles de composantes ayant $k = 1, \ldots, b-1$ sommets marqués. Par exemple, dans le cas où l'excès vaut $\ell = -1$, soit l'analyse de l'algorithme quand en entrée un hyperarbre est choisi uniformément, une telle décomposition se traduit par la proposition qui suit :

**Proposition 4.1.3.** *Le marquage d'une hyperarête d'un hyperarbre se traduit comme suit :*

$$\frac{1}{b-1}\left(-H_{-1} \circ T(x) + x\frac{\mathrm{d}}{\mathrm{d}\,x}H_{-1} \circ T(x)\right) = \frac{x^b}{b!}\exp\left(b\frac{T(x)^{b-1}}{(b-1)!}\right) . \tag{4.8}$$

**Proposition 4.1.4.** *Pour tout $\ell \geq 0$, la SGE $H_\ell$ des composantes d'excès $\ell$ satisfait la relation :*

$$\frac{1}{b-1}\left(\ell H_\ell \circ T(x) + x\frac{\mathrm{d}}{\mathrm{d}\,x}H_\ell \circ T(x)\right) = \frac{x^b}{b!}\exp\left(b\frac{T(x)^{b-1}}{(b-1)!}\right) \times \tag{4.9}$$

$$\times \left[\mathrm{Cyc}^{\ell+1}\right] \exp\left(\sum_{j=0}^{\ell}\sum_{k=1}^{b-1}\mathrm{Cyc}^{j+k}\beta_k(\mathrm{Cyc},x)x^k\frac{\mathrm{d}^k}{\mathrm{d}\,x^k}H_j \circ T(x) + \right. \tag{4.10}$$

$$\left. + \sum_{k=2}^{b-1}\mathrm{Cyc}^{k-1}\beta_k(\mathrm{Cyc},x)x^k\frac{\mathrm{d}^k}{\mathrm{d}\,x^k}H_{-1} \circ T(x)\right) , \tag{4.11}$$

*avec*

$$\beta_k(\mathrm{Cyc},x) = \tag{4.12}$$

$$\frac{1}{k!}\left(b\frac{T(x)^{b-2}}{(b-2)!}\right)^k \sum_{p=0}^{\ell}\frac{1}{p!}\left(pb\frac{T(x)^{b-2}}{(b-2)!}\mathrm{Cyc}\right)^p + \tag{4.13}$$

$$+\binom{b}{2}\frac{k}{(k-1)!}\left(b\frac{T(x)^{b-2}}{(b-2)!}\right)^{k-1}\left(\frac{T(x)^{b-3}}{(b-3)!}\right)\sum_{p=0}^{\ell-1}\frac{1}{p!}\left(pb\frac{T(x)^{b-2}}{(b-2)!}\right)^p \mathrm{Cyc}^{p+1} + \tag{4.14}$$

$$+\binom{k}{2}\frac{b}{(k-2)!}\left(b\frac{T(x)^{b-2}}{(b-2)!}\right)^{k-2}\left(\frac{T(x)^{b-3}}{(b-3)!}\right)\sum_{p=0}^{\ell-1}\frac{1}{p!}\left(pb\frac{T(x)^{b-2}}{(b-2)!}\right)^p \mathrm{Cyc}^{p+1} + \tag{4.15}$$

$$+\ldots \tag{4.16}$$

Les décompositions données dans ces deux propositions sont valides en notant que $T(x) = \frac{\mathrm{d}}{\mathrm{d}\,x}H_{-1} \circ T(x)$, puis en substituant les $H_\ell \circ T(x)$ par les SGEs $G^{(\ell)}(x,z)$ correspondantes (des dérivées droites en $x$ deviennent des dérivées partielles) à un facteur $z$ près car une hyperarête a été choisie et mise dans l'hypercouplage : la décomposition est en accord avec l'indépendance des variables aléatoires dans les différentes composantes. Nous avons donc :

**Proposition 4.1.5.** *Dans le cas des hyperarbres,*

$$\frac{1}{b-1}\left(-G^{(-1)}(x,z)+x\frac{\partial}{\partial x}G^{(-1)}(x,z)\right)=z\frac{x^b}{b!}\exp\left(b\frac{(x\frac{\partial}{\partial x}G^{(-1)}(x,z))^{b-1}}{(b-1)!}\right). \quad (4.17)$$

Entre autres motivations, afin d'alléger les notations, nous adoptons les notations des opérateurs de marquages suivantes :

**Définition 4.1.6.** Dans des structures, nous traduisons le marquage de $k$ sommets ordonnés par l'opérateur sur des SGEs

$${\vartheta_x}^k = x^k\frac{\partial^k}{\partial x^k}, \quad (4.18)$$

si $x$ désigne la variable liée aux sommets.

En particulier, pour le marquage d'un sommet

$$\vartheta_x = x\frac{\partial}{\partial x}. \quad (4.19)$$

Et nous adopterons cette notation de marquage selon la notation de la variable utilisée : $\vartheta_t = t\frac{\partial}{\partial t}$ ou encore $\vartheta_y = y\frac{\partial}{\partial y}$.

**Définition 4.1.7.** Dans le cas des structures d'excès $\ell$, nous traduisons le marquage d'une hyperarête par l'opérateur sur des SGEs

$${\vartheta_w}^{(\ell)} = \frac{\ell+\vartheta_x}{b-1}, \quad (4.20)$$

si $w$ désigne la variable liée aux hyperarêtes et $x$, celle liée aux sommets.

### 4.1.3 La performance gloutonne sur les hyperarbres

**Théorème 4.1.8.** *La SGE*

$$E^{(-1)}(x) = T(x)\left.\frac{g_{-1}(y)}{(1+y)^{1/(b-1)}}\right|_{y=\frac{T(x)^{b-1}}{(b-2)!}} \quad (4.21)$$

*de la moyenne de la variable aléatoire $Y_n^{(-1)}$ est déterminée par la récurrence*

$$(b-1)(1+y)^{(b-2)/(b-1)}{g_{-1}}'(y) = \frac{1}{b}, \quad (4.22)$$

*par laquelle $g_{-1}(y)$ est uniquement déterminée avec la condition $g_{-1}(0)=0$.*

*Preuve.* En différenciant par rapport à la variable $z$, l'équation de la proposition précédente, puis en fixant la valeur $z = 1$, nous obtenons

$$\frac{1}{b-1}\left(-\frac{\partial}{\partial z}G^{(-1)}(x,z) + \vartheta_x \frac{\partial}{\partial z}G^{(-1)}(x,z)\right)\Bigg|_{z=1} = \tag{4.23}$$

$$\left\{\frac{x^b}{b!}\exp\left(b\frac{(\vartheta_x G^{(-1)}(x,z))^{b-1}}{(b-1)!}\right) + \right. \tag{4.24}$$

$$+z\left(b\frac{(\vartheta_x G^{(-1)}(x,z))^{b-2}}{(b-2)!}\right)\left(\vartheta_x \frac{\partial}{\partial z}G^{(-1)}(x,z)\right) \times \tag{4.25}$$

$$\left.\times \frac{x^b}{b!}\exp\left(b\frac{\vartheta_x G^{(-1)}(x,z)^{b-1}}{(b-1)!}\right)\right\}\Bigg|_{z=1}, \tag{4.26}$$

soit

$$\vartheta_w{}^{(-1)}E^{(-1)}(x) = \tag{4.27}$$

$$= \frac{x^b}{b!}\exp\left(b\frac{T(x)^{b-1}}{(b-1)!}\right)\left(1 + \left(b\frac{T(x)^{b-2}}{(b-2)!}\right)\left(\vartheta_x E^{(-1)}(x)\right)\right) = \tag{4.28}$$

$$= \frac{T(x)^b}{b!}\left(1 + \left(b\frac{T(x)^{b-2}}{(b-2)!}\right)\left(\vartheta_x E^{(-1)}(x)\right)\right), \tag{4.29}$$

soit en passant les occurrences de $E^{(-1)}$ au premier membre

$$b\left(-E^{(-1)}(x) + \left(1 - \frac{T(x)^{2b-2}}{[(b-2)!]^2}\right)\vartheta_x E^{(-1)}(x)\right) = \frac{T(x)^b}{(b-2)!} \tag{4.30}$$

et

$$-\frac{E^{(-1)}(x)}{T(x)} + \left(1 - \left(\frac{T(x)^{b-1}}{(b-2)!}\right)^2\right)\frac{\vartheta_x E^{(-1)}(x)}{T(x)} = \frac{1}{b}\frac{T(x)^{b-1}}{(b-2)!}. \tag{4.31}$$

Déterminons alors $E^{(-1)}(x)$ sous la forme

$$E^{(-1)}(x) = T(x)\frac{g_{-1}(y)}{(1+y)^{1/(b-1)}}\Bigg|_{y=\frac{T(x)^{b-1}}{(b-2)!}}. \tag{4.32}$$

Notons que pour une fonction $f$, si $\tau(t) = t^{b-1}/(b-2)!$,

$$\vartheta_x\left(T(x) f\circ\tau\circ T(x)\right) = \tag{4.33}$$

$$= (\vartheta_x T(x)) \times \left(f\circ\tau\circ T(x) + T(x)\times\tau'\circ T(x)\times f'\circ\tau\circ T(x)\right) = \tag{4.34}$$

$$= T(x)\left(\frac{1+(b-1)\vartheta_y}{1-y}\right)f(y)\Bigg|_{y=\tau\circ T(x)}. \tag{4.35}$$

(4.32) permet le changement de variable

$$\begin{cases} \dfrac{T(x)^{b-1}}{(b-2)!} \to y \\ \dfrac{E^{(-1)}(x)}{T(x)} \to \dfrac{g_{-1}(y)}{(1+y)^{1/(b-1)}} \\ \dfrac{\vartheta_x E^{(-1)}(x)}{T(x)} \to \left(\dfrac{1+(b-1)\vartheta_y}{1-y}\right) \dfrac{g_{-1}(y)}{(1+y)^{1/(b-1)}} . \end{cases} \tag{4.36}$$

qui, si nous le portons dans l'équation (4.31), donne

$$-\frac{g_{-1}(y)}{(1+y)^{1/(b-1)}} + \left(1-y^2\right)\left(\frac{1+(b-1)\vartheta_y}{1-y}\right)\left(\frac{g_{-1}(y)}{(1+y)^{1/(b-1)}}\right) = \frac{y}{b} . \tag{4.37}$$

D'où nous déduisons

$$(b-1)(1+y)^{(b-2)/(b-1)} {g_{-1}}'(y) = \frac{1}{b} . \tag{4.38}$$

♢

La résolution de la récurrence, que définit l'équation différentielle (4.22), est immédiate et nous donne :

**Théorème 4.1.9.** *Dans le cas des hyperarbres, l'expression de la SGE, de la moyenne de la variable aléatoire $Y_n^{(-1)}$, est*

$$E^{(-1)}(x) = \frac{T(x)}{b}\left(1 - \frac{1}{\left(1+\frac{T(x)^{b-1}}{(b-2)!}\right)^{1/(b-1)}}\right) . \tag{4.39}$$

Notons que ce résultat généralise celui de [11] qui est le cas des arbres ($b = 2$), dans quel cas, l'expression de la SGE est

$$E^{(-1)}(x) = \frac{T(x)^2}{2(1+T(x))} . \tag{4.40}$$

Disposons de la SGE $E^{(-1)}(x)$, de la moyenne de la variable aléatoire $Y_n^{(-1)}$, qui s'exprime en la série $T(x)$ des hyperarbres enracinés, nous sommes en mesure de procéder à l'analyse du comportement asymptotique de cette moyenne via le comportement asymptotique du coefficient de la série.

**Théorème 4.1.10.** *L'équivalent asymptotique du coefficient $[x^n]\, E^{(-1)}(x)$ de la SGE liée à la moyenne de la variable aléatoire $Y_n^{(-1)}$ est :*

$$[x^n]\, E^{(-1)}(x) = \frac{e^{n/(b-1)}(1 - b/2^{b/(b-1)})}{nb[(b-2)!]^s\sqrt{2\pi s}} \left\{1 + O(\frac{1}{s^{1/6}})\right\} , \tag{4.41}$$

*avec $s$ le nombre d'hyperarêtes et $n = n(s) = s(b-1)+1$.*

*Preuve.* Notons $\hat{E}^{(-1)}(t)$ la SGE lisse correspondant à $E^{(-1)}(x)$ . Alors, $E^{(-1)}(x) = \hat{E}^{(-1)} \circ T(x)$ et

$$\hat{E}^{(-1)}(t) = \frac{t}{b}\left(1 - \frac{1}{\left(1 + \frac{t^{b-1}}{(b-2)!}\right)^{1/(b-1)}}\right). \tag{4.42}$$

La dérivée de la SGE $\hat{E}^{(-1)}(t)$ est

$$\frac{\mathrm{d}}{\mathrm{d}\,t}\hat{E}^{(-1)}(t) = \frac{1}{b} - \frac{1}{b\,(1+\tau(t))^{1/(b-1)}} + \frac{\tau(t)}{b\,(1+\tau(t))^{b/(b-1)}} \tag{4.43}$$

$$= \frac{1}{b} - \frac{1}{b\,(1+\tau(t))^{b/(b-1)}}, \tag{4.44}$$

avec $\tau(t) = t^{b-1}/(b-2)!$ . Notons alors

$$\hat{\hat{E}}^{(-1)}(y) = \tag{4.45}$$

$$= \frac{1}{b} - \frac{1}{b\,(1+y)^{b/(b-1)}} = \tag{4.46}$$

$$= \left(\frac{1}{b} - \frac{1}{2^{b/(b-1)}b}\right) + \frac{1}{(b-1)2^{1/(b-1)}2^2}(y-1) + \dots . \tag{4.47}$$

Soient $s$ , l'entier correspondant au nombre d'hyperarêtes et $n = n(s) = s(b-1)+1$ , le nombre de sommets, alors la formule d'inversion de Lagrange donne le coefficient

$$[z^n]\,E^{(-1)}(x) = \tag{4.48}$$

$$= \frac{1}{2i\pi n}\oint \hat{\hat{E}}^{(-1)} \circ \tau(t)\frac{\exp(n\tau(t)/(b-1))}{t^{n-1}}\frac{\mathrm{d}\,t}{t} = \tag{4.49}$$

$$= \frac{1}{2i\pi n[(b-2)!]^s}\oint \hat{\hat{E}}^{(-1)} \circ \tau(t)\frac{\exp(n\tau(t)/(b-1))}{(b-1)\tau(t)^s}\frac{\mathrm{d}\,\tau(t)}{\tau(t)}, \tag{4.50}$$

avec un contour d'intégration, suffisamment petit, qui encercle une fois l'origine dans le sens direct. Ainsi,

$$[x^n]\,E^{(-1)}(x) = \tag{4.51}$$

$$= \frac{1}{2i\pi n[(b-2)!]^s}\oint \hat{\hat{E}}^{(-1)}(t)\frac{\exp(nt/(b-1))}{t^s}\frac{\mathrm{d}\,t}{t} = \tag{4.52}$$

$$= \frac{1}{2i\pi n[(b-2)!]^s}\oint \hat{\hat{E}}^{(-1)}(t)\exp(t/(b-1))\frac{\exp(st)}{t^s}\frac{\mathrm{d}\,t}{t}. \tag{4.53}$$

L'expression de la dérivée (4.45) nous indique que l'ordre de grandeur asymptotique recherché est celui de l'intégrale

$$I = \frac{1 - 1/(2^{b/(b-1)})}{2nbi\pi[(b-2)!]^s}\oint \exp(t/(b-1))\frac{\exp(st)}{t^s}\mathrm{d}\,t, \tag{4.54}$$

qui correspond au facteur près à (3.60) . Aussi nous obtenons

$$I = \frac{e^{n/(b-1)}(1 - 1/2^{b/(b-1)})}{nb[(b-2)!]^s\sqrt{2\pi s}} \left\{1 + O(\frac{1}{s^{1/6}})\right\} , \tag{4.55}$$

soit l'ordre de grandeur recherché.

$\diamondsuit$

Connaissant le comportement asymptotique du coefficient de la SGE $E^{(-1)}(x)$ de la moyenne de la variable aléatoire $Y_n^{(-1)}$ et celui de la SGE $H_{-1}\circ T(x)$ , déduite du théorème 3.1.3 , par la définition (4.3) , il découle :

**Théorème 4.1.11.** *La moyenne $E_n^{(-1)}$ de la variable aléatoire $Y_n^{(-1)}$ correspondant à la taille de l'hypercouplage produit par l'algorithme glouton quand l'entrée de l'algorithme est choisie uniformément parmi les hyperarbres ayant n sommets, est telle que*

$$E_n^{(-1)} \sim \left(1 - 1/2^{b/(b-1)}\right) \frac{n}{b} . \tag{4.56}$$

### 4.1.4 La performance gloutonne sur les $\ell$-composantes ($\ell \geq 0$ fixé)

Introduisons, pour l'analyse dans le cas des structures d'excès $\ell \geq 0$ en particulier, les fonctions $F^{(\ell)}(x,z)$ , pour réduire la longueur des formules décrivant des décompositions.

**Définition 4.1.12.** Notons $F^{(\ell)}$ la SGE bivariée

$$F^{(\ell)}(x,z) = \tag{4.57}$$

$$\left[\mathrm{Cyc}^{\ell+1}\right] \exp\left(\sum_{j=0}^{\ell}\sum_{k=1}^{b-1} \mathrm{Cyc}^{j+k}\bar{\beta}_k \circ G^{(-1)}(x,z){\vartheta_x}^k G^{(j)}(x,z) + \right. \tag{4.58}$$

$$\left. + \sum_{k=2}^{b-1} \mathrm{Cyc}^{k-1}\bar{\beta}_k \circ G^{(-1)}(x,z){\vartheta_x}^k G^{(-1)}(x,z)\right) , \tag{4.59}$$

avec

$$\bar{\beta}_k(Z) = \tag{4.60}$$

$$\frac{1}{k!}\left(b\frac{Z^{b-2}}{(b-2)!}\right)^k \sum_{p=0}^{\ell} \frac{1}{p!}\left(pb\frac{Z^{b-2}}{(b-2)!}\mathrm{Cyc}\right)^p + \tag{4.61}$$

$$+\binom{b}{2}\binom{k}{1}\frac{(b-2)}{bZ}\frac{1}{(k-1)!}\left(b\frac{Z^{b-2}}{(b-2)!}\right)^{k-1}\sum_{p=0}^{\ell-1}\frac{1}{p!}\left(pb\frac{Z^{b-2}}{(b-2)!}\right)^p \mathrm{Cyc}^{p+1} + \tag{4.62}$$

$$+\binom{b}{1}\binom{k}{2}\frac{(b-2)}{bZ}\frac{1}{(k-1)!}\left(b\frac{Z^{b-2}}{(b-2)!}\right)^{k-1}\sum_{p=0}^{\ell-1}\frac{1}{p!}\left(pb\frac{Z^{b-2}}{(b-2)!}\right)^p \mathrm{Cyc}^{p+1} + \tag{4.63}$$

$$+\ldots \tag{4.64}$$

Cette définition des SGEs $F^{(\ell)}$ permet de formuler de manière concise les équations des propositions 4.1.3 et 4.1.4 en évaluant l'équation suivante en $z=1$ :

$$\vartheta_w{}^{(\ell)} G^{(\ell)}(w,z) = z\frac{x^b}{b!}\exp\left(b\frac{\vartheta_x G^{(-1)}(x,z)^{b-1}}{(b-1)!}\right) F^{(\ell)}(x,z)\,, \tag{4.65}$$

qui est l'écriture des décompositions relatées dans ces propositions avec les SGEs $G^{(j)}(x,z)$ .

**Théorème 4.1.13.** *Pour $\ell \geq 0$ , la SGE $E^{(\ell)}(x)$ de la performance moyenne de l'algorithme glouton de couplage, quand l'entrée est choisie uniformément parmi les $\ell$-composantes, est telle que*

$$E^{(\ell)}(x) = \frac{1}{T(x)^\ell} g_\ell(y)(1+y)^{\ell/(b-1)}\Big|_{y=T(x)^{b-1}/(b-2)!}\,, \tag{4.66}$$

*avec la fonction $g_\ell$ déterminée, uniquement avec la condition $g_\ell(0)=0$ , par une récurrence*

$$g_\ell{}'(y)(1+y)^{(\ell+b-1)/(b-1)} = J_\ell(y)\,. \tag{4.67}$$

*Preuve.* Nous obtenons, en différenciant (4.65) par rapport à $z$ puis en y fixant alors $z=1$ , une récurrence des $E^{(\ell)}(x)$ :

$$\vartheta_w{}^{(\ell)} E^{(\ell)}(x) = \frac{x^b}{b!}\exp\left(b\frac{T(x)^{b-1}}{(b-1)!}\right)\Bigg\{ \tag{4.68}$$

$$F^{(\ell)}(x,1) + b\vartheta_x E^{(-1)}(x)\frac{T(x)^{b-2}}{(b-2)!}F^{(\ell)}(x,1) + \frac{\partial}{\partial z}F^{(\ell)}(x,z)\Big|_{z=1}\Bigg\}\,, \tag{4.69}$$

soit

$$\vartheta_w{}^{(\ell)} E^{(\ell)}(x) = \tag{4.70}$$

$$\frac{T(x)^b}{b!}F^{(\ell)}(x,1)\left(1 + b\frac{\vartheta_x E^{(-1)}(x)}{T(x)}\tau\circ T(x)\right) + \frac{T(x)^b}{b!}\frac{\partial}{\partial z}F^{(\ell)}(x,z)\Big|_{z=1}\,. \tag{4.71}$$

Soit

$$\bar{F}^{(\ell-1)}(x) = \frac{\partial}{\partial z}F^{(\ell)}(x,z)\Big|_{z=1} - b\frac{\tau\circ T(x)}{T(x)}\vartheta_x E^{(\ell)}(x) \tag{4.72}$$

alors,

$$\vartheta_w{}^{(\ell)} E^{(\ell)}(x) - \frac{(\tau\circ T(x))^2}{b-1}\vartheta_x E^{(\ell)}(x) = \tag{4.73}$$

$$= \frac{T(x)^b}{b!}F^{(\ell)}(x,1)\left(1 + b\frac{\vartheta_x E^{(-1)}(x)}{T(x)}\tau\circ T(x)\right) + \frac{T(x)^b}{b!}\bar{F}^{(\ell-1)}(x) = \tag{4.74}$$

$$= \left(1 + b\frac{\vartheta_x E^{(-1)}(x)}{T(x)}\tau\circ T(x)\right)\vartheta_w{}^{(\ell)} H_\ell\circ T(x) + \frac{T(x)^b}{b!}\bar{F}^{(\ell-1)}(x) = \tag{4.75}$$

$$= \left(1 + b\frac{\vartheta_x E^{(-1)}(x)}{T(x)}\tau\circ T(x)\right)\vartheta_w{}^{(\ell)} H_\ell\circ T(x) + \frac{T(x)^{b-1}}{b!T(x)^\ell}\bar{\bar{F}}^{(\ell-1)}\circ\tau\circ T(x)\,. \tag{4.76}$$

avec la fonction $\bar{\bar{F}}^{(\ell-1)}$ telle que

$$\bar{\bar{F}}^{(\ell-1)} \circ \tau \circ T(x) = T(x)^{\ell+1} \bar{F}^{(\ell-1)}(x) \,. \tag{4.77}$$

Nous obtenons alors

$$\vartheta_w{}^{(\ell)} E^{(\ell)}(x) - \frac{(\tau \circ T(x))^2}{b-1} \vartheta_x E^{(\ell)}(x) = \tag{4.78}$$

$$\left(1 + b \frac{\vartheta_x E^{(-1)}(x)}{T(x)} \tau \circ T(x)\right) \vartheta_w{}^{(\ell)} H_\ell \circ T(x) + \tag{4.79}$$

$$+ \frac{\tau \circ T(x)}{b(b-1)T(x)^\ell} \bar{\bar{F}}^{(\ell-1)} \circ \tau \circ T(x) \,. \tag{4.80}$$

Ainsi, la forme

$$E^{(\ell)}(x) = \frac{1}{T(x)^\ell} g_\ell(y)(1+y)^{\ell/(b-1)} \Bigg|_{y = \frac{T(x)^{b-1}}{(b-2)!}} \tag{4.81}$$

permet le changement de variable

$$\begin{cases} \dfrac{T(x)^{b-1}}{(b-2)!} \to y \\ T(x)^\ell E^{(\ell)}(x) \to g_\ell(y)(1+y)^{\ell/(b-1)} \\ T(x)^\ell \vartheta_x E^{(\ell)}(x) \to \left(\dfrac{-\ell + (b-1)\vartheta_y}{1-y}\right) \left(g_\ell(y)(1+y)^{\ell/(b-1)}\right) \,, \end{cases} \tag{4.82}$$

pour avoir

$$y g_\ell{}'(y)(1+y)^{(\ell+b-1)/(b-1)} = \tag{4.83}$$

$$(1 + by \left(\frac{1 + (b-1)\vartheta_y}{1-y}\right) \frac{g_{-1}(y)}{(1+y)^{1/(b-1)}}) y \bar{J}_\ell(y) + \frac{y}{b(b-1)} \bar{\bar{F}}^{(\ell-1)}(y) \,, \tag{4.84}$$

avec

$$\bar{J}_\ell \circ \tau \circ T(x) = \frac{T(x)^\ell \vartheta_w{}^{(\ell)} H_\ell \circ T(x)}{\tau \circ T(x)} \,. \tag{4.85}$$

Ce qui permet de conclure sur la récurrence des $g_\ell$, donc des $E^{(\ell)}$ du théorème en identifiant

$$J_\ell(y) = (1 + by \left(\frac{1 + (b-1)\vartheta_y}{1-y}\right) \frac{g_{-1}(y)}{(1+y)^{1/(b-1)}}) \bar{J}_\ell(y) + \bar{\bar{F}}^{(\ell-1)}(y)/(b^2 - b) \,. \tag{4.86}$$

$\diamondsuit$

**Proposition 4.1.14.** *La fonction $g_0$, telle que*

$$E^{(0)}(x) = g_0(y) \Bigg|_{y = T(x)^{b-1}/(b-2)!} \,, \tag{4.87}$$

*admet un développement en $y = 1$ qui commence comme suit :*

$$g_0(y) = \frac{(3b-2)(1 - 1/2^{b/(b-1)})}{2^3 b((1-y)^2} + \ldots \tag{4.88}$$

*Preuve.* Pour $\ell = 0$, l'équation (4.86) donne $J_0$ telle que

$$g_0{}'(y) = \frac{J_0(y)}{(1+y)} \,. \tag{4.89}$$

Nous avons

$$J_0(y) = (1 + by\left(\frac{1+(b-1)\vartheta_y}{1-y}\right)\frac{g_{-1}(y)}{(1+y)^{1/(b-1)}})\bar{J}_0(y) + \frac{\bar{\bar{F}}^{(-1)}(y)}{b^2-b} \tag{4.90}$$

et à partir de (4.85),

$$\bar{J}_0(y) = \frac{y}{2(1-y)^2} = \frac{1}{2(1-y)^2} + \dots \tag{4.91}$$

En remarquant que

$$g_{-1}(y) = \frac{1}{b}\left((1+y)^{1/(b-1)} - 1\right) \,, \tag{4.92}$$

nous avons aussi le développement en $y = 1$ suivant :

$$by\left(\frac{1+(b-1)\vartheta_y}{1-y}\right)\frac{g_{-1}(y)}{(1+y)^{1/(b-1)}} = \frac{1-1/2^{b/(b-1)}}{(1-y)} + \dots \tag{4.93}$$

Le premier terme du second membre de (4.90) admet donc un développement en $y = 1$ qui commence comme suit :

$$(1 + by\left(\frac{1+(b-1)\vartheta_y}{1-y}\right)\frac{g_{-1}(y)}{(1+y)^{1/(b-1)}})\bar{J}_0(y) = \frac{1-1/2^{b/(b-1)}}{2(1-y)^3} + \dots \tag{4.94}$$

Voyons maintenant comment commence le développement en $y = 1$ du second terme, à savoir $\bar{\bar{F}}^{(-1)}(y)/(b^2-b)$, du second membre de (4.90). Si $\ell = 0$, dans les définitions (4.77) et (4.72) nous obtenons :

$$\bar{\bar{F}}^{(-1)} \circ \tau \circ T(x) = T(x)\bar{F}^{(-1)}(x) \tag{4.95}$$

$$= T(x)\frac{\partial}{\partial z}F^{(0)}(x,z)\Big|_{z=1} - b\tau \circ T(x)\vartheta_x E^{(0)}(x) \,. \tag{4.96}$$

Dans $T(x)\frac{\partial}{\partial z}F^{(0)}(x,z)\big|_{z=1}$, les termes (des recombinaisons) qui contribuent au coefficient de $(1/(1-y))$ de plus grande puissance dans le développement de $\bar{\bar{F}}^{(-1)}(y)$, en $y = 1$, se déduisent des recombinaisons qui contribuent le plus avec les SGEs univariées. Le passage aux SGEs $G^{(j)}$ bivariées, puis l'application de $\frac{\partial}{\partial z}\,()\,\big|_{z=1}$ ont pour effet de changer exactement un terme ${\vartheta_x}^k H_j \circ T(x)$ en ${\vartheta_x}^k E_j(x)$. Ici, pour $\ell = 0$, ces termes ou encore ces constructions proviennent de

$$T(x)\left(\frac{1}{2}\left(b\frac{\vartheta_x G^{(-1)}(x,z)^{b-2}}{(b-2)!}\right)^2 + \frac{b}{2}\frac{\vartheta_x G^{(-1)}(x,z)^{b-3}}{(b-3)!}\right){\vartheta_x}^2 G^{(-1)}(x,z) \tag{4.97}$$

et sont

$$\left(\frac{1}{2}\,(b\tau\circ T(x))^2+\frac{b(b-2)}{2}\tau\circ T(x)\right)\frac{{\vartheta_x}^2E^{(-1)}(x)}{T(x)}\,. \tag{4.98}$$

Par le changement de variable

$$\frac{T(x)^{b-1}}{(b-2)!}\to y\,, \tag{4.99}$$

nous avons

$$\frac{{\vartheta_x}^2E^{(-1)}(x)}{T(x)}\to\frac{(1-1/2^{b/(b-1)})(b-1)}{b(1-y)^3}+\ldots \tag{4.100}$$

Nous trouvons un développement en $y=1$ suivant :

$$\frac{\bar{\bar{F}}^{(0)}(y)}{b(b-1)}=\frac{(1-1/2^{b/(b-1)})(b-1)}{b(1-y)^3}+\ldots \tag{4.101}$$

Ce qui, avec (4.94) , donnent

$$J_0(y)=\frac{(3b-2)(1-1/2^{b/(b-1)})}{2b(1-y)^3}+\ldots \tag{4.102}$$

donc

$${g_0}'(y)=\frac{(3b-2)(1-1/2^{b/(b-1)})}{2^2b(1-y)^3}+\ldots \tag{4.103}$$

et

$$g_0(y)=\frac{(3b-2)(1-1/2^{b/(b-1)})}{2^3b((1-y)^2}+\ldots \tag{4.104}$$

◇

**Proposition 4.1.15.** *L'équivalent asymptotique du coefficient $[x^n]\,E^{(0)}(x)$ de la SGE liée à la moyenne de la variable aléatoire $Y_n^{(0)}$ est :*

$$[x^n]\,E^{(0)}(x)=\frac{(3b-2)(1-1/2^{b/(b-1)})(e/3)^{3/2}e^s}{2^3b\sqrt{\pi}[(b-2)!]^s}\left\{1+O(\frac{1}{s^{1/2}})\right\}\,, \tag{4.105}$$

*avec $s$ , le nombre d'hyperarêtes, et $n=n(s)=s(b-1)$ .*

*Preuve.* Notons $\bar{E}^{(0)}$ , la SGE lisse associée à la SGE $E^{(0)}$ . Alors,

$$\bar{E}^{(0)}(t)=g_0\circ\tau(t)=\frac{(3b-2)(1-1/2^{b/(b-1)})}{2^3b(1-\tau(t))^2}+\ldots\,. \tag{4.106}$$

et

$$\frac{\mathrm{d}}{\mathrm{d}\,t}\bar{E}^{(0)}(t)=\frac{(3b-2)(1-1/2^{b/(b-1)})}{2^2b(1-\tau(t))^3}\frac{(b-1)\tau(t)}{t}+\ldots\,. \tag{4.107}$$

La formule d'inversion de Lagrange donne

$$[x^n]\,E^{(0)}(x) = \tag{4.108}$$
$$= \frac{(b-1)(3b-2)(1-1/2^{b/(b-1)})}{2^3 bi\pi n} \oint \left(\frac{\tau(t)}{(1-\tau(t))^3} + \ldots\right) \frac{e^{s\tau(t)}}{t^n}\frac{\mathrm{d}\,t}{t} = \tag{4.109}$$
$$= \frac{(b-1)(3b-2)(1-1/2^{b/(b-1)})}{2^3 bi\pi n[(b-2)!]^s(b-1)} \oint \left(\frac{\tau(t)}{(1-\tau(t))^3} + \ldots\right) \frac{e^{s\tau(t)}}{\tau(t)^s}\frac{\mathrm{d}\,\tau(t)}{\tau(t)}. \tag{4.110}$$

Nous en déduisons que

$$[x^n]\,E^{(0)}(x) = \tag{4.111}$$
$$= \frac{(b-1)(3b-2)(1-1/2^{b/(b-1)})}{2^3 bi\pi n[(b-2)!]^s} \oint \left(\frac{1}{(1-t)^3} + \ldots\right) \frac{\exp(st)}{t^s}\mathrm{d}\,t. \tag{4.112}$$

L'asymptotique de ce coefficient est donc porté par

$$\frac{(b-1)(3b-2)(1-1/2^{b/(b-1)})}{2^3 bi\pi n[(b-2)!]^s} \oint \frac{1}{(1-t)^3} \frac{\exp(st)}{t^s}\mathrm{d}\,t. \tag{4.113}$$

Cette expression a été déjà rencontrée au facteur près dans la preuve du lemme 3.3.3. Par ce lemme, nous trouvons avec les valeurs $\ell = 0$ et $m = 3$, l'équivalent asymptotique suivant :

$$\frac{(b-1)(3b-2)(1-1/2^{b/(b-1)})}{2^3 bi\pi n[(b-2)!]^s} \oint \frac{1}{(1-t)^3} \frac{\exp(st)}{t^s}\mathrm{d}\,t = \tag{4.114}$$
$$= \frac{(3b-2)(1-1/2^{b/(b-1)})(e/3)^{3/2}e^s}{2^3 b\sqrt{\pi}[(b-2)!]^s}\left\{1 + O(\frac{1}{s^{1/2}})\right\}. \tag{4.115}$$

Ainsi, nous concluons l'équivalent asymptotique du coefficient :

$$[x^n]\,E^{(0)}(x) = \frac{(3b-2)(1-1/2^{b/(b-1)})(e/3)^{3/2}e^s}{2^3 b\sqrt{\pi}[(b-2)!]^s}\left\{1 + O(\frac{1}{s^{1/2}})\right\}. \tag{4.116}$$

$\diamondsuit$

Nous avons le coefficient $[x^n]\,E^{(0)}(x)$ par cette proposition et l'asymptotique du coefficient $[x^n]\,H_0 \circ T(x)$ par la proposition 3.2.1. Nous trouvons alors par la définition de la SGE $E^{(0)}(x)$ :

**Théorème 4.1.16.** *La moyenne $E_n^{(0)}$ de la variable aléatoire $Y_n^{(0)}$ correspondant à la taille de l'hypercouplage produit par l'algorithme glouton, quand l'entrée de l'algorithme est choisie uniformément parmi les hypercycles à $n$ sommets, est telle que*

$$E_n^{(0)} \sim \frac{(3b-2)(1-1/2^{b/(b-1)})(e/3)^{3/2}}{2b\sqrt{\pi}}s, \tag{4.117}$$

*avec $s$, tel que $n = s(b-1)$.*

**Proposition 4.1.17.** *Pour $\ell \geq 1$ , la fonction $g_\ell$ , telle que*

$$E^{(\ell)}(x) = \frac{1}{T(x)^\ell} g_\ell(y)(1+y)^{\ell/(b-1)} \Big|_{y=T(x)^{b-1}/(b-2)!}, \tag{4.118}$$

*admette un développement en $y = 1$ qui commence comme suit :*

$$g_\ell(y) = \frac{\hat{b}_\ell (b-1)^{2\ell}}{(3\ell+2)(1-y)^{3\ell+2}} + \dots, \tag{4.119}$$

*où $\hat{b}_\ell$ admet la récurrence*

$$\hat{b}_0 = \frac{(3b-2)(1-1/2^{b/(b-1)})}{2^2 b}, \tag{4.120}$$

$$\hat{b}_\ell = \frac{(1-1/2^{b/(b-1)})\lambda_\ell}{2^{\ell/(b-1)}2} + \tag{4.121}$$

$$+ \frac{(3\ell+1)\hat{b}_{\ell-1}2^{(\ell-1)/(b-1)} + 2\sum_{p=0}^{\ell-1} \lambda_{\ell-1-p}\hat{b}_p 2^{p/(b-1)}}{2^{\ell/(b-1)}2}, \tag{4.122}$$

*$\lambda_j$ étant la notation définie dans le théorème 2.3.37.*

*Preuve.* Dans la récurrence (4.83) de la preuve du théorème 4.1.13 , nous remarquons que pour déterminer $E^{(\ell)}$ , il suffit d'intégrer

$$g_\ell{}'(y) = \frac{(1 + by\left(\frac{1+(b-1)\vartheta_y}{1-y}\right)\frac{g_{-1}(y)}{(1+y)^{1/(b-1)}})\bar{J}_\ell(y)}{(1+y)^{(\ell+b-1)/(b-1)}} + \frac{\bar{\bar{F}}^{(\ell-1)}(y)}{b(b-1)(1+y)^{(\ell+b-1)/(b-1)}}. \tag{4.123}$$

Or, dans cette équation, le premier terme du second membre admet un développement en $y = 1$ qui commence comme suit :

$$\frac{(1 + by\left(\frac{1+(b-1)\vartheta_y}{1-y}\right)\frac{g_{-1}(y)}{(1+y)^{1/(b-1)}})\bar{J}_\ell(y)}{(1+y)^{(\ell+b-1)/(b-1)}} = \frac{(1-1/2^{b/(b-1)})\lambda_\ell(b-1)^{2\ell}}{2^{\ell/(b-1)}2(1-y)^{3\ell+3}} + \dots \tag{4.124}$$

car

$$(1 + by\left(\frac{1+(b-1)\vartheta_y}{1-y}\right)\frac{g_{-1}(y)}{(1+y)^{1/(b-1)}}) = \frac{1-1/2^{b/(b-1)}}{(1-y)} + \dots \tag{4.125}$$

et

$$\bar{J}_\ell(y) = \frac{\lambda_\ell(b-1)^{2\ell}}{(1-y)^{3\ell+2}} + \dots, \tag{4.126}$$

avec la notation $\lambda_\ell$ du théorème 2.3.37 . Nous trouvons ensuite que le début du développement en $y = 1$ du second terme du second membre de (4.123) est :

$$\frac{\bar{\bar{F}}^{(\ell-1)}(y)}{b(b-1)(1+y)^{(\ell+b-1)/(b-1)}} = \tag{4.127}$$

$$\frac{(b-1)^{2\ell}\left\{(3\ell+1)\hat{b}_{\ell-1}2^{(\ell-1)/(b-1)} + 2\sum_{p=0}^{\ell-1}\lambda_{\ell-1-p}\hat{b}_p 2^{p/(b-1)}\right\}}{2^{\ell/(b-1)}2(1-y)^{3\ell+3}} + \dots \tag{4.128}$$

Comme dans le lemme 2.3.36 , si une composante $\ell$ subit la décomposition (suppression d'hyperarêtes) suggérée par l'algorithme 3 produisant une famille, indexée par $i$ , de composantes d'excès $j_i$ et ayant $k_i$ marquages (comptés avec l'ordre de multiplicité défini par le nombre d'hyperarêtes supprimées auxquelles un sommet marqué appartenait) alors, en supposant que $(j_i, k_i) \neq (-1, 1)$ , le degré maximum de $(1/(1-y))$ du développement en $y = 1$ d'une SGE liée à la description de telles familles, vaut au plus

$$\sum_i (3j_i + 2k_i) = (3\ell + 3) - \sum_i k_i \,. \tag{4.129}$$

Comme la SGE $\bar{\bar{\bar{F}}}$ , (4.77) , est liée à de telles familles sauf à celle réduite à une composante d'excès $\ell$ et ayant un seul marquage, le degré est au plus, avec $\sum_i k_i = 2$ ,

$$3\ell + 3 - 2 + 2 = 3\ell + 3 \,, \tag{4.130}$$

où le 2 rajouté est à cause d'un facteur en ${\vartheta_x}^k E^{(j)}(x)$ remplaçant un facteur en ${\vartheta_x}^k H_j \circ T(x)$ par la dérivation $\frac{\partial}{\partial z}$ dans (4.72) . Si $b \geq 3$ , nous distinguons deux types de familles telles que $\sum_i k_i = 2$ avec des $k_i$ d'ordre 1 (cette restriction est motivée parce que nous ne sommes intéressés que par le coefficient du terme en $(1/(1-y))$ avec la plus grande puissance) :

– Si les deux marques appartiennent à une même composante, alors les familles sont liées à des constructions décrites dans

$$\begin{aligned} &{C_1}^{(\ell-1)}(x,z) = && (4.131)\\ &T(x)^{\ell+1}\left(\frac{1}{2}\left(b\frac{\vartheta_x G^{(-1)}(x,z)^{b-2}}{(b-2)!}\right)^2 + \frac{b}{2}\frac{\vartheta_x G^{(-1)}(x,z)^{b-3}}{(b-3)!}\right)\times && (4.132)\\ &\times {\vartheta_x}^2 G^{(\ell-1)}(x,z)\,. && (4.133)\end{aligned}$$

– Si les deux marques appartiennent à deux composantes distinctes, alors les familles sont liées à des constructions décrites dans

$$\begin{aligned} &{C_2}^{(\ell-1)}(x,z) = && (4.134)\\ &T(x)^{\ell+1}\left\{\frac{1}{2}\left(b\frac{\vartheta_x G^{(-1)}(x,z)^{b-2}}{(b-2)!}\right)^2 + \frac{b}{2}\frac{\vartheta_x G^{(-1)}(x,z)^{b-3}}{(b-3)!}\right\}\times && (4.135)\\ &\times \sum_{p=0}^{\ell-1} \vartheta_x G^{(p)}(x,z) \times \vartheta_x G^{(\ell-1-p)}(x,z)\,. && (4.136)\end{aligned}$$

D'un coté, l'application de $\frac{\partial}{\partial z}\,()\,|_{z=1}$ à (4.131) nous permet de préciser les constructions contribuant au coefficient de $(1/(1-y))$ portant la plus grande puissance. À partir de (4.131) , nous identifions alors les constructions décrites dans l'expression suivante

$$\left(\frac{1}{2}\,(b\tau \circ T(x))^2 + \frac{b(b-2)}{2}\tau \circ T(x)\right) T(x)^{\ell-1} {\vartheta_x}^2 E^{(\ell-1)}(x)\,. \tag{4.137}$$

Notons par $\hat{b}_j(b-1)^{2j}/(3j+2)$, le coefficient de $(1/(1-y))$ portant la plus grande puissance de $g_j(y)$ dans le développement de ce dernier, ce que nous avons écrit

$$g_j(y) = \frac{\hat{b}_j(b-1)^{2j}}{(3j+2)(1-y)^{3j+2}} + \dots , \tag{4.138}$$

avec

$$\hat{b}_0 = \frac{(3b-2)(1-1/2^{b/(b-1)})}{2^2 b} . \tag{4.139}$$

Ces constructions, identifiées dans (4.137), contribuent donc au coefficient par

$$b(b-1)^{2\ell+1}(3\ell+1)\hat{b}_{\ell-1}2^{(\ell-1)/(b-1)} . \tag{4.140}$$

De l'autre coté, l'application de $\frac{\partial}{\partial z}()|_{z=1}$ à (4.134) nous permet d'obtenir les autres constructions qui permettent de déterminer ce coefficient de $(1/(1-y))$ portant la plus grande puissance. Aussi, nous identifions les constructions décrites dans l'expression suivante :

$$T(x)^{\ell-1}\left(\frac{1}{2}(b\tau \circ T(x))^2 + \frac{b(b-2)}{2}\tau \circ T(x)\right) \times \tag{4.141}$$

$$\times 2\sum_{p=0}^{\ell-1} \vartheta_x E^{(p)}(x) \times \vartheta_x H_{\ell-1-p} \circ T(x) . \tag{4.142}$$

Ces constructions contribuent au coefficient par

$$2b(b-1)^{2\ell+1}\sum_{p=0}^{\ell-1} \lambda_{\ell-1-p}\hat{b}_p 2^{p/(b-1)} , \tag{4.143}$$

où nous adoptons la notation $\lambda_\ell$, du théorème 2.3.37, relative au coefficient du terme en $(1-\tau \circ T(x))^{-3\ell}$ portant l'asymptotique du coefficient de $H_\ell \circ T(x)$ :

$$H_\ell(t) = \frac{1}{t^\ell}\left(\frac{\lambda_\ell(b-1)^{2\ell}}{3\ell}\frac{1}{(1-\tau(t))^{3\ell}} + \dots\right) , \tag{4.144}$$

avec

$$\begin{cases} \lambda_0 = \dfrac{1}{2} , \\ \lambda_1 = \dfrac{5}{8} , \\ \lambda_\ell = \dfrac{3\ell}{2}\lambda_{\ell-1} + \dfrac{1}{2}\displaystyle\sum_{p=1}^{\ell-2}\lambda_p\lambda_{\ell-1-p} , \qquad \text{pour } \ell = 2, 3 \dots \end{cases} \tag{4.145}$$

La contribution de $\hat{\hat{F}}(y)/(b(b-1)(1+y)^{(\ell+b-1)/(b-1)})$ au coefficient est donc

$$\frac{(b-1)^{2\ell}\left\{(3\ell+1)\hat{b}_{\ell-1}2^{(\ell-1)/(b-1)} + 2\sum_{p=0}^{\ell-1}\lambda_{\ell-1-p}\hat{b}_p 2^{p/(b-1)}\right\}}{2^{\ell/(b-1)}2} . \tag{4.146}$$

Nous obtenons alors par (4.124) et (4.127) la récurrence

$$\hat{b}_\ell = \frac{(1-1/2^{b/(b-1)})\lambda_\ell}{2^{\ell/(b-1)}2} + \tag{4.147}$$

$$+\frac{(3\ell+1)\hat{b}_{\ell-1}2^{(\ell-1)/(b-1)} + 2\sum_{p=0}^{\ell-1}\lambda_{\ell-1-p}\hat{b}_p 2^{p/(b-1)}}{2^{\ell/(b-1)}2}. \tag{4.148}$$

◊

Ainsi, par exemple, nous trouvons :

**Exemple 4.1.18.** Pour $\ell = 1$,

$$\hat{b}_1 = \frac{(1-1/2^{b/(b-1)})\lambda_1 + 5\hat{b}_0}{2^{1/(b-1)}2}, \tag{4.149}$$

soit

$$\hat{b}_1 2^{1/(b-1)} = (1-1/2^{b/(b-1)})(7b-4)\frac{\lambda_1}{2b}. \tag{4.150}$$

**Exemple 4.1.19.** Et pour $\ell = 2$,

$$\hat{b}_2 2^{2/(b-1)} = \frac{(1-1/2^{b/(b-1)})\lambda_2}{2} + \frac{8\hat{b}_1 2^{1/(b-1)} + 2\lambda_1\hat{b}_0}{2}, \tag{4.151}$$

soit

$$\hat{b}_2 2^{2/(b-1)} = (1-1/2^{b/(b-1)})(65b-34)\frac{\lambda_2}{12b}. \tag{4.152}$$

**Exemple 4.1.20.** Pour $\ell = 3$,

$$\hat{b}_3 2^{3/(b-1)} = (1-1/2^{b/(b-1)})(293b-144)\frac{11\lambda_3}{442b}. \tag{4.153}$$

**Exemple 4.1.21.** Pour $\ell = 4$,

$$\hat{b}_4 2^{4/(b-1)} = (1-1/2^{b/(b-1)})(7081b-3314)\frac{7\lambda_4}{5424b}. \tag{4.154}$$

**Proposition 4.1.22.** *La récurrence des* $\hat{b}_\ell$,

$$g_\ell(y) = \frac{\hat{b}^\ell (b-1)^{2\ell}}{(3\ell+2)(1-y)^{3\ell+2}} + \dots, \tag{4.155}$$

*s'écrit encore*

$$\frac{2\hat{b}_\ell 2^{\ell/(b-1)}}{(1-1/2^{b/(b-1)})} = \sigma_\ell(b)3\lambda_\ell, \tag{4.156}$$

*avec* $\lambda_\ell$ *, tel que*

$$\begin{cases} \lambda_0 = \frac{1}{2}\,, \\ \lambda_\ell = \frac{1}{2}\lambda_{\ell-1}(3\ell-1) + \frac{1}{2}\sum_{p=0}^{\ell-1}\lambda_p\lambda_{\ell-1-p}\,, \qquad \textit{pour } \ell = 1,2\ldots \end{cases} \tag{4.157}$$

*et*

$$\begin{cases} \sigma_0(b) = \frac{3b-2}{3b}\,, \\ \sigma_\ell(b) = \frac{1}{3} + \frac{1}{2}\lambda_{\ell-1}(3\ell+1)\frac{\sigma_{\ell-1}(b)}{3\lambda_\ell} + \\ \qquad + \frac{1}{2}\sum_{p=0}^{\ell-1}\lambda_{\ell-1-p}\lambda_p\frac{\{\sigma_p(b)+\sigma_{\ell-1-p}(b)\}}{3\lambda_\ell}\,, \qquad \textit{pour } \ell = 1,2\ldots \end{cases} \tag{4.158}$$

**Proposition 4.1.23.** *Le coefficient de la SGE* $E^{(\ell)}(x)$ *qui code la moyenne de la variable aléatoire* $Y_n^{(\ell)}$ *admet l'équivalent asymptotique suivant :*

$$[x^n]\,E^{(\ell)}(x) = \tag{4.159}$$

$$\frac{(b-1)^{2\ell+1}(1-1/2^{b/(b-1)})\sigma_\ell(b)3\lambda_\ell e^{s+\frac{3\ell+3}{2}-\frac{\ell}{b-1}}(s/(3\ell+3))^{(3\ell+3)/2}}{4\sqrt{s\pi}[(b-2)!]^s} \times \tag{4.160}$$

$$\times\left\{1+O(\frac{1}{s^{1/2}})\right\}. \tag{4.161}$$

*Preuve.* Comme la SGE lisse $\hat{E}^{(\ell)}$ correspondant à la SGE $T(x)^\ell E^{(\ell)}(x)$ est telle que

$$\hat{E}(t) = g_\ell \circ \tau(t)(1+\tau(t))^{\ell/(b-1)} = \frac{(b-1)^{2\ell}(1-1/2^{b/(b-1)})\sigma_\ell(b)3\lambda_\ell}{2(3\ell+2)(1-\tau(t))^{3\ell+2}} + \ldots \tag{4.162}$$

et

$$\frac{\mathrm{d}}{\mathrm{d}\,t}\hat{E}(t) = \frac{(b-1)^{2\ell}(1-1/2^{b/(b-1)})\sigma_\ell(b)3\lambda_\ell}{2(1-\tau(t))^{3\ell+3}}\frac{(b-1)\tau(t)}{t} + \ldots \tag{4.163}$$

Par la formule d'inversion de Lagrange, l'asymptotique du coefficient $[x^n]\,E^{(\ell)}(x)$ est porté par celui de

$$I = \frac{(b-1)^{2\ell+1}(1-1/2^{b/(b-1)})\sigma_\ell(b)3\lambda_\ell}{2n}\left[t^{n+\ell}\right]\left\{\frac{\tau(t)}{(1-\tau(t))^m}\Phi(t)^n\right\}, \tag{4.164}$$

avec $\tau(t) = t^{b-1}/(b-2)!$ , $\Phi(t) = \exp(\tau(t)/(b-1))$ et $m = 3\ell+3$ . En notant $n = n(s) =$

$s(b-1)-\ell$, nous obtenons le résultat par le lemme 3.3.3 car

$$I = \tag{4.165}$$
$$= \frac{(b-1)^{2\ell+1}(1-1/2^{\frac{b}{(b-1)}})\sigma_\ell(b)3\lambda_\ell e^{s+\frac{m}{2}-\frac{\ell}{b-1}}(s/m)^{\frac{m}{2}}}{4n\sqrt{s\pi}[(b-2)!]^s}\left\{1+O(\frac{1}{s^{1/2}})\right\} = \tag{4.166}$$
$$= \frac{(b-1)^{2\ell+1}(1-1/2^{b/(b-1)})\sigma_\ell(b)3\lambda_\ell e^{s+\frac{3\ell+3}{2}-\frac{\ell}{b-1}}(s/(3\ell+3))^{(3\ell+3)/2}}{4n\sqrt{s\pi}[(b-2)!]^s} \times \tag{4.167}$$
$$\times \left\{1+O(\frac{1}{s^{1/2}})\right\}. \tag{4.168}$$

♢

Par les asymptotiques des coefficients $[x^n]\,E^{(\ell)}(x)$ de la proposition précédente et $[x^n]\,H_\ell \circ T(x)$ de la proposition 3.3.4 , nous trouvons :

**Théorème 4.1.24.** *La moyenne $E_n^{(\ell)}$ de la variable aléatoire $Y_n^{(\ell)}$ correspondant à la taille de l'hypercouplage produit par l'algorithme glouton, quand l'entrée de l'algorithme est choisie uniformément parmi les hypercycles à $n$ sommets est telle que*

$$E_n^{(\ell)} \sim \left(\frac{\sigma_\ell(b)e(1-1/2^{b/(b-1)})(\frac{3\ell+1}{3\ell+3})^{\frac{3\ell+1}{2}}}{2(\ell+1)}\right)s\,, \tag{4.169}$$

*avec $s$, le nombre d'hyperarêtes quand $n$ est le nombre de sommets, soit $n = n(s) = s(b-1)-\ell$, et les $\sigma_\ell$ définis dans l'énoncé de la proposition 4.1.22 .*

## 4.2 Hypergraphes évoluants

Dans cette section nous généralisons un résultat de [14] sur les graphes évoluants. Plusieurs résultats sur les graphes aléatoires [5] seraient aussi accessibles pour hypergraphes par l'approche suggérée dans [14] .

---

**Algorithme 4** : Génération d'hypergraphe jusqu'à l'apparition du premier cycle

---

**Entrées** : Entier $n$, le nombre des sommets étiquetés.
**Sorties** : Forêt d'hyperarbre non enraciné et un cycle avec une hyperarête intérieure marquée.
**début**
  Initialiser avec un hypergraphe à $n$ sommets sans hyperarête.
  **répéter**
    Ajouter une *nouvelle* hyperarête aléatoire
  **jusqu'à** *une composante d'excès $\geq 0$ a été formée*
**fin**
Marquer la dernière hyperarête ajoutée dans l'hypergraphe.
**retourner** *l'hypergraphe*

---

### 4.2.1 Définitions et notions

Dans cette section, nous disposons de la SGE $\hat{H}_{-1}(w,z)$ bivariée des hyperarbres :

$$\hat{H}_{-1}(w,z) = \frac{H_{-1} \circ T(w^{1/(b-1)}z)}{w^{1/(b-1)}}, \tag{4.170}$$

avec

- $z$ : dénote la variable liée au nombre de sommets de l'hypergraphe,
- $w$ : dénote la variable liée au nombre de ses hyperarêtes,
- $T$ : la SGE des hyperarbres enracinés

$$T(z) = z \exp\left(\frac{T(z)^{b-1}}{(b-1)!}\right), \tag{4.171}$$

- $H_{-1}$ : la SGE “lisse” des hyperarbres

$$H_{-1}(t) = t\left(1 - \frac{\tau(t)}{b}\right), \tag{4.172}$$

où $\tau(t) = \frac{t^{b-1}}{(b-2)!}$.

**Définition 4.2.1.** Dénotons $\Psi_n$ l'opérateur

$$\Psi_n F = \sum_{m\geq 1} \frac{f_{m,n}}{m\binom{N}{m}}, \qquad \text{avec } N = \binom{n}{b}, \tag{4.173}$$

et $F(w,z)$ la SGE bivariée

$$F(w,z) = \sum_{n,m\geq 0} f_{m,n} w^m \frac{z^n}{n!}. \tag{4.174}$$

Cet opérateur $\Psi_n$ admet une formulation intégrale suivante :

**Définition 4.2.2.** l'opérateur $\Psi_n$ est

$$\Psi_n F = \int_0^\infty \frac{1}{(1+t)^N} F_n(t) \frac{\mathrm{d}\,t}{t(1+t)}, \tag{4.175}$$

où

$$N = \binom{n}{b} \tag{4.176}$$

et

$$F_n(w) = n!\,[z^n]\,F(w,z) = \frac{n!}{2i\pi} \oint \frac{F(w,z)}{z^{n+1}} \mathrm{d}\,z. \tag{4.177}$$

L'équivalence des deux définitions se justifie, comme souligné dans [14], par la $\beta$ fonction intégrale : La substitution $u = t/(1+t)$ transforme $\int_0^\infty t^{m-1}\mathrm{d}\,t/(1+t)^{N+1}$ en

$$\int_0^1 u^{m-1}(1-u)^{N-m}\mathrm{d}\,u = B(N+1-m,m) = \frac{1}{m\binom{N}{m}}. \tag{4.178}$$

**Définition 4.2.3.** Un hypergraphe évoluant est un hypergraphe dont les hyperarêtes apparaîssent une par une.

Ces hypergraphes évoluants sont à priori différents des hypergraphes où chacune des hyperarêtes existe avec une certaine probabilité $p$ (modèle binomial d'hypergraphe aléatoire étudier dans [10]) .

### 4.2.2 L'attente moyenne de l'apparition du premier cycle

Pour obtenir une expression de l'espérance du nombre d'hyperarêtes ajoutées par l'algorithme 4 , nous considérons la SGE bivariée liées aux structures rencontrées lors du déroulement de l'algorithme telles que le processus de création d'une nouvelle hyperarête continu. Ces structures rencontrées ne sont autres que des forêts d'hyperarbres, chacune de ces structures ayant $m$ hyperarêtes peut apparaître en suivant $m!$ historiques distinctes selon l'ordre de création des $m$ hyperarêtes. Sachant qu'un hypergraphe ayant $n$ sommets admet $m$ hyperarêtes ordonnées, la probabilité qu'il s'agit d'une forêt d'hyperarbres est

$$\frac{m! f_{m,n}}{m! \binom{N}{m}} = \frac{f_{m,n}}{\binom{N}{m}}, \tag{4.179}$$

avec $N = \binom{n}{b}$ et avec $f_{m,n}$ le nombre de forêts d'hyperarbres chacun ayant $n$ sommets et $m$ hyperarêtes. Nous obtenons alors le nombre moyen d'hyperarêtes ajoutées en sommant le terme (4.179) pour $m \geq 0$ :

$$E_n = \sum_{m \geq 0} \frac{f_{m,n}}{\binom{N}{m}}. \tag{4.180}$$

Notons $F$ la SGE des forêts d'hyperarbres

$$F(w, z) = \exp\left(\frac{H_{-1} \circ T(w^{1/(b-1)} z)}{w^{1/(b-1)}}\right) = \sum_{m \geq 0, n \geq 1} f_{m,n} w^m \frac{z^n}{n!}, \tag{4.181}$$

alors l'application de l'opérateur $\Psi_n$ à $F$ donne presque l'expression de la moyenne $E_n$ :

$$\Psi_n F = \sum_{m \geq 1} \frac{f_{m,n}}{m \binom{N}{m}}, \tag{4.182}$$

une expression plus proche de la moyenne $E_n$ est

$$\Psi_n \vartheta_w F = \sum_{m \geq 1} \frac{f_{m,n}}{\binom{N}{m}}. \tag{4.183}$$

Ainsi, nous voyons que la moyenne $E_n$ s'exprime avec l'opérateur $\Psi_n$ comme :

$$E_n = 1 + \Psi_n \vartheta_w F. \tag{4.184}$$

Notons alors que

$$\vartheta_w F(w,z) = F(w,z)\frac{T(w^{1/(b-1)}z)^b}{w^{1/(b-1)}b!} \quad (4.185)$$

$$= \frac{T(w^{1/(b-1)}z)^b}{w^{1/(b-1)}b!}\exp\left(\frac{H_{-1}\circ T(w^{1/(b-1)}z)}{w^{1/(b-1)}}\right). \quad (4.186)$$

Notons $\hat{f}_n(w)$, le coefficient $[z^n]\,\vartheta_w F(w,z)$ :

$$\hat{f}_n(w) = [z^n]\,\frac{T(w^{1/(b-1)}z)^b}{w^{1/(b-1)}b!}\exp\left(\frac{H_{-1}\circ T(w^{1/(b-1)}z)}{w^{1/(b-1)}}\right). \quad (4.187)$$

Une application d'une version de la formule d'inversion de Lagrange 2.3.2, réadaptée à cette SGE bivariée avec une seconde variable relative au nombre d'hyperarêtes, donne

$$\hat{f}_n(w) = \quad (4.188)$$

$$= \frac{1}{n}\left[t^{n-1}\right]\left(\frac{\mathrm{d}}{\mathrm{d}\,t}\left(w\frac{t^b}{b!}\exp\left(\bar{H}_{-1}(w,t)\right)\right)\exp\left(nw\frac{\tau(t)}{b-1}\right)\right) = \quad (4.189)$$

$$= \frac{1}{2in\pi}\oint w\frac{\mathrm{d}}{\mathrm{d}\,t}\left(\frac{t^b}{b!}e^{\bar{H}_{-1}(w,t)}\right)e^{nw\frac{\tau(t)}{b-1}-n\ln(t)}\mathrm{d}\,t, \quad (4.190)$$

avec $\tau(t) = t^{b-1}/(b-2)!$,

$$\bar{H}_{-1}(w,t) = t\left(1 - w\frac{\tau(t)}{b}\right). \quad (4.191)$$

Nous utiliserons une autre formulation de la version de l'inversion de Lagrange qui ne comptabilise pas les hyperarêtes en mettant la variable $w$ du coté des forêts, mais du coté de la structure centrale. Échanger la manière de comptabiliser les hyperarêtes pour les comptabiliser du coté de la structure centrale, se traduira par une multiplication par $w^{n/(b-1)}$, $n/(b-1)$ étant le nombre d'hyperarêtes dans une forêt à $n$ sommets pour garantir la présence d'au moins un cycle, et un changement de $t$ en $tw^{-1/(b-1)}$. Nous obtenons alors une version de l'inversion de Lagrange suivant :

$$\hat{f}_n(w) = \quad (4.192)$$

$$= \frac{w^{\frac{n}{b-1}}}{n}\left[t^{n-1}\right]\left(\frac{\mathrm{d}}{\mathrm{d}\,t}\left(\frac{1}{w^{\frac{1}{(b-1)}}}\frac{t^b}{b!}\exp\left(\bar{H}_{-1}(w,\frac{t}{w^{\frac{1}{b-1}}})\right)\right)e^{n\frac{\tau(t)}{b-1}}\right) = \quad (4.193)$$

$$= \frac{w^{\frac{n-1}{b-1}}}{n}\left[t^{n-1}\right]\left(\frac{\mathrm{d}}{\mathrm{d}\,t}\left(\frac{t^b}{b!}\exp\left(\bar{H}_{-1}(w,\frac{t}{w^{\frac{1}{b-1}}})\right)\right)\exp\left(n\frac{\tau(t)}{b-1}\right)\right). \quad (4.194)$$

De manière analytique, cela s'écrit

$$\hat{f}_n(w) = \frac{w^{\frac{n-1}{b-1}}}{2i\pi n}\times \quad (4.195)$$

$$\times\oint\left(\frac{\mathrm{d}}{\mathrm{d}\,t}\left(\frac{t^b}{b!}\exp\left(\bar{H}_{-1}(w,\frac{t}{w^{\frac{1}{b-1}}})\right)\right)\exp\left(n\frac{\tau(t)}{b-1} - n\ln(t)\right)\right)\mathrm{d}\,t. \quad (4.196)$$

Nous avons

$$\frac{\mathrm{d}}{\mathrm{d}\,t}\left(\frac{t^b}{b!}\exp\left(\bar{H}_{-1}(w,\frac{t}{w^{\frac{1}{b-1}}})\right)\right) = \tag{4.197}$$

$$= \frac{\mathrm{d}}{\mathrm{d}\,t}\left(\frac{t^b}{b!}\exp\left(\frac{t}{w^{\frac{1}{b-1}}}\left(1-\frac{\tau(t)}{b}\right)\right)\right) = \tag{4.198}$$

$$= \left(\frac{t^{b-1}}{(b-1)!} + \frac{t^b}{b!w^{\frac{1}{b-1}}}\left(1-\tau(t)\right)\right)\exp\left(\frac{t}{w^{\frac{1}{b-1}}}\left(1-\frac{\tau(t)}{b}\right)\right) = \tag{4.199}$$

$$= \frac{\tau(t)}{b-1}\left(1+\frac{t}{bw^{\frac{1}{b-1}}}\left(1-\tau(t)\right)\right)\exp\left(\frac{t}{w^{\frac{1}{b-1}}}\left(1-\frac{\tau(t)}{b}\right)\right). \tag{4.200}$$

Et $\hat{f}_n(w)$ se met sous la forme suivante :

$$\hat{f}_n(w) = \frac{w^{\frac{n-1}{b-1}}}{2i\pi n(b-1)}\times \tag{4.201}$$

$$\times \oint \tau(t)\left(1+\frac{t}{bw^{\frac{1}{b-1}}}\left(1-\tau(t)\right)\right)e^{tw^{\frac{-1}{b-1}}\left(1-\frac{\tau(t)}{b}\right)}e^{n\frac{\tau(t)}{b-1}-n\ln(t)}\mathrm{d}\,t\,. \tag{4.202}$$

Pour évaluer $\Psi_n\vartheta_w F$ qui s'écrit sous la forme intégrale suivante :

$$\Psi_n\vartheta_w F = \int_0^\infty \frac{1}{(1+w)^N}F_n(w)\frac{\mathrm{d}\,w}{w(1+w)}\,, \tag{4.203}$$

avec $N = \binom{n}{b}$ et

$$F_n(w) = n!\hat{f}_n(w)\,, \tag{4.204}$$

nous compensons ce facteur $n!$ dans l'estimation de $\hat{f}_n$ par le changement de variable

$$\begin{cases} w \to \dfrac{e^{b-1}}{(wn)^{b-1}} \\ \dfrac{\mathrm{d}\,w}{w} \to -(b-1)\dfrac{\mathrm{d}\,w}{w} \\ n!\hat{f}_n(w) \to n!\hat{f}_n(\dfrac{e^{b-1}}{(wn)^{b-1}}) \end{cases} \tag{4.205}$$

et nous obtenons

$$\Psi_n\vartheta_w F = \int_0^\infty (b-1)n!\frac{\hat{f}_n(\frac{e^{b-1}}{(wn)^{b-1}})}{w(1+\frac{e^{b-1}}{(wn)^{b-1}})^{N+1}}\mathrm{d}\,w\,. \tag{4.206}$$

Dans ce changement de variable (4.205), nous avons

$$\hat{f}_n(\frac{e^{b-1}}{(wn)^{b-1}}) = \frac{e^{n-1}}{2i\pi n^n(b-1)w^{n-1}}\oint g(w,t)e^{nh(t)}\mathrm{d}\,t\,, \tag{4.207}$$

avec

$$g(w,t) = \tau(t)\left(1+\frac{nwt}{eb}\left(1-\tau(t)\right)\right) \tag{4.208}$$

et

$$h(t) = \frac{\tau(t)}{b-1} - \ln(t) + \frac{wt}{e}(1 - \frac{\tau(t)}{b}). \tag{4.209}$$

La dérivée de cette fonction $h$ est

$$h'(t) = \frac{\tau(t)}{t} - \frac{1}{t} + \frac{w}{e}(1 - \tau(t)) = (1 - \tau(t))(\frac{w}{e} - \frac{1}{t}). \tag{4.210}$$

La dérivée seconde de la fonction $h$ est

$$h''(t) = -(b-1)\frac{\tau(t)}{t}(\frac{w}{e} - \frac{1}{t}) + (1 - \tau(t))(\frac{1}{t^2}). \tag{4.211}$$

Nous distinguons alors deux cas :

1. Si $w \leq \frac{e}{[(b-2)!]^{1/(b-1)}}$ , alors la dérivée seconde au point col $t_0 = [(b-2)!]^{1/(b-1)}$ est positive

$$h''(t_0) = -(b-1)\frac{1}{[(b-2)!]^{1/(b-1)}}(\frac{w}{e} - \frac{1}{[(b-2)!]^{1/(b-1)}}). \tag{4.212}$$

   Alors nous faisons passer le contour d'intégration verticalement par le point col $t_0$ .

2. Si $w \geq \frac{e}{[(b-2)!]^{1/(b-1)}}$ , alors la dérivée seconde au point col $t_1 = e/w$ est positive

$$h''(t_1) = +(1 - \frac{e^{b-1}}{w^{b-1}(b-2)!})\frac{w^2}{e^2}. \tag{4.213}$$

   Alors nous faisons passer le contour d'intégration verticalement par le point col $t_1$ .

Dans chaque cas, nous utilisons le contour d'intégration suivant selon $t_j$ :

$$\gamma \ : \begin{cases} \gamma_1 : t = t_j + iv/n\,, & v \uparrow\in [-3nt_j, 3nt_j] \\ \gamma_0 : t = t_j + 3t_j \exp(i\alpha)\,, & \alpha \uparrow\in [\pi/2, 3\pi/2]\,. \end{cases} \tag{4.214}$$

Notons $\hat{\gamma}_1$ , une portion du chemin $\gamma_1$ :

$$\hat{\gamma}_1 \ : t = t_j + iv/n\,, \quad v \uparrow\in [-\kappa_s, \kappa_s]\,, \tag{4.215}$$

avec $\kappa_s$ , un nombre positif qui sera précisé plus tard. Soit alors la valeur $I_1$ de l'intégrale (4.207) restreinte à cette portion $\hat{\gamma}_1$ :

$$I_1 = \frac{e^{n-1}}{2i\pi n^n (b-1) w^{n-1}} \int_{\hat{\gamma}_1} g(w,t) e^{nh(t)} \mathrm{d}\, t\,. \tag{4.216}$$

Ce qui, si l'intégrande est approchée au point col, donne

$$I_1 = \frac{e^{n-1} g(w,t_j) e^{nh(t_j)}}{2i\pi n^n (b-1) w^{n-1}} \int_{\hat{\gamma}_1} \exp\left(nh''(t_j)(t-t_j)^2/2 + O(nh^{(4)}(t_j)(t-t_j)^4\right) \mathrm{d}\, t\,, \tag{4.217}$$

soit, avec la variable d'intégration $v$

$$I_1 = \frac{e^{n-1} g(w,t_j) e^{nh(t_j)}}{2\pi n^{n+1}(b-1)w^{n-1}} \int_{-\kappa_s}^{\kappa_s} \exp\left(-h''(t_j)\frac{v^2}{2n} + O(h^{(4)}(t_j)\frac{v^4}{n^3})\right) \mathrm{d}\, v\,. \tag{4.218}$$

Soit, encore par le changement de variable

$$\begin{cases} v \to \sqrt{\frac{2n}{h''(t_j)}} r \\ \mathrm{d}\, v \to \sqrt{\frac{2n}{h''(t_j)}} \mathrm{d}\, r \end{cases} \tag{4.219}$$

et en notant

$$C_{w,n} = \sqrt{h''(t_j)/(2n)}\,, \tag{4.220}$$

$$I_1 = \tag{4.221}$$

$$= \frac{e^{n-1} g(w,t_j) e^{nh(t_j)}}{\pi n^n (b-1) w^{n-1} \sqrt{2nh''(t_j)}} \int_{-\kappa_s C_{w,n}}^{\kappa_s C_{w,n}} e^{-r^2 + O(\frac{h^{(4)}(t_j)}{h''(t_j)^2}\frac{r^4}{n})} \mathrm{d}\, r = \tag{4.222}$$

$$= \frac{e^{n-1} g(w,t_j) e^{nh(t_j)}}{\pi n^n (b-1) w^{n-1} \sqrt{2nh''(t_j)}} \int_{-\kappa_s C_{w,n}}^{\kappa_s C_{w,n}} e^{-r^2} \left(1 + O(\frac{h^{(4)}(t_j)}{h''(t_j)^2}\frac{r^4}{n})\right) \mathrm{d}\, r = \tag{4.223}$$

$$= \frac{e^{n-1} g(w,t_j) e^{nh(t_j)}}{\pi n^n (b-1) w^{n-1} \sqrt{2nh''(t_j)}} \times \tag{4.224}$$

$$\times \left\{ \int_{-\kappa_s C_{w,n}}^{\kappa_s C_{w,n}} e^{-r^2} \mathrm{d}\, r + O(\frac{h^{(4)}(t_j)}{h''(t_j)^2}\frac{(\kappa_s C_{w,n})^5}{n}) \right\}\,. \tag{4.225}$$

Alors l'intégrale $I_1$ devient

$$I_1 = \frac{e^{n-1} g(w,t_j) e^{nh(t_j)}}{\pi n^n (b-1) w^{n-1} \sqrt{2nh''(t_j)}} \times \tag{4.226}$$

$$\times \left\{ \int_{-\kappa_s C_{w,n}}^{\kappa_s C_{w,n}} e^{-r^2} \mathrm{d}\, r + O(h^{(4)}(t_j)\sqrt{h''(t_j)}\frac{{\kappa_s}^5}{n^{7/2}}) \right\}\,. \tag{4.227}$$

En prenant $\kappa_s = n^{2/3}$, nous obtenons

$$I_1 = \frac{e^{n-1} g(w,t_j) e^{nh(t_j)}}{n^n (b-1) w^{n-1} \sqrt{\pi 2nh''(t_j)}} \left\{ 1 + O(h^{(4)}(t_j)\sqrt{h''(t_j)}\frac{1}{n^{1/6}}) \right\}\,. \tag{4.228}$$

Sur le contour restant, la contribution étant exponentiellement petit, nous obtenons que $I_1$ est l'ordre asymptotique de (4.207) : si $w \leq e/[(b-2)!]^{1/(b-1)}$, alors on prendra $j = 0$ soit le point col $t_0 = [(b-2)!]^{1/(b-1)}$, sinon on prendra $j = 1$ soit le point col $t_1 = e/w$. Et l'ordre asymptotique de (4.207) est

$$\hat{f}_n(\frac{e^{b-1}}{(wn)^{b-1}}) = \frac{e^{n-1} g(w,t_j) e^{nh(t_j)}}{n^n (b-1) w^{n-1} \sqrt{\pi 2nh''(t_j)}} \left\{ 1 + O(h^{(4)}(t_j)\sqrt{h''(t_j)}\frac{1}{n^{1/6}}) \right\}\,. \tag{4.229}$$

Afin d'estimer asymptotiquement l'espérance, nous portons le terme asymptotique principal de l'équation précédente dans (4.206) et nous avons alors à considérer l'intégrale ainsi obtenue suivante :

$$I = \frac{n!e^{n-1}}{n^n\sqrt{\pi 2n}} \int_0^\infty \frac{g(w,t_j)}{(1+\frac{e^{b-1}}{(wn)^{b-1}})^{N+1}} \frac{e^{nh(t_j)-n\ln(w)}}{\sqrt{h''(t_j)}} \mathrm{d}\, w\,, \tag{4.230}$$

avec

$$t_j = \begin{cases} [(b-2)!]^{1/(b-1)} & \text{si } w \leq \frac{e}{[(b-2)!]^{1/(b-1)}} \\ e/w & \text{si } w \geq \frac{e}{[(b-2)!]^{1/(b-1)}}\,, \end{cases} \tag{4.231}$$

$$g(w,t) = \tau(t)\left(1 + \frac{nwt}{eb}\,(1-\tau(t))\right)\,, \tag{4.232}$$

$$h(t) = \frac{\tau(t)}{b-1} - \ln(t) + \frac{w}{e}t(1-\frac{\tau(t)}{b})\,, \tag{4.233}$$

la dérivée seconde de cette dernière étant

$$h''(t) = -(b-1)\frac{\tau(t)}{t}(\frac{w}{e}-\frac{1}{t}) + (1-\tau(t))(\frac{1}{t^2})\,. \tag{4.234}$$

Pour évaluer cette intégrale (4.230), nous distinguerons les deux intervalles $]0,\lambda_0[$ et $]\lambda_0,\infty[$ avec $\lambda_0 = e/((b-2)!)^{1/(b-1)}$. Soit alors $I_a$ la valeur de l'intégrale sur $]0,\lambda_0[$ :

$$I_a = \frac{n!e^{n-1}}{n^n\sqrt{\pi 2n}} \int_0^{\lambda_0} \frac{1}{(1+\frac{e^{b-1}}{(wn)^{b-1}})^{N+1}} \frac{e^{nh(t_j)-n\ln(w)}}{\sqrt{h''(t_j)}} \mathrm{d}\, w\,, \tag{4.235}$$

où

$$h(t_j) = \frac{1}{b-1} - \ln(t_j) + \frac{(b-1)t_jw}{be} = -\frac{b-2}{b-1} + \ln(\lambda_0) + \frac{(b-1)w}{b\lambda_0} \tag{4.236}$$

et

$$h''(t_j) = (b-1)\frac{1}{{t_j}^2}(1-\frac{wt_j}{e}) = (b-1)\left(\frac{\lambda_0}{e}\right)^2(1-\frac{w}{\lambda_0})\,. \tag{4.237}$$

Cette dernière admet une racine en $w=\lambda_0$. Par les équivalents suivants :

$$N+1 \sim n^{b-1}\frac{n}{b!} \tag{4.238}$$

et

$$((1+\frac{e^{b-1}}{(wn)^{b-1}})^{n^{b-1}})^{n/b!} \sim e^{\frac{n}{b!}\left(\frac{e}{w}\right)^{b-1}}\,, \tag{4.239}$$

nous obtenons

$$I_a \sim \frac{n! e^{n-1}}{n^n \sqrt{\pi 2n}} \int_0^{\lambda_0} \frac{e^{-\frac{n}{b!}\left(\frac{e}{w}\right)^{b-1} + nh(t_j) - n\ln(w)}}{\sqrt{h''(t_j)}} \mathrm{d}\,w \tag{4.240}$$

$$\sim \frac{n! e^{n-1} [(b-2)!]^{\frac{1}{(b-1)}} \lambda_0}{n^n (b-1)^{\frac{1}{2}} \sqrt{\pi 2n}} \times \tag{4.241}$$

$$\times \int_0^{\lambda_0} \frac{e^{-\frac{n}{b(b-1)}\left(\frac{\lambda_0}{w}\right)^{b-1} + n\frac{b-1}{b}\frac{w}{\lambda_0} - n\frac{b-2}{b-1} - n\ln\left(\frac{w}{\lambda_0}\right)}}{\sqrt{1 - \frac{w}{\lambda_0}}} \frac{\mathrm{d}\,w}{\lambda_0} \tag{4.242}$$

$$\sim \frac{n! e^{n-1} [(b-2)!]^{\frac{1}{(b-1)}} \lambda_0}{n^n (b-1)^{\frac{1}{2}} \sqrt{\pi 2n}} \int_0^1 \frac{e^{-\frac{n}{b(b-1)}\left(\frac{1}{w}\right)^{b-1} + n\frac{b-1}{b} w - n\frac{b-2}{b-1} - n\ln(w)}}{\sqrt{1-w}} \mathrm{d}\,w \tag{4.243}$$

$$\sim \frac{n! e^n}{n^n (b-1)^{\frac{1}{2}} \sqrt{\pi 2n}} \int_0^1 \frac{e^{n\hat{h}(w)}}{\sqrt{1-w}} \mathrm{d}\,w\,, \tag{4.244}$$

avec $\hat{h}$, définie par

$$\hat{h}(w) = -\frac{1}{b(b-1)}\left(\frac{1}{w}\right)^{b-1} + \frac{b-1}{b} w - \frac{b-2}{b-1} - \ln(w)\,, \tag{4.245}$$

qui s'annule pour $w = 1$. La dérivée de cette fonction est

$$\hat{h}'(w) = \frac{1}{b}\left(\frac{1}{w}\right)^b + \frac{b-1}{b} - \frac{1}{w}\,, \tag{4.246}$$

qui s'annule pour la valeur de $w = 1$. La dérivée seconde de $\hat{h}$ est

$$\hat{h}''(w) = -\left(\frac{1}{w}\right)^{b+1} + \frac{1}{w^2}\,, \tag{4.247}$$

qui s'annule pour la valeur $w = 1$ donc le point col est d'ordre $2$. La dérivée troisième de $\hat{h}$ est

$$\hat{h}'''(w) = (b+1)\left(\frac{1}{w}\right)^{b+2} - \frac{2}{w^3}\,, \tag{4.248}$$

donc $\hat{h}'''(1) > 0$ (1 est un point d'inflexion de $\hat{h}$ qui est croissante) . Notons alors l'intégrale $I'_a$, la restriction de de l'intégrale $I_a$ dans un voisinage de $1$, suivante :

$$I'_a = \frac{n! e^n}{n^n (b-1)^{\frac{1}{2}} \sqrt{\pi 2n}} \int_{1-\epsilon_n}^1 \frac{e^{-n((b-1)(1-w)^3/6 + O((1-w)^4))}}{\sqrt{1-w}} \mathrm{d}\,w \tag{4.249}$$

$$= \frac{n! e^n}{n^n (b-1)^{\frac{1}{2}} \sqrt{\pi 2n}} \int_0^{\epsilon_n} \frac{e^{-n((b-1)w^3/6 + O(w^4))}}{\sqrt{w}} \mathrm{d}\,w \tag{4.250}$$

$$= \frac{n! e^n}{n^n (b-1)^{\frac{1}{2}} \sqrt{\pi 2n}} \int_0^{\epsilon_n} \frac{e^{-n(b-1)w^3/6}}{\sqrt{w}} \left(1 + O(nw^4)\right)\right) \mathrm{d}\,w\,. \tag{4.251}$$

Ceci, par le changement de variable

$$\left\{ \begin{array}{l} w \to \frac{r}{(n(b-1)/6)^{1/3}} \\ \mathrm{d}\, w \to \frac{\mathrm{d}\, r}{(n(b-1)/6)^{1/3}}\,, \end{array} \right. \tag{4.252}$$

donne

$$I'_a \;=\; \frac{n! e^n 6^{\frac{1}{6}}}{n^n (b-1)^{\frac{2}{3}} n^{\frac{1}{6}} \sqrt{\pi 2n}} \int_0^{\epsilon_n (n(b-1)/6)^{1/3}} \frac{e^{-r^3}}{\sqrt{r}} \left(1 + O(\frac{r^4}{n^{1/3}})\right) \mathrm{d}\, r\,. \tag{4.253}$$

En prenant $\epsilon_n = n^{-1/4}$ nous obtenons

$$I'_a = O(\frac{1}{n^{1/6}})\,. \tag{4.254}$$

Nous en déduisons que la valeur asymptotique de $I_a$ est :

$$I_a = O(\frac{1}{n^{1/6}})\,. \tag{4.255}$$

La moyenne recherchée est alors déterminée par l'évaluation de la valeur de l'intégrale (4.230) restreinte à l'intervalle $]\lambda_0, \infty[$. Notons $\hat{I}_a$ , cette intégrale sur cet intervalle :

$$\hat{I}_a = \frac{n! e^{n-1}}{n^n \sqrt{\pi 2n}} \int_{\lambda_0}^{\infty} \frac{g(w, e/w)}{(1 + \frac{e^{b-1}}{(wn)^{b-1}})^{N+1}} \frac{e^{nh(t_j) - n\ln(w)}}{\sqrt{h''(t_j)}} \mathrm{d}\, w\,, \tag{4.256}$$

où

$$g(w, e/w) = \frac{e^{b-1}}{w^{b-1}(b-2)!} \left(1 + \frac{n}{b}\left(1 - \frac{e^{b-1}}{w^{b-1}(b-2)!}\right)\right)\,, \tag{4.257}$$

$$h(t_j) = \frac{e^{b-1}}{w^{b-1} b!} + \ln(w)\,, \tag{4.258}$$

et

$$h''(t_j) = +(1 - \frac{e^{b-1}}{w^{b-1}(b-2)!})(\frac{w^2}{e^2})\,. \tag{4.259}$$

Cette dernière admet une racine en $w = \lambda_0 = e/((b-2)!)^{1/(b-1)}$ . Par les équivalents suivants

$$N + 1 \sim n^{b-1} \frac{n}{b!} \tag{4.260}$$

et

$$((1 + \frac{e^{b-1}}{(wn)^{b-1}})^{n^{b-1}})^{n/b!} \sim e^{\frac{n}{b!}\left(\frac{e}{w}\right)^{b-1}}\,, \tag{4.261}$$

nous obtenons

$$\hat{I}_a \;\sim\; \frac{n! e^{n-1}}{n^n \sqrt{\pi 2n}} \int_{\lambda_0}^{\infty} \frac{g(w, e/w)}{\sqrt{h''(t_j)}} e^{-\frac{n}{b!}\left(\frac{e}{w}\right)^{b-1} + nh(t_j) - n\ln(w)} \mathrm{d}\, w \tag{4.262}$$

$$\sim\; \frac{n! e^{n}}{n^n \sqrt{\pi 2n}} \int_{\lambda_0}^{\infty} \frac{g(w, e/w)}{\sqrt{1 - \frac{e^{b-1}}{w^{b-1}(b-2)!}}} e^{-\frac{n}{b!}\left(\frac{e}{w}\right)^{b-1} + n\frac{e^{b-1}}{w^{b-1} b!}} \frac{\mathrm{d}\, w}{w}\,. \tag{4.263}$$

Ceci, par le changement de variable

$$\begin{cases} w \to \dfrac{e}{(w(b-2)!)^{1/(b-1)}} \\ \dfrac{\mathrm{d}\, w}{w} \to -\dfrac{\mathrm{d}\, w}{(b-1)w} \\ g(w, e/w) \to w\left(1 + \dfrac{n}{b}\,(1-w)\right)\,, \end{cases} \tag{4.264}$$

donne

$$\hat{I}_a \sim \frac{n!e^n}{n^n(b-1)\sqrt{\pi 2n}} \int_0^1 \frac{1 + \frac{n}{b} - \frac{n}{b}w}{\sqrt{1-w}} \mathrm{d}\, w = \frac{n!e^n}{n^n(b-1)\sqrt{\pi 2n}} \left(\frac{6b+2n}{3b}\right)\,. \tag{4.265}$$

Soit,

**Théorème 4.2.4.** *Le nombre moyen d'hyperarêtes au moment de l'apparition du premier cycle dans un hypergraphe évoluant à $n$ sommet est asymptotiquement*

$$\hat{I}_a \sim \frac{2n}{3b(b-1)}\,. \tag{4.266}$$

# Chapitre 5

# Annexe

## 5.1 Preuves des deux identités combinatoires

**Lemme 5.1.1.** *Pour $j, a \in \mathbb{N}^*$ (donc $j + a > 0$),*

$$\frac{1}{\theta^j} = \sum_{i=0}^{j-1} \binom{j+a}{i} \frac{(1-\theta)^{j+a-i}}{\theta^{j-i}} + \sum_{i=0}^{a} \binom{j+a-i-1}{j-1} (1-\theta)^{a-i} \; . \tag{5.1}$$

*Preuve.*

$$\begin{aligned}
\frac{1}{\theta^j} &= \sum_{i=0}^{j-1} \binom{j+a}{i} \frac{(1-\theta)^{j+a-i}}{\theta^{j-i}} + \sum_{i=0}^{a} \binom{j+a-i-1}{j-1} (1-\theta)^{a-i} & (5.2)\\
&= (1-\theta)^a \left(\frac{1-\theta}{\theta}\right)^j \sum_{i=0}^{j+a} \binom{j+a}{i} \left(\frac{\theta}{1-\theta}\right)^i - \sum_{i=j}^{j+a} \binom{j+a}{i} \frac{(1-\theta)^{j+a-i}}{\theta^{j-i}} + \sum_{i=0}^{a} \binom{j+a-i-1}{j-1} (1-\theta)^{a-i} & (5.3)\\
&= (1-\theta)^a \left(\frac{1-\theta}{\theta}\right)^j \left(1 + \frac{\theta}{1-\theta}\right)^{j+a} - \sum_{i=j}^{j+a} \binom{j+a}{i} \frac{(1-\theta)^{j+a-i}}{\theta^{j-i}} + \sum_{i=0}^{a} \binom{j+a-i-1}{j-1} (1-\theta)^{a-i} & (5.4)\\
&= (1-\theta)^a \left(\frac{1-\theta}{\theta}\right)^j \left(\frac{1}{1-\theta}\right)^{j+a} - \sum_{i=j}^{j+a} \binom{j+a}{i} \frac{(1-\theta)^{j+a-i}}{\theta^{j-i}} + \sum_{i=0}^{a} \binom{j+a-i-1}{j-1} (1-\theta)^{a-i} & (5.5)\\
&= \frac{1}{\theta^j} - \sum_{i=j}^{j+a} \binom{j+a}{i} \frac{(1-\theta)^{j+a-i}}{\theta^{j-i}} + \sum_{i=0}^{a} \binom{j+a-i-1}{j-1} (1-\theta)^{a-i} & (5.6)\\
&= \frac{1}{\theta^j} - (1-\theta)^a \sum_{i=j}^{j+a} \binom{j+a}{i} \frac{(1-\theta)^{j-i}}{\theta^{j-i}} + (1-\theta)^a \sum_{i=0}^{a} \binom{j+a-i-1}{j-1} (1-\theta)^{-i} \; . & (5.7)
\end{aligned}$$

Il suffit de montrer

$$\begin{aligned}
\sum_{i=j}^{j+a} \binom{j+a}{i} \frac{(1-\theta)^{j-i}}{\theta^{j-i}} &= \sum_{i=0}^{a} \binom{j+a-i-1}{j-1} (1-\theta)^{-i} & (5.8)\\
\sum_{k=0}^{a} \binom{j+a}{j+k} \frac{(1-\theta)^{-k}}{\theta^{-k}} &= \sum_{i=0}^{a} \binom{j+a-i-1}{j-1} (1-\theta)^{-i} & (5.9)\\
\sum_{k=0}^{a} \binom{j+a}{j+k} \left(\frac{\theta}{1-\theta}\right)^k &= \sum_{i=0}^{a} \binom{j+a-i-1}{j-1} \left(\frac{1}{1-\theta}\right)^i \; . & (5.10)
\end{aligned}$$

Comme

$$\sum_{i=0}^{a} \binom{j+a-i-1}{j-1} \left(\frac{1}{1-\theta}\right)^{i} = \sum_{i=0}^{a} \binom{j+a-i-1}{j-1} \left(1+\frac{\theta}{1-\theta}\right)^{i} \tag{5.11}$$

$$= \sum_{i=0}^{a} \left\{ \binom{j+a-i-1}{j-1} \sum_{t=0}^{i} \binom{i}{t} \left(\frac{\theta}{1-\theta}\right)^{t} \right\} \tag{5.12}$$

et

$$\binom{j+a}{j+k} = \binom{j+a-k-1}{j-1} + \binom{j+a-k-2}{j} + \binom{j+a-k-2}{j+1} + \cdots + \binom{j+a-k-2}{j+k} \tag{5.13}$$

$$\binom{j+a}{j+k} = \binom{j+a-k-1}{j-1} + \sum_{r=0}^{k} \binom{j+a-k+r-1}{j+r} \tag{5.14}$$

nous obtenons l'identité. ♢

**Lemme 5.1.2.** *Si* $a-j \geq 0$ *alors*

$$\theta^{j} = \sum_{i=0}^{a} \binom{a-j-i-1}{-j-1} (1-\theta)^{a-i} , \tag{5.15}$$

*où si* $k \in \mathbb{N}$ *et* $t \in \mathbb{Z}$ *alors*

$$\binom{t}{k} = \binom{t}{t-k} = \frac{t(t-1)\cdots(t-k+1)}{k!} . \tag{5.16}$$

*Preuve.*

$$\theta^{j} = \sum_{i=0}^{a} \binom{a-j-i-1}{a-i} (1-\theta)^{a-i} \tag{5.17}$$

$$= \sum_{r=0}^{a} \binom{r-j-1}{r} (1-\theta)^{r} \tag{5.18}$$

$$= \sum_{r=0}^{a} \binom{j}{r} (-1)^{r} (1-\theta)^{r} \tag{5.19}$$

$$= \sum_{r=0}^{j} \binom{j}{r} (-1)^{r} (1-\theta)^{r} \tag{5.20}$$

$$= (1-(1-\theta))^{j} . \tag{5.21}$$

♢

## 5.2 L'hypergraphe utilisé pour le déroulement de l'algorithme glouton d'hypercouplage

```
#les 55 hyperaretes de l'hypergraphe 4-uniforme
 27  29  34  10
 22  12  32  30
 29  18  4  27
 22  15  3  18
 18  3  13  2
 4  23  8  33
 26  13  28  3
 32  17  31  1
 11  7  32  21
 6  26  4  22
 18  26  3  33
 21  2  12  23
 6  12  8  4
 4  22  10  19
 19  27  15  20
 15  0  9  26
 32  30  0  34
 20  19  2  0
 17  24  33  19
 13  33  26  25
 7  29  6  17
 14  2  9  6
 34  25  18  8
 28  27  3  26
 28  14  23  12
 17  23  6  21
 26  31  19  29
 21  3  13  16
 9  31  7  11
 17  13  11  19
 32  31  13  24
 0  4  17  14
 4  6  31  27
 13  14  25  16
 10  9  22  13
 0  2  34  31
 9  23  14  22
 11  33  31  7
 23  21  7  4
```

| | | | |
|---|---|---|---|
| 15 | 34 | 8 | 22 |
| 30 | 0 | 12 | 22 |
| 25 | 28 | 9 | 12 |
| 15 | 30 | 2 | 27 |
| 32 | 1 | 24 | 6 |
| 24 | 15 | 6 | 0 |
| 25 | 2 | 8 | 14 |
| 0 | 27 | 18 | 28 |
| 26 | 4 | 15 | 34 |
| 4 | 33 | 7 | 32 |
| 19 | 24 | 15 | 33 |
| 16 | 12 | 34 | 17 |
| 31 | 1 | 9 | 2 |
| 13 | 0 | 16 | 33 |
| 26 | 17 | 21 | 10 |
| 29 | 25 | 2 | 5 |

# Bibliographie

**Hypergraphes aléatoires et algorithmiques**

**Résumé :** Les hypergraphes sont des structures décomposables ou descriptibles donc peuvent être énumérés récursivement. Ici, avec les fonctions génératrices exponentielles, nous obtenons des résultats d'énumérations exactes et asymptotiques des hypergraphes connexes à nombre de sommets et à nombre d'hyperarêtes donnés. Dans un cadre combinatoire, par un raisonnement d'inclusion exclusion, nous aboutissons à un encadrement des nombres des composantes d'hypergraphes : c'est une généralisation de l'encadrement de Wright pour les graphes. Pour obtenir les résultats asymptotiques, la méthode du point col permet, en passant par l'analyse complexe, d'obtenir des démonstrations qui sont au final très lisibles grâce à l'utilisation des fonctions génératrices. Soulignons que nous avons ainsi caractérisé :

- les composantes à nombre de sommets et à nombre d'hyperarêtes donnés par rapport à la taille moyenne d'un hypercouplage aléatoire de ces structures,
- les hypergraphes aléatoires (évoluant hyperarête par hyperarête) par rapport au nombre moyen d'hyperarêtes pour l'apparition du premier cycle.

Cette thèse laisse envisager la possibilité de mieux connaître les phénomènes de seuil avec des hypergraphes, ceci en s'inspirant des lignes de preuves qui s'y trouvent.

---

**Random hypergraphs and algorithmics**

**Abstract :** Hypergraphs are structures that can be decomposed or described; in other words they are recursively countable. Here, we get exact and asymptotic enumeration results on hypergraphs by means of exponential generating functions. The number of hypergraph components is bounded, as a generalisation of Wright inequalities for graphs : the proof is a combinatorial understanding of the structure by inclusion exclusion. Asymptotic results are obtained, proofs are at the end very easy to read thanks to generating functions, through complex analysis by saddle point method. We characterized :

- the components with a given number of vertices and of hyperedges by the expected size of a random hypermatching in these structures.
- the random hypergraphs (evolving hyperedge by hyperedge) according to the expected number of hyperedges when the first cycle appears in the evolving structure.

This work is an open road to further works on random hypergraphs such as threshold phenomenon, for which tools used here seem to be sufficient at first sight.